\documentclass[12pt]{iopart}
\usepackage{color}

\def\ket#1{|#1\rangle}

\usepackage{bm}
\usepackage{amsfonts}
\usepackage{graphicx}
\usepackage{caption}
\usepackage[font={small,it}]{caption}
\usepackage{ifpdf}
\ifpdf
\usepackage{epstopdf}
\fi
\usepackage{multirow}
\usepackage[toc,page]{appendix}
\usepackage{xcolor}
\colorlet{ColorPink}{red!10}
\colorlet{ColorPink}{purple!10}
\colorlet{ColorPink}{magenta!10}
\usepackage{amsfonts, amssymb}
\usepackage{tikz}
\usetikzlibrary{patterns}

\usepackage{fancyhdr}
\usepackage{color}   
\usepackage[linktocpage=true]{hyperref}
\usepackage{hyperref} 
\hypersetup{colorlinks, citecolor=green, filecolor=black, linkcolor=blue, urlcolor=blue }

\expandafter\let\csname equation*\endcsname\relax
\expandafter\let\csname endequation*\endcsname\relax
\usepackage{amsmath}
\numberwithin{equation}{section}
\begin{document}
\title[]
{Entanglement entropy after selective  measurements  in  quantum chains}

\author{Khadijeh Najafi }
\address{ Department of Physics, Georgetown University, 37th and O Sts. NW, Washington, DC 20057, USA}

\author{M.~A.~Rajabpour}
\address{ Instituto de F\'isica, Universidade Federal Fluminense, Av. Gal. Milton Tavares de Souza s/n, Gragoat\'a, 24210-346, Niter\'oi, RJ, Brazil}

\date{\today{}}

\begin{abstract}

We study bipartite post measurement entanglement entropy after selective measurements in  quantum chains. We first study the quantity for the critical systems
that can be described by conformal field theories. We find a connection between post measurement entanglement entropy and the Casimir energy of floating objects. Then
we provide formulas for the post measurement entanglement entropy for open and finite temperature systems. We also comment on the 
Affleck-Ludwig boundary entropy in the context of the post measurement entanglement entropy. Finally, we also provide some formulas regarding modular hamiltonians and entanglement spectrum
in the after measurement systems. After through discussion regarding CFT systems we also provide some predictions regarding massive field theories. We then discuss
a generic method to calculate the post measurement entanglement entropy in the free fermion systems. Using the method we study the post measurement entanglement entropy in the XY spin chain.
We check numerically  the CFT and the massive field theory results in the transverse field Ising chain and the XX model. In particular, we study the post meaurement entanglement
entropy in the infinite, periodic and open critical transverse field Ising chain and the critical XX model. The effect of the temperature and the gap is also discussed in these models.

\end{abstract}
\maketitle
\tableofcontents

\section{Introduction}

 Entanglement entropy of many body systems has been a very useful tool and a fundamental concept in the last three decades in the vast majority of areas of research in physics. It has been studied
 in the context of free field theories \cite{Bombelli1986,Srednicki,Casini2009a}, conformal field theories \cite{Holzey1994,Calabrese2004,Calabrese2009}, holographic theories \cite{Ryu,Ryu2},
 integrable models \cite{Doyon2007,Doyon2009} and many other branches of the condensed matter physics \cite{EE}.
 
 It is a useful concept to classify field theories, especially the massless conformal field theories and ultimately it can be used to extract a lot of information regarding the universality class of
the critical systems. It is now well-known that the bipartite entanglement entropy of the ground states of the quantum systems follow the area-law \cite{Bombelli1986,Srednicki}, for review see \cite{Eisert2010}.
The most famous exception to this law appears in the critical $1+1$ dimensional systems. The bipartite entanglement entropy of the ground state of an infinite critical chain has a logarithmic behaviour
with respect to the size of the subsystem with a coefficient which is dependent on the central charge of the underlying conformal field theory \cite{Holzey1994}. This behavior opened a new way to classify the universality classes of
systems at and near quantum critical points using entanglement entropy \cite{Calabrese2009}. Since the bipartite entanglement entropy of the ground state of the system
does not determine the universality class uniquely, there has been an intense research to  calculate quantities like the entanglement entropy of two disjoint intervals \cite{one dimension MI theory,one dimension MI  numeric}
and the entanglement entropy of excited states \cite{Chico}. Although the bipartite entanglement entropy of the ground state of the quantum chains has been studied thoroughly there are not many studies
regarding tripartite systems. There are few entanglement measures for tripartite systems, such as negativity \cite{Vidal} and localizable entanglement \cite{localizable,localizable E references,Cirac2012}.
Negativity has recently  been the subject of intense studies in the context of many body systems \cite{negativity} and references therein. However, as we will comment in the next section,
because of the nature of the definition of the localizable entanglement it has been very difficult to make progress in that direction. Recently, we introduced a new setup for tripartite systems
which is although intimately related to the localizable entanglement it has the advantage of being calculable \cite{Rajabpour2015b}. The setup which will be further elaborated in the next section is as follows: 
take a many body entangled state and make a partial projective measurement of an observable in part of the system. After the measurement that part of the system
is decoupled from the rest of the system, however, the remaining part still has an entangled state. When the result of the measurement is known  the final state is a pure state 
and we call the measurement "selective measurement". When the result of the measurement is not known the final state is a mixed state and we call the process "non-selective measurement". The goal is  the investigation of the bipartite entanglement entropy in the remaining state.
  
In \cite{Rajabpour2015b}, we studied the post measurement entanglement entropy after selective measurement in the $1+1$ dimensional conformal field theories. It was argued that one can use the conformal field theory techniques as far as
one does the measurement in particular bases, so-called "conformal bases". The conformal bases have been studied intensely in recent years in the context of 
Shannon information \cite{AR2013,Stephan2014,AR2014,AR2015,Alcaraz2016} and formation probabilities \cite{Stephan2013,Najafi,RajabCasimir}. The important result of these studies is that there are some bases that
if one makes the measurement in those bases the final system has a boundary which is conformally invariant and so one can use the techniques of boundary conformal field theory (BCFT) to calculate
the entanglement entropy. The technique used in \cite{Rajabpour2015b} was based on the well-known method of twist operators introduced in \cite{Calabrese2004}. However, this technique is not much useful in those cases that
after the projective measurement the two parts of the remaining region are completely decoupled. In \cite{Rajabpour2016} we introduced a new method of calculation of the entanglement entropy which
has a close connection to the Casimir energy of floating objects. The idea was inspired by the earlier works on the  entanglement entropy  \cite{Holzey1994,Cardy2015} and the Casimir energy of
floating objects \cite{Machta,Kardar}. The method suggests that the  R\'enyi entropy can be considered as the ratio of the Casimir energy of two floating objects on the Reimann surfaces.
Although this connection might have some deep consequences in the study of the entanglement entropy of field theories in this paper we focus on its practical use 
in calculating the post measurement entanglement entropy in conformal and massive quantum field theories. The effect of the measurement on the area-law in higher dimensions has been also studied numerically in \cite{Rajabpourarea}.
It is worth mentioning that the post measurement entanglement entropy setup
has found recently many interesting applications in the study of  quantum teleportation in holography \cite{Takayanagi}. In the same work the authors also study the evolution of the entanglement
entropy after the projective measurement.

In this paper we extend the results of \cite{Rajabpour2015b,Rajabpour2016} and \cite{Rajabpourarea} in few more directions. In the next section, we first define the setup
and fix some notations. In section ~3, we first review the method introduced in \cite{Rajabpour2016}. Using this method we find the post measurement entanglement entropy in different situations
such as, semi-infinite system and finite temperature. We also study the Affleck-Ludwig boundary entropy. We then provide the entanglement Hamiltonian of the post measurement systems in different cases and finally, we discuss 
post measurement entanglement spectrum and entanglement gaps. In section ~4, we make some predictions regarding post measurement entanglement entropy in massive systems. Most of the results
in this section are based on physical arguments and not some concrete mathematical calculations. In section ~5, we provide an efficient method to calculate the post measurement entanglement entropy
in free fermions. Although the method can be used in any dimension in this paper we focus on just $1+1$ dimension. The rest of the article is almost exclusively dedicated to the numerical study of the post measurement entanglement entropy
in the well-known XY chain. The XY-chain provides a perfect laboratory to check numerically the CFT formulas derived in the earlier sections. In section ~6, we provide all the necessary ingredients regarding XY chain including
the partition functions on the annulus and the conformal bases and the conformal configurations. In section ~7, we study throughly the post measurement entanglement entropy in the critical transverse field Ising chain 
as  an especial limit of the XY-chain. Then in section ~8, we focus on the critical XX-chain. The reason that we dedicate two separate sections
for these two models will be clear throughout the paper. In section ~9, we numerically study the gapped Ising chain. In section ~10, we will study numerically the effect of the finite temperature on the post measurement entanglement entropy.
In the section ~11 we will briefly comment on the possible experimental setup to study the post measurement entanglement entropy.
Finally, in the last section, we will conclude the paper with some general remarks about the results and future directions.

\section{Setup and definitions}

Consider   a quantum system in a generic dimension  and divide the system
into two subsystems $D$ and $\bar{D}$. The von Neumann entanglement entropy of $D$ with respect to $\bar{D}$
is defined as follows:
\begin{eqnarray}\label{von Neumann}
S[D,\bar{D}]=-\tr\rho_D\ln\rho_D,
\end{eqnarray}
where $\rho_D$ is the reduced density matrix of the subsystem $D$. There is a generalization of the von Neumann
entanglement entropy  called R\'enyi entropy and is defined as
\begin{eqnarray}\label{Renyi}
S_{\alpha}[D,\bar{D}]=\frac{1}{1-\alpha}\ln\tr\rho_D^{\alpha}.
\end{eqnarray}
The limit $\alpha\to 1$ gives back the von Neumann entropy. Note that when there is no danger of confusion,
we replace $S_{\alpha}[D,\bar{D}]$ with $S_{\alpha}$. The setup of our problem is as follows: 
consider a  quantum system in its ground state and then choose an observable. Finally,  make local projective measurements of the
chosen observable 
in a subsystem $A$ of the total system. Note that $A$ does not need to be a simply connected domain. After partial projective measurement,  the subsystem $A$ gets
disentangled from its complement $\bar{A}$. However, the subsystem $\bar{A}$ has a state which is in principle entangled. If after the projective
measurement we know the outcome then the post measurement state will be a pure state which can have a definite wave function. In this case, we call
the procedure "selective measurement". However, it is quite possible that after partial projective measurement we do not know
exactly the outcome of the measurement. In this case, the system can have different wave functions with different probabilities. In other words
\begin{eqnarray}\label{Rho mixed}
\rho_{ns}[\bar{A}]=\sum p_i |\psi_i\rangle\langle \psi_i|,
\end{eqnarray}
where $p_i$ is the probability of collapsing to the wave function $|\psi_i\rangle$.
The system is in a 
mixed state and we call the procedure "non-selective measurement".

Now divide the subsystem
$\bar{A}$ to two new subsystems $B$ and $\bar{B}$. Note that $B$ and $\bar{B}$ do not need to be connected to each other. We are interested in the entanglement entropy between $B$ and $\bar{B}$. 
When the measurement is selective one is left with a pure state and so one can use von Neumann entanglement entropy as the entanglement measure as before.
However, in the case of non-selective measurement, the situation is  more complicated. Although still, the von Neumann entropy
is an interesting quantity to calculate  it is not a measure of entanglement. There are a few entanglement measures for mixed states, such as 
entanglement witnesses, partial transposition and negativity \cite{Cirac2012}, however, they are all  difficult quantities to calculate. For the non-selective
measurement it is possible to show that 
\begin{eqnarray}\label{non-selective Abol}
\bar{\rho}[\bar{A}]=\tr_A\rho_{ns}[\bar{A}]=\tr_A\rho
\end{eqnarray}
where $\rho$ is the initial density matrix of the total system. Note that $\bar{\rho}[\bar{A}]$ is a mixed state and for the CFTs its entanglement content  is already studied in the context
of the entanglement negatvity in  \cite{negativity}.

The setup defined above is reminiscent of a concept called localizable entanglement, see \cite{EE,localizable E references,Cirac2012}.
It is a useful quantity when one is interested in a tripartite system as our setup. The localizable entanglement
between the two parts $B$ and $\bar{B}$ after doing local projective measurement in the rest of the system $A$ is defined as
\begin{eqnarray}\label{localizable}
E_{loc}(B,\bar{B})=sup_{\mathcal{E}}\sum_i p_i E(\ket{\psi_{i}}_{B\bar{B}}),
\end{eqnarray}
where $\mathcal{E}$ is the set of all possible outcomes $(p_i,E(\ket{\psi_{i}}_{B\bar{B}})$ of the measurements and $E$ is the chosen entanglement measure.
The maximization is done with respect to all the possible  observables to make the quantity independent of the observable.
Because of the maximization over all the possible measurements, the localizable entanglement is a very difficult quantity to calculate \cite{Huang}.
Note that in our setup we take $E()$ to be the von Neumann or the R\'enyi entropy and in principle, we calculate just
$ E(\ket{\psi_{i}}_{B\bar{B}})$ for just one observable.  Consequently knowing $p_i$  in our setup can in principle  provide a lower-bound for the localizable entanglement. 
A complete discussion about this point will appear in a future work \cite{Najafi-nonselective}. Finally note that, as we will discuss in more detail in section 11, in our setup
we do not consider the evolution of the entanglement entropy after selective measurement as it is discussed in \cite{Takayanagi}. Apart from the discussion in section
11 there is also another reason behind this: 
as we discussed in this section one of the motivation of this study  is the definition of a tripartite setup for the entanglement entropy. From this perspective
one can actually forget about projective measurement and talks about conditional entanglement entropy. From this perspective one does not need to worry about the evolution of the system
after projective measurement.

\section{Conformal field theory results: 1+1 dimensions}

The R\'enyi entropy in the Euclidean
 languge can be derived as \cite{Holzey1994,Calabrese2009}:
 \begin{eqnarray}\label{Reimann surfaces}
S_{\alpha}=\frac{1}{1-\alpha}\ln\frac{Z_{\alpha}}{Z_1^{\alpha}},
\end{eqnarray}
where $Z_{\alpha}$ is the partition function of the system on $\alpha$-sheeted surfaces. If the short-range interacting system 
is at the critical point, then it is expected that one can replace $Z_{\alpha}$ of the discrete critical system with
the partition function of the CFT on the $\alpha$-sheeted Riemann surfaces. Then using the CFT techniques one can calculate the entanglement entropy
exactly \cite{Holzey1994,Calabrese2009}.
As we already stated  before, the
bipartite Von Neumann and R\'enyi entropies after partial projective measurements are dependent on both the basis (observable) 
that one chooses to perform the measurement and also to the outcome of the measurement. After partial measurement,  the $A$ part of the system decouples and one is left with the $\bar{A}$ part. In the Euclidean language, 
one can still use the equation
 (\ref{Reimann surfaces}) but with a slit on the $A$ part. Depending on the chosen basis for the measurement and the outcome of the measurement
 the boundary condition on the slit can be different. Consider that the chosen basis and the outcome of the measurement
 are in a way that the induced boundary condition on the slit is conformally invariant. In this particular case, which as we will comment with
 more detail later is a very frequent 
 scenario for quantum critical chains \cite{AR2013,AR2015,Rajabpour2015b,Rajabpour2016}, one can use CFT techniques to calculate the equation (\ref{Reimann surfaces}). 
 Since these particular bases do not destroy the conformal structure of the system we will call them conformal bases. In these
 particular circumstances interestingly one can even go further and calculate the probability of occurrence of particular
 configuration as the result of the projective measurement \cite{Stephan2013,Najafi,RajabCasimir}. We will come back to this point when we discuss
 localizable entanglement \cite{Najafi-nonselective}. In the following sections we will first summarize the results 
 of \cite{Rajabpour2015b} and \cite{Rajabpour2016} for the infinite and the periodic systems. Then using the same
 method as \cite{Rajabpour2016} we will derive  the formula for the post measurement entanglement entropy for the open systems. After presenting the formulas
 for the post measurement entanglement entropy in different conditions we will comment on the entanglement gaps and entanglement Hamiltonians.

\begin{figure} [htb] 
\centering
\begin{tikzpicture}[scale=1]
\draw [fill=white ,ultra thick, pink!50] (0,0) rectangle (9,5);
\draw [fill=magenta!50,  thick, magenta!50] (0.5,4.5) circle [radius=0.4];
\draw [darkgray,ultra thick] (0,2.5) -- (1,2.5);
\draw [fill=white] (1,2.44) rectangle (3,2.64);
\draw [blue!70,ultra thick] (1,2.44) rectangle (3,2.64);
\draw [darkgray,ultra thick] (8,2.5) -- (9,2.5);
\draw [fill=white] (6.5,2.44) rectangle (8,2.64);
\draw [teal!70,ultra thick] (6.5,2.44) rectangle (8,2.64);
\draw [darkgray,ultra thick,dashed] (3,2.5) -- (6.5,2.5);
\node [above right, black] at (0.2,4.25){\large$z$};
\node [above right, black] at (0.25,2.5){\large$\bar{B}$};
\node [above right, black] at (1.7,2.65){\large$A$};
\node [above right, black] at (4.3,2.5){\large$B$};
\node [above right, black] at (6.9,2.65){\large$A$};
\node [above right, black] at (8.2,2.5){\large$\bar{B}$};
\node [above right, black] at (1.7,1.9){\large$s_{1}$};
\node [above right, black] at (4.5,2){\large$l$};
\node [above right, black] at (6.9,1.9){\large$s_{2}$};
\draw [fill=pink!30 ,ultra thick, pink!50] (4.5,-4) circle [radius=2.5];
\draw [ ultra thick,teal!70 ] (4.5,-4) circle [radius=2.5];
\draw [fill=magenta!50,  thick, magenta!50] (2.8,-3) circle [radius=0.4];
\draw [fill=white ] (4.5,-4) circle [radius=1];
\draw [ultra thick,blue!70] (4.5,-4) circle [radius=1];
\node [above right, blue] at (2.5,-3.3){\large$w$};
\draw [ultra thick,->] (4.5,-0.2) -- (4.5,-1.2);
\node [above right, black] at (4.7,-1.){\large$w(z)$};
\draw [black,thick,<->] (4.5,-4) -- (4.5,-3);
\draw [black,thick,<->] (4.5,-4) -- (2,-4);
\node [above right, black] at (4.5,-4){\small$\bf{e^{-h/\alpha}}$};
\node [above right, black] at (2.7,-4){\large$1$};

\end{tikzpicture}
\caption{(Color online) Mapping between different regions. The whole plane with two slits $A$ and a branch cut (dashed line) on $B$ can be mapped to
an annulus  by the conformal map $w_{\alpha}(z)$.  } 
\end{figure}
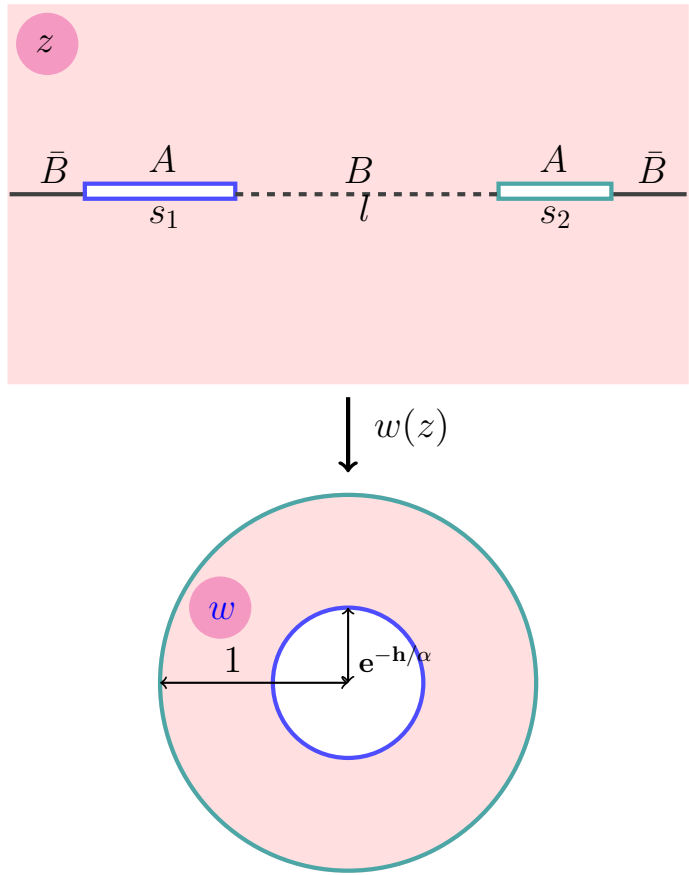

 

\subsection{Entanglement entropy after selective measurements and the Casimir effect}

In this subsection we summarize the results of \cite{Rajabpour2015b,Rajabpour2016} regarding the post measurement
entanglement entropy in the $1+1$-dimensional CFT's. The results concerning the open boundary conditions and finite temperature  are new. For later convenience, 
consider that the measurement 
region $A$ is made of
two disconnected sections with the lengths $s_1$ and $s_2$ and the distance $l$ as it is shown in the figure ~1. The branch cut on $B$ part is needed 
to produce Riemann surfaces. It is quite obvious that this setup is  related to the Casimir energy 
of two slits on the Riemann surfaces. In other words based on (\ref{Reimann surfaces}) to calculate the entanglement entropy one just needs to calculate
the Casimir free energy of two slits on the Riemann surfaces. This simple connection helps us to hire the techniques used in
the study of the Casimir energy to calculate the entanglement entropy. Using the techniques of \cite{Holzey1994,Cardy2015} and \cite{Machta,Kardar} it was shown in
\cite{Rajabpour2016} that one can calculate the partition function on the Riemann surfaces by mapping the system to the annulus. On the annulus, the partition function of the CFT is known so one just needs to  consider  an extra term which comes from the conformal mapping.
The final result is as following \cite{Rajabpour2016}:
\begin{eqnarray}\label{CFT result}
\ln Z_{\alpha}=\ln Z^{geom}_{\alpha}+\ln Z^{annu}_{\alpha},
\end{eqnarray}
where $Z^{annu}_{\alpha}$ is the partition function on the annulus and $Z^{geom}_{\alpha}$ is the geometric term coming from the conformal mapping.
The annulus part of the partition function which is dependent on the full operator content of the CFT  can be written in two equivalent forms as follows\cite{Cardy2004}:
\begin{eqnarray}\label{annulus part1}
\ln Z^{annu}_{\alpha}&=&\ln [q_{\alpha}^{-c/24}(1+\sum_jn_jq_{\alpha}^{\Delta_j})]-c\frac{h}{12\alpha},\\
\label{annulus part2}
\ln Z^{annu}_{\alpha}&=&\ln [\tilde{q}_{\alpha}^{-c/24}(b_0^2+\sum_jb^2_j\tilde{q}_{\alpha}^{\Delta_j})]-c\frac{h}{12\alpha},
\end{eqnarray}
where $n_j$ and $b_j$ are numbers and $\Delta_j$ in the first formula is the boundary scaling dimension and in the second formula is the
bulk scaling dimension. Here $r=e^{-\frac{h}{\alpha}}$ is the inner radius of the annulus. Finally $q_{\alpha}$ and $\tilde{q}_{\alpha}$ are defined as
\begin{eqnarray}\label{q tilde q}
q_{\alpha}=e^{-\pi\frac{2\pi\alpha}{h}},\hspace{1cm}\tilde{q}_{\alpha}=e^{-\frac{2h}{\alpha}}.
\end{eqnarray}
The geometric part of the partition function which is only dependent on the central charge can be written as
\begin{eqnarray}\label{geometric}
\frac{\delta\ln Z^{geom}_{\alpha}}{\delta l}=- \frac{ic}{12\pi}\oint_{\partial S_2}\{w_{\alpha},z\}dz,
\end{eqnarray}
where $w_{\alpha}$ is the conformal map from the original $\alpha$-sheeted Riemann surface with slits to the annulus and
$\{f,z\}=\frac{ f'''}{f'}-\frac{3}{2}(\frac{f''}{f'})^2$ is the Schwartzian derivative and the integral
is around one of the slits (here the second one). Later, for notational convenience, we will also
use $S(f)=\{f,z\}$ for the Schwartzian derivative. Note that the above formulas
are correct even for finite size systems as far as the Riemann surface is topologically equivalent to an annulus.

\subsubsection{Infinite systems:}

This case is already discussed in full detail in the \cite{Rajabpour2016}\footnote{The corresponding conformal map is written explicitly in the Appendix.}. 
When $s_1$ and $s_2$ are much smaller than $l$ and is in the order of the lattice spacing 
one is left with the bipartite entanglement entropy without any measurement. This is the well-known case and it is fully studied in the last two decades,
see for example \cite{Holzey1994,Calabrese2009}. When  $s_2$ is in the order of lattice spacing but $s_1$ and $l$ are macroscopically big
the setup corresponds to the post measurement entanglement entropy of the connected regions $B$ and $\bar{B}$. The formula, in this case, is
\cite{Rajabpour2015b,Rajabpour2016}: 
\begin{eqnarray}\label{rajab2015}
S_{\alpha}=\frac{c}{12}(\frac{1+\alpha}{\alpha})\ln\frac{l(l+s_1)}{s_2s_1}+2\ln b_0+\frac{b_1^2}{b_0^2}(\frac{s_2s_1}{2l(l+s_1)})^{2\Delta_1/\alpha}+ ...,
\end{eqnarray}
where  $\Delta_1$ is the smallest scaling dimension present in the spectrum of the system
and the second term is the  Affleck-Ludwig boundary term \cite{Affleck-Ludwig} studied already in the context of the 
entanglement entropy in the \cite{Calabrese2009}. When we have just one simply connected measurement domain,
we should substitute $2\ln b_0$ with $\ln b_0$. We will discuss this issue in more detail in the later sections.
Note that when $s_1$ goes to infinity the above result goes to the entanglement entropy of a domain at the beginning of a
semi-infinite chain. 
For later use, we also report here the approximate value for $h$  when  $s_2$ is in the order of lattice spacing but $s_1$ and $l$ are macroscopically big
as follows\cite{Rajabpour2016}:

\begin{eqnarray}\label{h infinite connected}
h=-\ln\frac{s_2s_1}{16l(l+s_1)}+....
\end{eqnarray}
Finally, when $s_1$, $s_2$ and   $l$ are all
much bigger than the lattice spacing one is left with the two regions $B$
and $\bar{B}$ that are effectively disconnected. 
For $l\ll s_1=s_2=s$ one can  find \cite{Rajabpour2016}:
\begin{equation}\label{power-law decay1}
{S_{\alpha}}\asymp \left\{
\begin{array}{c l}      
    \frac{1}{\alpha-1} (\frac{l}{8s})^{2\alpha\Delta_1}, & \alpha<1\\
        (\frac{l}{8s})^{2\Delta_1}\ln\frac{8s}{l}, &\alpha=1\\
        \frac{\alpha}{\alpha-1} (\frac{l}{8s})^{2\Delta_1}, & \alpha>1,
\end{array}\right.
\end{equation}
where $\Delta_1$ is the smallest boundary scaling dimension in the spectrum of the system.
The above formula is an example of entanglement entropy of two disconnected regions. For later use we also report the approximate
value of $h$ in the above limit as follows \cite{Rajabpour2016}:
\begin{eqnarray}\label{h infinite disconnected}
h=\frac{\pi^2}{\ln\frac{8s}{l}}+....
\end{eqnarray}

\subsubsection{Finite periodic systems:}

One can follow the above procedure also for a system with the periodic boundary conditions with the total size $L$. The corresponding conformal map
which is already discussed in \cite{Rajabpour2016} can be found in the Appendix A.
As before when $s_1,s_2\ll l$ we have just the bipartite 
entanglement entropy without the projective measurement, see \cite{Holzey1994,Calabrese2009}. The case $s_2\ll l,s_1$ is
the post measurement entanglement entropy of two connected regions $B$ and $\bar{B}$. The first leading term, in this case, has the following
form \cite{Rajabpour2015b,Rajabpour2016}\footnote{The corresponding conformal map is written explicitly in the Appendix A.}:
\begin{eqnarray}\label{SB for PBC}
S_{\alpha}=\frac{c}{12}(1+\frac{1}{\alpha})\ln \Big{(}\frac{L}{\pi}\frac{\sin\frac{\pi}{L}(l+s_1)\sin\frac{\pi}{L}l}{s_2\sin\frac{\pi}{L}s_1}\Big{)}+...,
\end{eqnarray}
where the first important term in the dots is the Affleck-Ludwig term that we will discuss with more details later.
The $h$ in this limit is
\begin{eqnarray}\label{h finite disconnected}
h=- \ln\frac{\pi s_2\sin[\frac{\pi s_1}{L}]}{16L\sin[\frac{\pi l}{L}]\sin[\frac{\pi (l+s_1)}{L}]}+....
\end{eqnarray}
Finally when $l\ll s_1=s_2=s$ (in a way that we have $s=\frac{L-2l}{2}$) one can derive the following formula \cite{Rajabpour2016}:
\begin{equation}\label{power-law decay 2}
{S_{\alpha}}\asymp\left\{
\begin{array}{c l}      
    \frac{1}{\alpha-1} (\frac{\pi l}{4L})^{4\alpha\Delta_1}, & \alpha<1\\
        (\frac{\pi l}{4L})^{4\Delta_1}\ln\frac{\pi l}{4L}, &\alpha=1\\
        \frac{\alpha}{\alpha-1} (\frac{\pi l}{4L})^{4\Delta_1}, & \alpha>1,
\end{array}\right.
\end{equation}
The above formula is the second example of the post measurement entanglement entropy of two disconnected regions. The value
of $h$ in this limit is \cite{Rajabpour2016}
\begin{eqnarray}\label{h finite disconnected2}
h=\frac{-\pi^2}{2\ln\frac{\pi l}{4L}}+....
\end{eqnarray}

\subsubsection{Semi-infinite open systems:} 

This case has not been  addressed in the previous works. The  setup that we would like to study is
shown in the Figure ~2. As before the projective measurement is done on the $A$ part, and we would like to 
calculate the  entanglement entropy of $B$ with respect to $\bar{B}$. To derive the R\'enyi entropy one needs to calculate the partition
function of the Riemann surfaces shown in the Figure ~2.

\begin{figure} [htb] \label{fig2}
\centering
\begin{tikzpicture}[scale=1]
\draw [fill=pink!30 ,ultra thick, pink!50] (0,0) rectangle (12,6);
\draw [fill=magenta!50,  thick, magenta!50] (0.5,5.5) circle [radius=0.4];
\node [above right, blue] at (0.2,5.25){\large$z$};
\draw [line width=0.1cm ,teal!70] (0,0) -- (12,0);
\draw [darkgray,ultra thick] (6,6) -- (6,4);
\draw [fill=white] (6.10,1.5) rectangle (5.90,4.);
\draw [blue!70,ultra thick] (6.10,1.5) rectangle (5.90,4.);
\draw [darkgray,ultra thick,dashed] (6,1.5) -- (6,0);
\node [above right, black] at (6.15,4.8){\large$\bar{B}$};
\node [above right, black] at (6.15,2.5){\large$A$};
\node [above right, black] at (6.15,0.7){\large$B$};
\node [above right, black] at (5.25,3.80){\small$P_{2}$};
\node [above right, black] at (5.25,2.60){\large$s$};
\node [above right, black] at (5.25,1.3){\small$P_{1}$};
\node [above right, black] at (5.3,0.6){\large$l$};

\draw [ultra thick,->] (3.5,-0.5) -- (3.5,-1.4);
\node [above right, black] at (2.8,-1.1){$I$};

\draw [fill=pink!30 ,ultra thick, pink!50] (3.7,-4.2) circle [radius=2.2];
\draw [ ultra thick,teal!70 ] (3.7,-4.2) circle [radius=2.2];
\draw [fill=white] (2.1,-4.2) rectangle (4.4,-4);
\draw [blue!70,ultra thick] (2.1,-4.2) rectangle (4.4,-4);
\draw [ultra thick,->] (3.7,-4.) -- (3.7,-2.5);
\draw [ultra thick,dashed,->] (4.4,-4.1) -- (5.9,-4.1);
\node [above right, black] at (1.9,-4.8){\small$P_{1}$};
\node [above right, black] at (4.,-4.8){\small$P_{2}$};

\draw [ultra thick,->] (6.1,-4.1) -- (7.,-4.1);
\node [above right, black] at (6.2,-3.9){\large$II$};

\draw [fill=pink!30 ,ultra thick, pink!50] (9.5,-4.2) circle [radius=2.2];
\draw [ ultra thick,teal!70 ] (9.5,-4.2) circle [radius=2.2];
\draw [fill=white] (8.3,-4.2) rectangle (10.6,-4);
\draw [blue!70,ultra thick] (8.3,-4.2) rectangle (10.6,-4);
\draw [ultra thick,->] (9.5,-4.) -- (9.5,-2.5);
\draw [ultra thick,dashed,->] (10.6,-4.1) -- (11.7,-4.1);
\node [above right, black] at (8,-4.8){\small$P_{2}$};
\node [above right, black] at (10.2,-4.8){\small$P_{1}$};
\node [above right, black] at (8.8,-4.){\small$d$};
\node [above right, black] at (9.8,-4){\small$d$};

\draw [ultra thick,->] (9.5,-6.8) -- (9.5,-7.7);
\node [above right, black] at (8.5,-7.5){\large$III$};

\draw [fill=pink!30 ,ultra thick, pink!50] (9.5,-10.2) circle [radius=2.2];
\draw [ ultra thick,teal!70 ] (9.5,-10.2) circle [radius=2.2];
\draw [fill=white ] (9.5,-10.2) circle [radius=1];
\draw [ultra thick,blue!70 ] (9.5,-10.2) circle [radius=1];
\draw [ultra thick,<->] (9.5,-10.2) -- (9.5,-9.2);
\draw [ultra thick,<->] (9.5,-10.2) -- (7.3,-10.2);
\draw [ultra thick,dashed] (10.5,-10.2) -- (11.7,-10.2);
\node [above right, black] at (9.5,-10.){\small$\bf{e^{-h}}$};
\node [above right, black] at (7.8,-10.){\large$\bf{1}$};

\draw [ultra thick,->] (7,-10.2) -- (6.1,-10.2);
\node [above right, black] at (6.2,-10){\large$IV$};

\draw [fill=pink!30 ,ultra thick, pink!50] (3.7,-10.2) circle [radius=2.2];
\draw [ ultra thick,teal!70 ] (3.7,-10.2) circle [radius=2.2];
\draw [fill=white ] (3.7,-10.2) circle [radius=1];
\draw [ultra thick,blue!70 ] (3.7,-10.2) circle [radius=1];
\draw [ultra thick,<->] (3.7,-10.2) -- (3.7,-9.2);
\draw [ultra thick,<->] (3.7,-10.2) -- (1.5,-10.2);
\node [above right, black] at (3.7,-10.){\small$\bf{e^{-h/\alpha}}$};
\node [above right, black] at (2.,-10.){\large$\bf{1}$};
\draw [fill=magenta!50,  thick, magenta!50] (2.7,-8.8) circle [radius=0.4];
\node [above right, blue] at (2.3,-9){\large$w$};

\end{tikzpicture}
\caption{(Color online) Mapping between different regions. The upper half plane with  slit $A$ and branch cut (dashed line) on $B$ can be mapped to
annulus  by the conformal map $w_{\alpha}(z)$ in four steps.  } 
\end{figure}
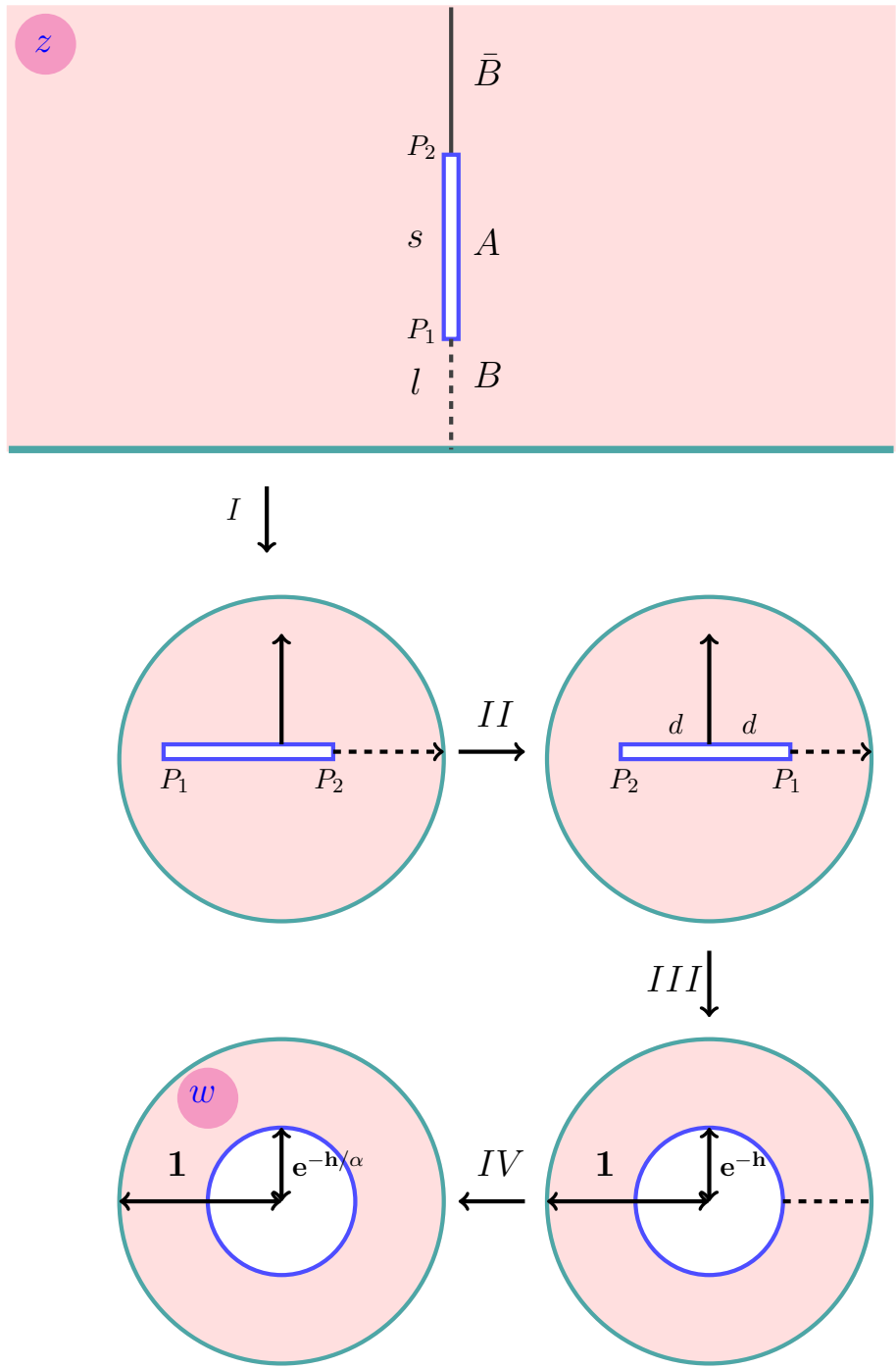



The corresponding conformal map from the upper half plane with one slit and a branch cut on $B$ to an annulus can be derived as follows:

Step I: we first map the upper half plane to a unit disc by the conformal map 
\begin{equation}\label{half plane to disc}
z_1=z_1=\frac{z-i}{z+i}.
\end{equation}
The coordinates of $P_1$  and  $P_2$ are now $(b,0)$ and $(a,0)$ respectively with
\begin{eqnarray}\label{a and b}
a=\frac{l-1}{l+1},\\
b=\frac{l+s-1}{l+s+1}.
\end{eqnarray}

Step III: The unit disc with unsymmetric slit can be mapped to a unit disc with symmetric slit by
the conformal map
\begin{eqnarray}\label{unit to unit}
z_2=\frac{g-z_1}{1-gz_1},\\
g=\frac{1+ab-\sqrt{(a^2-1)(b^2-1)}}{a+b}.
\end{eqnarray}
The length of the slit is now $2d$ with
\begin{equation}\label{d}
d=\frac{-1+ab+\sqrt{(a^2-1)(b^2-1)}}{a-b}.
\end{equation}

Step III and IV: The remaing disc with slit can now be mapped to the annulus by using the conformal map
$w_1(z_2)$ provided in \cite{Nehari}. Finally, one needs to uniformize the 
surface by the map $(w_1(z))^{\frac{1}{\alpha}}$. The final result is
\begin{equation}\label{conformal map OBC}
w_{\alpha}(z)=\Big{(}i e^{-h}e^{\frac{\pi}{2i\mathcal{K}(k^2)}\mbox{sn}^{-1}\frac{z_2}{d}}\Big{)}^{\frac{1}{\alpha}};
\end{equation}
where
\begin{eqnarray}\label{k and h OBC}
k&=&d^2,\\
h&=&\frac{\pi}{4}\frac{\mathcal{K}(1-k^2)}{\mathcal{K}(k^2)}.
\end{eqnarray}
Note that the equation (\ref{conformal map OBC}) is valid just for $Im z>0$ and in principle for $Im z<0$ one needs to use 
\begin{equation}\label{conformal map OBC2}
w_{\alpha}(z)=\Big{(}-i e^{-h}e^{\frac{\pi}{2i\mathcal{K}(k^2)}\mbox{sn}^{-1}-\frac{z_2}{d}}\Big{)}^{\frac{1}{\alpha}};
\end{equation}
This subtility does not affect the upcoming calculations.

To calculate the Schwartzian derivative we need the following  chain rule 
\begin{eqnarray}\label{chain rule}
S(f\circ g)=\big{(}S(f)\circ g\Big{)}(g')^2+S(g).
\end{eqnarray}
The first two steps do not contribute to the Schwartzian derivative because they are both Mobius transformations.
The  Schwartzian derivative has two poles at $z=il$ and $z=i(l+s)$. After calculating the integral in (\ref{geometric}) we have

\begin{eqnarray}\label{geometric open}
\frac{\delta\ln Z^{geom}_{\alpha}}{\delta l}= \frac{c}{12\pi i}2\pi \alpha \frac{(-1+g)(\pi^2-4(1+k^2)\alpha^2\mathcal{K}^2(k^2))}
{8\alpha^2(1+g)(-1+k^2)\mathcal{K}^2(k^2)},
\end{eqnarray}
We are interested to study two limits: the first interesting limit $s\ll l$  is the problem of the
entanglement entropy of a subsystem without any projective measurement. In this limit we have 
\begin{eqnarray}\label{tilde q and h}
h=\ln\frac{8l}{s}+...,\\
\tilde{q}=(\frac{s}{8l})^{\frac{2}{\alpha}}+.... 
\end{eqnarray}
Since in this limit $\tilde{q}$ is the small parameter as far as $\alpha$ is not too big we use the equation (\ref{annulus part2}). We have
\begin{eqnarray}\label{expansion of annulus part}
 \ln Z^{annu}_{\alpha}=2\ln b_0+\frac{b_1^2}{b_0^2}(\frac{s}{8l})^{\frac{2\Delta_1}{\alpha}}+...,
\end{eqnarray}
where $\Delta_1$ is the smallest dimension in the conformal tower.
In addition after expanding (\ref{geometric open}) and integrating with respect to $l$ we have
\begin{eqnarray}\label{z geometric open}
\ln Z^{geom}_{\alpha}=\frac{c}{12}\frac{1-\alpha^2}{\alpha}\ln\frac{l}{a}+..., 
\end{eqnarray}
where the dots are the subleading terms. Putting all the terms together we have
\begin{eqnarray}\label{EE open}
S_{\alpha}=\frac{c}{12}\frac{(1+\alpha)}{\alpha}\ln\frac{l}{a}+2\ln b_0+\frac{b_1^2}{b_0^2}(\frac{s}{8l})^{\frac{2\Delta_1}{\alpha}}+....
\end{eqnarray}
 The above result is the standard result of the entanglement entropy of a subsystem \cite{Calabrese2009}. The second term
is the Affleck-Ludwig boundary term and the third term is the unusual correction to the entanglement entropy discussed in
\cite{CardyCalabreseCorrections}. Note that in the limit of no measurement region one needs to 
replace $2\ln b_0$ with $\ln b_0$.

The next interesting limit is $l\ll s$  which is the third example of the post measurement entanglement entropy of disconnected regions.
In this case, $q$ is the small parameter and we have
\begin{eqnarray}\label{ q and h}
h=\frac{\pi^2}{2\ln\frac{4s}{l}}+...,\\
q=(\frac{l}{4s})^{4\alpha}+.... 
\end{eqnarray}
Then after a bit of algebra we have
\begin{eqnarray}\label{z geometric and annulus open}
\ln Z^{geom}_{\alpha}=\frac{c}{24}\alpha(\frac{\ln\frac{a}{l}}{2}+\frac{\pi^2}{\alpha^2\ln\frac{4s}{l}}) +...,\\
\ln Z^{annu}_{\alpha}=-\frac{c}{24}\alpha(4\ln\frac{l}{4s}+\frac{\pi^2}{\alpha^2\ln\frac{4s}{l}})+n_1(\frac{l}{4s})^{4\alpha\Delta_1}+....
\end{eqnarray}
Summing over all the terms  gives
\begin{equation}\label{power-law decay 3}
{S_{\alpha}}\asymp \left\{
\begin{array}{c l}      
    \frac{1}{\alpha-1} (\frac{l}{4s})^{4\alpha\Delta_1}, & \alpha<1\\
        (\frac{l}{4s})^{2\Delta_1}\ln\frac{l}{8s}, &\alpha=1\\
        \frac{\alpha}{\alpha-1} (\frac{l}{4s})^{4\Delta_1}, & \alpha>1,
\end{array}\right.
\end{equation}
where $\Delta_1$ is the smallest boundary scaling dimension in the spectrum of the system.

\subsubsection{Finite open systems:} 
\begin{figure} [hthp!] 
\centering
\begin{tikzpicture}[scale=1]
\draw [line width=0.1cm ,black] (0,0) -- (9,0);
\draw [ ultra thick , <-> ] (0,-0.7) -- (9,-0.7);
 \draw [fill=blue] (0,0) circle [radius=0.15];
\draw [fill=blue] (9,0) circle [radius=0.15];
\draw [line width=0.05cm ,blue] (3,0.15) -- (3,-0.15);
\draw [line width=0.05cm ,blue] (6,0.15) -- (6,-0.15);
\node [above right, black] at (4.5,-1.5){\large$\bf{L}$};
\node [above right, black] at (1.5,0.3){\large$\bf{A}$};
\node [above right, black] at (4.5,0.3){\large$\bf{B}$};
\node [above right, black] at (7.5,0.3){\large$\bf{\bar{B}}$};
\end{tikzpicture}
\caption{(Color online) The setup for the post measurement entanglement entropy in  an open finite system. } 
\label{OBCFinite}
\end{figure}
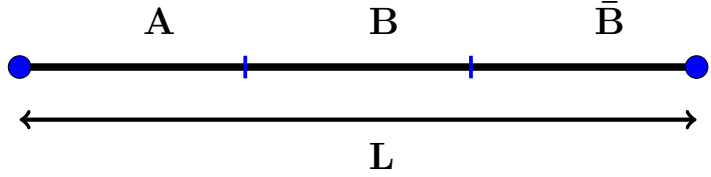

In this case, we consider a finite total system with length $L$ and  make a projective measurement
in the part $A$ which is a connected subsystem with length $s$ starting from one side of the system, see Figure (\ref{OBCFinite}). Then we calculate the entanglement entropy of
the simply connected subsystems $B$  and $\bar{B}$ with lengths $l$ and $L-l-s$ respectively. In this setup $B$ and $\bar{B}$ are connected and the formula of the post measurement
entanglement entropy is already calculated in \cite{Rajabpour2015b} by using the twist operator technique. Although in principle  the formula 
can be  re-derived with 
the method of the beginning of this section, we will just report the final result \cite{Rajabpour2015b}:
\begin{eqnarray}\label{SB for OBC}
S_{\alpha}=\frac{c}{12}(\frac{1+\alpha}{\alpha})\ln \Big{(}\frac{2L}{\pi}\frac{\cos\frac{\pi s}{L}-\cos\pi\frac{l+s}{L}}{a\cos^2\frac{\pi s}{2L}}\cot\frac{\pi(l+s)}{2L}\Big{)}+...,
\end{eqnarray}
where the dots are the subleading terms. In the limit $L\to\infty$ we have the following simple formula
\begin{eqnarray}\label{SB for OBC semi-infinite}
S_{\alpha}=\frac{c}{12}(\frac{1+\alpha}{\alpha})\ln \frac{4l(l+2s)}{l+s}+....
\end{eqnarray}

Note that the above results are correct as far as the measurement induces the same boundary condition as the natural boundary condition
of the open system. When the conformal boundary condition on the slit is different from the  boundary conditions of the open system
one needs to consider the effect of the boundary changing operator. Although these boundary condition changing operators can appear frequently for technical reasons
we leave the proper treatment of them to another
work.



\subsubsection{Finite temperature:}

It is quite straightforward to extend the above results to a system with the finite temperature. In principle, with a finite temperature, we mean that one first starts with a Gibbs state for the entire system $e^{-\frac{H}{T}}$ and then by tracing out one part of the system derives the reduced density matrix. Then the R\'enyi entropy can be derived as before. Technically one just needs to study
the two-dimensional cylinder with a base circumferences $\beta$ with two slits and a branch cut in the direction of the axes of the cylinder.
In principle, one can use the results of the finite periodic system to extract the results for the finite temperature. This can be done by just replacing
 $L$ with $i\beta$. For example, when the system is infinite and $s_2$ is small we will have

\begin{eqnarray}\label{SB for finite temperature 1}
S_{\alpha}(\beta)=
\frac{c}{12}(1+\frac{1}{\alpha})\ln \Big{(}\frac{\beta}{\pi}\frac{\sinh\frac{\pi}{\beta}(l+s_1)\sinh\frac{\pi}{\beta}l}{s_2\sinh\frac{\pi}{\beta}s_1}\Big{)}+...,
\end{eqnarray}
In the limit of small $s_1$, we recover  the result of the finite temperature R\'enyi entropy for a system without projective measurement
\cite{Calabrese2009}, i.e.
\begin{eqnarray}\label{SB for finite temperature 1 no measurement}
S_{\alpha}=
\frac{c}{6}(\frac{1+\alpha}{\alpha})\ln \Big{(}\frac{\beta}{\pi}\sinh\frac{\pi l}{\beta}\Big{)}+....
\end{eqnarray}
It is easy to see that for a large temperature one can simply derive
\begin{eqnarray}\label{SB for infinite temperature}
S_{\alpha}(\beta)=
\frac{\pi c}{6}(1+\frac{1}{\alpha})\frac{l}{\beta}+...,
\end{eqnarray}
In this limit, the entropy is extensive as it is expected. When $s_1=s_2=s$  is much bigger than $l$
one can use the formulas of the appendix and find
\begin{eqnarray}\label{finite temperature h }
h=-\frac{\pi^2}{\ln\Big{[}\frac{\pi l}{8\beta}\coth\frac{\pi s}{\beta}\Big{]}}+...,
\end{eqnarray}
In the limit of small $l$ when $\frac{\pi l}{8\beta}\coth\frac{\pi s}{\beta}\ll 1$ we have $h\to 0$ which means that the $q$ is the small parameter and we have
\begin{eqnarray}\label{finite temperature  q}
q=(\frac{\pi l}{8\beta}\coth\frac{\pi s}{\beta})^{2\alpha}+....
\end{eqnarray}
Then after a bit of algebra we have

\begin{equation}\label{power-law decay finite temperature}
{S_{\alpha}}(\beta)\asymp\left\{
\begin{array}{c l}      
    \frac{1}{\alpha-1} (\frac{\pi l}{8\beta}\coth\frac{\pi s}{\beta})^{2\alpha\Delta_1}, & \alpha<1\\
        (\frac{\pi l}{8\beta}\coth\frac{\pi s}{\beta})^{2\Delta_1}\ln(\frac{\pi l}{8\beta}\coth\frac{\pi s}{\beta}), &\alpha=1\\
        \frac{\alpha}{\alpha-1} (\frac{\pi l}{8\beta}\coth\frac{\pi s}{\beta})^{2\Delta_1}, & \alpha>1,
\end{array}\right.
\end{equation}
In the limit of the zero temperature, we are back again to the formula (\ref{power-law decay1}). However, when 
$\frac{\pi s}{\beta}\gg 1\gg \frac{\pi l}{8\beta}$ one can write
\begin{eqnarray}\label{finite temperature  q 2}
q=(\frac{\pi l}{8\beta})^{2\alpha}+....
\end{eqnarray}
and consequently we have
\begin{equation}\label{power-law decay finite temperature2}
{S_{\alpha}}(\beta)\asymp\left\{
\begin{array}{c l}      
    \frac{1}{\alpha-1} (\frac{\pi l}{8\beta})^{2\alpha\Delta_1}, & \alpha<1\\
        (\frac{\pi l}{8\beta})^{2\Delta_1}\ln(\frac{\pi l}{\beta}), &\alpha=1\\
        \frac{\alpha}{\alpha-1} (\frac{\pi l}{8\beta})^{2\Delta_1}, & \alpha>1,
\end{array}\right.
\end{equation}
The above result  interestingly  shows that as far as the measurement region is big and the size of the isolated subsystem small
the  entropy increases like a power-law with respect to the temperature with a power which is dependent on the smallest scaling dimension in the 
spectrum of the system.
When $\beta$ is small we need to use the expansion with respect to $\tilde{q}$ and we are back again to the formula (\ref{SB for infinite temperature}).

The result for the connected regions can be also extended to the semi-infinite system at finite temperature. In this case one just needs to replace
$L$ with $\frac{\beta}{2}$ in the equation (\ref{SB for OBC}). The final result is 
\begin{eqnarray}\label{SB finite temperature for semi-infinite }
S_{\alpha}=
\frac{c}{12}(\frac{1+\alpha}{\alpha})\ln \Big{(}\frac{\beta}{\pi}\frac{
\cosh2\pi\frac{l+s}{\beta}-\cosh\frac{2\pi s}{\beta}}{a\cosh^2\frac{\pi s}{\beta}}\coth\frac{\pi(l+s)}{\beta}\Big{)}+....
\end{eqnarray}
In the limit of  $\beta\to\infty$, we redrive the formula (\ref{SB for OBC semi-infinite}) and
when $\beta\to 0$ we are back again to the formula (\ref{SB for infinite temperature}). When we do not have any measurement
region we are back to the well-known result of \cite{Calabrese2009}, i.e. 
\begin{eqnarray}\label{SB finite temperature for semi-infinite no measurement}
S_{\alpha}=
\frac{c}{12}(\frac{1+\alpha}{\alpha})\ln \Big{(}\frac{\beta}{\pi}\sinh\frac{2\pi l}{\beta}\Big{)}+....
\end{eqnarray}
The above results can not be extended easily to the finite periodic systems. We leave the proper treatment of this case to a future work.

\subsubsection{Affleck-Ludwig boundary entropy:} 

In this subsection, we make some further comments regarding  the Affleck-Ludwig boundary entropy term. So far we have been concentrating on the projective measurements
in a way that the measurement on the two slits are done on the same observables and the results are also the same. However, the more general case is when
the measurements are done on different observables or they are done on the same observables but with different outcomes. Depending on the observables
and the outcomes the boundary conditions on the two slits might be different. Note that even choosing the same observable on both slits does not
mean that the corresponding boundary conditions on the two slits are the same.  When there are two different conditions on the boundaries of the annulus, i.e. $A$ and $B$, the equations
(\ref{annulus part1}) and (\ref{annulus part2}) have the following more general forms \cite{Cardy2004}:
\begin{eqnarray}\label{annulus part3}
\ln Z^{annu}_{\alpha}(q)&=&\ln [q_{\alpha}^{-c/24}(1+\sum_jn^{AB}_jq_{\alpha}^{\Delta_j})]-c\frac{h}{12\alpha},\\
\label{annulus part4}
\ln Z^{annu}_{\alpha}(\tilde{q})&=&\ln [\tilde{q}_{\alpha}^{-c/24}(b^{A}_{0}b^B_{0}+\sum_jb^{A}_{j}b^B_{j}\tilde{q}_{\alpha}^{\Delta_j})]-c\frac{h}{12\alpha},
\end{eqnarray}
where $n^{AB}_j$ are the non-negative integers and $b^A_{j}=\langle A|j\rangle\rangle$ and $b^B_{j}=\langle\langle j|B\rangle$. 
$|A(B)\rangle $ and $|j\rangle\rangle$ are Cardy
and Ishibashi states respectively. Different coefficients are related to each other with the formula $n_j^{AB}=\sum_{j'}S_j^{j'}b_j^{A}b_{j'}^B$,
where $S_j^{j'}$ is the element of the modular matrix $S$, see \cite{Cardy2004}.  Using (\ref{annulus part4}) one can now write
the Affleck-Ludwig boundary entropy as \cite{Rajabpour2016}:
\begin{eqnarray}\label{ludwig-affleck}
S^{AL}=\ln b_0^A+\ln b_0^B
\end{eqnarray}
In the presence of one boundary, we need to consider just one of the above terms. The above result is correct also in the presence of the boundary, for example in the case of the semi-infinite system.
Note that since all of the derived formulas have also an extra non-universal constant contributions all the comments regarding
The Affleck-Ludwig term is meaningless if we do not factor out the unwanted non-universal terms. This can be done following 
\cite{Calabrese2009,Doyon2009,Doyon2008,Zhou2006,Laflorencie2006} as follows: we first write the entanglement entropy of a region without any measurement
domain for a periodic system as follows: 

\begin{eqnarray}\label{SB for no-measurement PBC}
S_{\alpha}=\frac{c}{6}(1+\frac{1}{\alpha})\ln \Big{(}\frac{L}{\pi}\sin\frac{\pi l}{L}\Big{)}+a_{\alpha},
\end{eqnarray}
Then for the case with the measurement we have
\begin{eqnarray}\label{SB for PBC affleck-ludwig}
S_{\alpha}=\frac{c}{12}(1+\frac{1}{\alpha})\ln \Big{(}\frac{4L}{\pi}\frac{\sin\frac{\pi}{L}(l+s_1)\sin\frac{\pi}{L}l}{s_2\sin\frac{\pi}{L}s_1}\Big{)}+
S^{AL}+\frac{a_{\alpha}}{2}.
\end{eqnarray}
With the above procedure, the definition of the $S^{AL}$ has no ambiguity. Note that in both equations the $a_{\alpha}$
is the same non-universal constant and we also introduced a factor of $4$ inside the logarithm in the second equation. At the moment we have
no concrete argument why that factor should be ~4 but as we will see in the upcoming sections its presence is dictated by the numerical calculations.
One way to see that a non-trivial factor should be there is just by realizing that the ultra-violet cut-off is different in the two
cases. However to fix the number exactly one possibly needs to start from the massive case and go to the massless regime 
as it was argued in \cite{Doyon2009,Doyon2008}.
The result for the infinite chain can be derived by just
sending $L$ to infinity. A similar result is also valid in the presence of the natural boundary of the system. For example for the finite open system
with one measurement domain we have
\begin{eqnarray}\label{SB for OBC affleck-ludwig}
S_{\alpha}=
\frac{c}{12}(\frac{1+\alpha}{\alpha})\ln \Big{(}\frac{2L}{\pi}\frac{\cos\frac{\pi s}{L}-
\cos\pi\frac{l+s}{L}}{a\cos^2\frac{\pi s}{2L}}\cot\frac{\pi(l+s)}{2L}\Big{)}+S^{AL}+\frac{a_{\alpha}}{2}.
\end{eqnarray}
The results can be extended also to non-critical systems. When we have a finite temperature infinite size critical system the corresponding formula for
the post measurement R\'enyi entropy is
\begin{eqnarray}\label{SB for finite temperature 1 affleck-ludwig}
S_{\alpha}(\beta)=
\frac{c}{12}(1+\frac{1}{\alpha})\ln \Big{(}\frac{4\beta}{\pi}\frac{\sinh\frac{\pi}{\beta}(l+s_1)\sinh\frac{\pi}{\beta}l}{s_2\sinh\frac{\pi}{\beta}s_1}\Big{)}
+S^{AL}+\frac{a_{\alpha}}{2},
\end{eqnarray}
Finally, for semi-infinite system at finite temperature we have
\begin{eqnarray}\label{SB finite temperature for semi-infinite affleck-ludwig}
S_{\alpha}&=&
\frac{c}{12}(\frac{1+\alpha}{\alpha})\ln \Big{(}\frac{\beta}{\pi}\frac{
\cosh2\pi\frac{l+s}{\beta}-\cosh\frac{2\pi s}{\beta}}{a\cosh^2\frac{\pi s}{\beta}}\coth\frac{\pi(l+s)}{\beta}\Big{)}\nonumber\\&+&S^{AL}+ \frac{a_{\alpha}}{2}.
\end{eqnarray}
Since the Affleck-Ludwig term is dependent on  the corresponding boundary conditions one can use it to identify
the  nature of the conformal boundary condition. We will use extensively this fact to identify the boundary conditions induced by the
configurations in the later sections. It is worth mentioning  that all of the above equations will change if the boundary condition 
changing operators are present in the system.

In all of the above equations, we assumed that one of the measurement regions is big and the other one is very small (or effectively does not exist)
in a way that
$\tilde{q}_{\alpha}$ is small. However, it is obvious that the situation would change if both of the measurement regions are big enough. In this case, 
one needs to consider the most general formulas of $Z^{annu}$ and $Z^{geom}$ and try to extract the universal $b_0$ terms. In this case, one might be 
even able to go further and detect all of the $b_j$ with $j>0$. However, since $\tilde{q}_{\alpha}^{\Delta_j}$ in the partition function expansion
is accompanied with non-universal
constants it might be really hard to detect them numerically. 

Finally, we close this subsection with some remarks regarding the $g$-theorem
which states that for a fixed bulk conformal theory, boundary conditions introduce  the $\ln b_0$ in a way that $b_0$
decreases to the infrared under the renormalization group \cite{gtheorm1}. This theorem is proved in a field theory context but  
there is no proof of it in the context of the entanglement entropy, see  \cite{gtheorm2}. In the context of the 
post measurement entanglement entropy, there might be two ways  to look at this theorem. The important point about g-theorem is that
the bulk theory is conformal but the boundary is flowing. This means that whatever measurement which induces non-conformal boundary condition
can lead to different value for the $b_0$. Basically, a measurement of different outcomes might lead to the same or different
conformal (non-conformal) boundary conditions. This means that one might derive different values for $b_0$ depending on the outcome of the measurement.
Of course, the same argument goes for also the post measurement entanglement entropy done in the other basis. The bottom line is that one might
interpret different results for the measurement or doing the measurement in different basis as some sort of boundary renormalization group flow.
We leave a more elaborate analysis of this point for a future work.

\subsubsection{Lattice effects :} 

In this section we briefly address the effect of the lattice on the CFT results. The effect in the presence of one slit is already studied in \cite{StephanDubail} and here we apply the results
to the post measurement entanglement entropy. As it is argued in \cite{StephanDubail} the effect of the lattice can be simulated by perturbing the CFT action by the energy momentum tensor as follows:
\begin{eqnarray}\label{perturbation}
S\to S+\frac{\xi}{2\pi}\int_{slit}dx T_{xx}
\end{eqnarray}
where $T_{xx}$ is the element of the energy-momentum tensor in the $x$ direction and $\xi$ is called extrapolation length and it usually  plays the role of
the UV cutoff in the presence of the boundary. To study the effect of the above perturbation on the post measurement entanglement entropy it is much easier to work with the twist operator
technique. Here we discuss the simplest setup of an infinite chain with one slit, in other words in the Figure ~1 we take $s_2=0$. 
Based on the Calabrese-Cardy technique \cite{Calabrese2004} the entanglement entropy of the region $B$ is given by 
\begin{eqnarray}\label{twist1}
S_{\alpha}=\frac{1}{1-\alpha}\ln \langle\mathcal{T}_{\alpha}\rangle_{slit},
\end{eqnarray}
where $\mathcal{T}_{\alpha}$ is the twist operator with the conformal weight $\delta_{\alpha}=\frac{c\alpha}{24}(1-\frac{1}{\alpha^2})$
sitting at the bundary between $B$ and $\bar{B}$. Finally $\langle\mathcal{T}_{\alpha}\rangle_{slit}$ is the expectation value of
the twist operator in the geometry of infinite plane minus a slit which can be calculated by mapping the whole space minus a slit to the upper half plane by the conformal  map $z(w)=\sqrt{\frac{s+2w}{s-2w}}$, see \cite{Rajabpour2015b}.
In \cite{StephanDubail}  the effect of the perturbation (\ref{perturbation}) on the one point function of an arbitrary primary operator is studied. Applying the result to the
twist operators for fixed $\frac{l}{s}$ we
have
\begin{eqnarray}\label{twist-correction}
\langle\mathcal{T}_{\alpha}\rangle_{slit}=(\frac{sa}{4l(l+s)})^{2\delta_{\alpha}}\Big{(}1-\frac{\xi}{\pi}\delta_{\alpha}(2+\frac{1}{\frac{l}{s}(1+\frac{l}{s})})\frac{\ln\frac{s}{a}}{s}\Big{)},
\end{eqnarray}
The entanglement entropy can now be calculated by pluging (\ref{twist-correction}) into (\ref{twist1}). For example, for the von Neumann entanglement entropy we find
\begin{eqnarray}\label{von neumann log l}
S=\frac{c}{6}\ln \frac{l(l+s)}{as}+\frac{c\xi}{12\pi}(2+\frac{1}{\frac{l}{s}(1+\frac{l}{s})})\frac{\ln\frac{s}{a}}{s}, 
\end{eqnarray}
where the first term is the usual term appeared already in the section (3.1.1) and the second term is the  $\frac{\log s}{s}$ correction coming from the lattice effects. Although
in many numerical
calculations these kinds of lattice corrections to the CFT results are the leading corrections, since in our numerical calculations we are going to investigate just the leading term
we will not explore further this interesting effect.

\subsection{Entanglement hamiltonians}

The entanglement hamiltonian which is also called modular hamiltonian $K_B$ is defined as follows:
\begin{eqnarray}\label{entanglement hamiltonian definition}
\rho_B=e^{-2\pi K_B},
\end{eqnarray}
where $\rho_B$ is as before the reduced density matrix of the subsystem. To calculate $K_B$ 
we recall the partition function of CFT on the cylinder $Z^{cyl}$ which has the following form:
\begin{eqnarray}\label{Z Cylinder}
Z^{cyl}=\tr q^{L_0-\frac{c}{24}}.
\end{eqnarray}
Using the definition of $q$ and the relation between $L_0$ and the energy-momentum tensor $T(z)$ one can simply write \cite{Cardy2015}, see also \cite{CardyTonni}:
\begin{eqnarray}\label{redused cylinder}
\rho_B\approx e^{-2\pi\int_0^h T(\tilde{x})d\tilde{x}},
\end{eqnarray}
where $h$ is the length of the cylinder with the base-circumference $2\pi$. Having the above result on the cylinder one just needs to come back to
the original geometry that has two slits on it. This can be done simply by first mapping the 
cylinder to the annulus by the map $w=e^{\tilde{w}}$ where $\tilde{w}$  and $w$ represent the cylinder and the annulus respectively.
After moving to the annulus  we can now use just the inverse of the conformal maps that we introduced before
to map the annulus to the geometry with the two slits stretched on the intervals $(0,s_1)$ and $(s_1+l,s_1+l+s_2)$. Since after the conformal map $f$ the 
energy-momentum tensor changes as $T(z)=(\partial_zf)^2T(f)+(c/12)\{f,z\}$ one can finally write
\begin{eqnarray}\label{redused cylinder2}
\rho_B\approx e^{-2\pi\int_{s_1}^{l+s_1} \Big{(}T(z)(\frac{\partial\tilde{w}(z)}{\partial z})^{-1}\Big{)}\Big{|}_{z=x}dx},
\end{eqnarray}
Having the above formula one can simply identify the modular hamiltonian as
\begin{eqnarray}\label{modular hamiltonian}
K_B= \int_{s_1}^{l+s_1}\Big{(}T(z)(\frac{\partial\tilde{w}(z)}{\partial z})^{-1}\Big{)}\Big{|}_{z=x}dx.
\end{eqnarray}
It is common to call the space dependent coefficient of the energy-momentum tensor the inverse of the temperature
$\beta(x)$. In other words we have:
\begin{eqnarray}\label{beta}
\beta(x)=2\pi(\frac{\partial\tilde{w}(z)}{\partial z})^{-1}\Big{|}_{z=x}.
\end{eqnarray}
The formula (\ref{modular hamiltonian}) is valid for all the cases that we studied so far. One just needs to calculate the derivative of $\tilde{w}(z)=\ln w(z)$
with respect to $z$ and plug it into the above formula. In the next subsections, we will list the entanglement hamiltonian of few interesting 
cases
such as the infinite system, finite periodic system and finite temperature.

\subsubsection{Infinite systems:}

Consider the infinite system with two measurement regions as the figure ~1.
Using the conformal map provided in the appendix we can simply write
\begin{eqnarray}\label{beta-infinite}
\beta(x)=2\pi\frac{1}{\pi}2kx(1-ax+bx)\mathcal{K}(1-k^2)\mbox{cd}[\mbox{sn}^{-1}[\frac{1-2ax+bx}{k+bkx},k^2],k^2],
\end{eqnarray}
where  $\mbox{cd}$ and $\mbox{sn}^{-1}$ are
the   Jacobi and inverse Jacobi functions. $a$, $b$ and $k$ are defined in the appendix. One can study the above formula in many different interesting limits. 
When $s_1=s_2=s\ll l$ we can simply find:
\begin{eqnarray}\label{EH1}
\beta(x)=2\pi\frac{x(l-x)}{l}.
\end{eqnarray}
If we symmetrize the above formula by putting $l=2R$ and $x\to x+R$ we reach to the well-known result 
of \cite{CHM}, see also \cite{entanglementHamiltonian1,Klich} and references therein. When $s_1=s\ll s_2,l$
we can again expand the formula (\ref{beta-infinite}) and find:
\begin{eqnarray}\label{EH2}
\beta(x)=2\pi x\sqrt{\frac{(l-x)(l+s_2-x)}{l(l+s_2)}}.
\end{eqnarray}
which is a generalized form of the equation (\ref{EH1}). It is very interesting to note that
one can now derive the entanglement entropy  by integrating the equilibrium thermal entropy per unite length as follows \cite{Holzey1994,Klich}:
\begin{eqnarray}\label{S and beta}
S_{1}=2\pi\frac{ c}{6}\int_{s_1}^{s_1+l}\frac{1}{\beta(x)}dx.
\end{eqnarray} 
Putting (\ref{EH2}) in the above formula and expanding it with respect to $s_1$ one can derive the leading term of the equation (\ref{rajab2015}).
Note that the subleading terms that are unusual corrections coming from the relevant operators sitting on the 
conical singularities \cite{CardyCalabreseCorrections} can not be derived by using (\ref{S and beta}). This is simply because this equation 
does not take into account the contributions coming from   the two very ends of the subsystem. 
Finally, when $l\ll s_1=s_2=s$ we first make a change of coordinates $z\to z+s+\frac{l}{2}$ and put also $l=2R$ then we expand the equation (\ref{beta-infinite})
for large $s$. Finally, we have
\begin{eqnarray}\label{EH3}
\beta(x)=2 \sqrt{R^2-x^2}\ln \frac{8s}{l}.
\end{eqnarray}
As it is expected one can not derive the equation (\ref{power-law decay1}) using the equations (\ref{S and beta}) and (\ref{EH3}).
However, the above equation has some of the expected properties such as: it is zero at the two ends of the subsection and it grows with increasing $s$.
It is important to mention that in the above limit although strictly speaking the $q$ is not the small parameter we used the expansion of
the partition function with respect to $q$ to derive the above formula. This means that the validity of the above equation might break down
for very large $s$. The right way to study the entanglement Hamiltonian in this limit might be working with the expansion with respect to $\tilde{q}$.
In all of the upcoming calculations, we will just use the expansion
of the cylinder partition function with respect to 
the $q$.

\subsubsection{Finite periodic systems:}

The entanglement hamiltonian for a finite system can also be derived following the same method. one just needs to use the conformal map
introduced in the appendix in the equation (\ref{modular hamiltonian}). Using Mathematica one can derive:
\begin{eqnarray}\label{beta-PBC}
\beta(x)=-\frac{iL}{\pi(b_0-a_0b_1)} e^{-\frac{2\pi i x}{L}}(b_0+b_1e^{\frac{2\pi i x}{L}})^2 \mathcal{K}(1-k^2)\times\nonumber\\    
\mbox{cd}[\mbox{sn}^{-1}[\frac{1}{b_1+\frac{b_0-a_0b_1}{a_0+e^{\frac{2\pi i x}{L}}}},k^2],k^2]\times
\mbox{dn}[\mbox{sn}^{-1}[\frac{1}{b_1+\frac{b_0-a_0b_1}{a_0+e^{\frac{2\pi i x}{L}}}},k^2],k^2],
\end{eqnarray}
where $a_0$, $b_0$, $b_1$ and $k$ are all defined in the appendix and $\mbox{cd}$ and $\mbox{dn}$
 are
the   Jacobi  functions and $\mbox{sn}^{-1}$ is the inverse Jacobi function.
One can study the above equation in different limits. For example, when $s_1=s_2=s\ll l$ one can derive
\begin{eqnarray}\label{beta-PBC 1}
\beta(x)=2L\frac{\sin[\frac{\pi(l-x)}{L}]\sin[\frac{\pi x}{L}]}{\sin[\frac{\pi l}{L}]}.
\end{eqnarray}
If we symmetrize the above formula by putting $l=2R$ and $x\to x+R$ we reach to the known result 
of \cite{Klich}. The other interesting case is when $s_1=s\ll s_2,l$. In this limit we have
\begin{eqnarray}\label{beta-PBC 2}
\beta(x)=2L\sin[\frac{\pi x}{L}]\sqrt{\frac{\sin[\frac{\pi}{L}(l-x)]\sin[\frac{\pi}{L}(l+s_2-x)]}{\sin[\frac{\pi l}{L}]\sin[\frac{\pi}{L}(l+s_2)]}}.
\end{eqnarray}
The above formula is the generalization of the formula (\ref{beta-PBC 1}) for the post measurement systems.
Finally, one can also study the  limit $l\ll s_1=s_2=s=\frac{L-2l}{2}$. In this case, we first symmetrize the system by change
of variables $z\to z+s+\frac{l}{2}$ and $l=2R$. Then we expand  the  formula (\ref{beta-PBC}) with respect to $R$ and find
\begin{eqnarray}\label{beta-PBC 3}
\beta(x)=\sqrt{1-(\frac{L}{\pi R}\tan\frac{\pi x}{L})^2}.
\end{eqnarray}
Note that for  $R\ll L$ we have $\beta(R)=\beta(-R)=0$ as it is expected.



\subsubsection{Finite temperature:}

Entanglement hamiltonian for an infinite system with the finite temperature  can be derived simply by replacing
$L$ with $i\beta$ in the formulas of the finite periodic system. For example for the case $s_1=s\ll s_2,l$ we have
\begin{eqnarray}\label{beta-Finite T}
\beta(x)=2\beta\sinh[\frac{\pi x}{\beta}]\sqrt{\frac{\sinh[\frac{\pi}{\beta}(l-x)]
\sinh[\frac{\pi}{\beta}(l+s_2-x)]}{\sinh[\frac{\pi l}{\beta}]\sinh[\frac{\pi}{\beta}(l+s_2)]}}.
\end{eqnarray}
When $s_2\ll l$ one can rederive the formula of \cite{Klich} concerning the entanglement hamiltonian of a system
without any projective measurement. i.e.
\begin{eqnarray}\label{beta-Finite T without measurement}
\beta(x)=2\beta\frac{\sinh[\frac{\pi x}{\beta}]\sinh[\frac{\pi}{\beta}(l-x)]
}{\sinh[\frac{\pi l}{\beta}]}.
\end{eqnarray}
It is worth mentioning that the  formula (\ref{beta-Finite T}) in the limit of large temperatures goes to $\beta(x)=2\beta$
which is a constant. This is expected from physical arguments because in the large temperature limit we expect to have just a Gibbs ensemble.

\subsection{Entanglement spectrum and entanglement gaps}

In this section, we study the entanglement spectrum of the system after partial projective measurement. To calculate
this quantity we follow the method of \cite{CalabreseLefevre}. First of all, we note that in the most general case one can write
\begin{eqnarray}\label{tr rho 1}
R_{\alpha}=\tr \rho^{\alpha}=\sum_{i}\lambda_i^{\alpha}=\frac{Z_{\alpha}}{Z_1^{\alpha}}=
\frac{Z_{\alpha}^{geom}Z_{\alpha}^{ann}}{(Z_{1}^{geom}Z_{1}^{ann})^{\alpha}},
\end{eqnarray}
where $\lambda_i$ is the eigenvalue of the reduced density matrix. We first note that when the two regions $B$ and $\bar{B}$ are connected the leading term of the above formula comes 
from the geometric part of the partition function. However, the subleading terms come from the annulus part and one needs to use the
expansion with respect to $q$. Another crucial point is that for the connected cases $s_1$ or $s_2$ is always in the order of lattice spacing which means that 
for sufficiently small or big  $\alpha$'s one can use the extracted formulas. Having all the $S_{\alpha}$'s one can hope to find the
distribution of the eigenvalues of the reduced density matrix. This is the method which has been used in \cite{CalabreseLefevre}
to derive the distribution of the eigenvalues in the case of the no projective measurement and we will also use the same method, for other related study see \cite{LSV2011}. 
The situation is different when the two regions $B$ and $\bar{B}$ are disconnected in a way that $s_1$, $s_2$ and $l$ are all bigger than the lattice spacing.
In this case, the leading term comes from the annulus part of the partition function and one needs to use the expansion with respect to $q$.
However, one should be careful that the expansion can break down for very small $\alpha$. 
For further details see \cite{Rajabpour2016}. This in principle means that  one can not rely on the equations (\ref{power-law decay1})
, (\ref{power-law decay 2}) and (\ref{power-law decay 3}) to get the distribution of eigenvalues. We leave the calculation
of the distribution of the eigenvalues of the non-connected cases as an open problem. We now consider the case of connected regions and write
\begin{eqnarray}\label{tr rho 1 connected}
R_{\alpha}\approx a_{\alpha}L_{eff}^{-\frac{c}{6}(\alpha-\frac{1}{\alpha})}
=a_{\alpha}e^{-b(\alpha-\frac{1}{\alpha})},
\end{eqnarray}
where here we adopted the notation of \cite{CalabreseLefevre} and defined $L_{eff}$ which  have the following form
in the case of the periodic boundary condition, see equation (\ref{SB for PBC}):
\begin{eqnarray}\label{L eff}
L_{eff}=\sqrt{\frac{L}{\pi}\frac{\sin\frac{\pi}{L}(l+s_1)\sin\frac{\pi}{L}l}{s_2\sin\frac{\pi}{L}s_1}}.
\end{eqnarray}
Similar $L_{eff}$ can be also defined  for the semi-infinite case. In addition, we also defined $b=\frac{c}{6}\ln L_{eff}$. 
Having the above formulas the rest of the calculation is identical to \cite{CalabreseLefevre}. 
We are interested to calculate $P(\lambda)=\sum_i\delta(\lambda-\lambda_i)$ which can be derived out of the formula
$\lambda P(\lambda)=\lim_{\epsilon\to 0}\mbox{Im}f(\lambda-i\epsilon)$, where $f(\eta)=\frac{1}{\pi}\sum_{n=1}^{\infty}R_n\eta^{-n}$.
Finally, after some calculations one has
\begin{eqnarray}\label{distribution of eigenvalues}
P(\lambda)=\delta(\lambda_{max}-\lambda)+\frac{b\theta(\lambda_{max}-\lambda)}{\lambda\sqrt{b\ln\frac{\lambda_{max}}{\lambda}}}
I_1\Big{(}2\sqrt{b\ln\frac{\lambda_{max}}{\lambda}}\Big{)},
\end{eqnarray}
where $b=-\ln\lambda_{max}$ and $I_1$ is the modified Bessel function. The above formula is
identical to the result of \cite{CalabreseLefevre} one just needs to consider that we have a new $L_{eff}$. The asymptotic
behavior of the above formula can be derived for the large values of the argument of the modified Bessel function as
\begin{eqnarray}\label{distribution of eigenvalues assymptotic}
P(\lambda)\asymp \frac{1}{\lambda b\ln\frac{\lambda_{max}}{\lambda}}e^{2\sqrt{b\ln\frac{\lambda_{max}}{\lambda}}}.
\end{eqnarray}
It is worth mentioning that
the above results are valid as far as $a_{\alpha}=a^{\frac{c}{6}(\alpha-1/\alpha)}f$, where $f$ is a constant. However, we know that the Affleck-Ludwig
term does not have such kind of form. Considering the Affleck-Ludwig term we have

\begin{eqnarray}\label{distribution of eigenvalues with AL term}
P(\lambda)\asymp \frac{b_0^Ab_0^B}{\lambda b\ln\frac{\lambda_{max}}{\lambda}}e^{2\sqrt{b\ln\frac{\lambda_{max}}{\lambda}}}.
\end{eqnarray}
The above formula shows the interesting physical meaning of $b_0^{A(B)}$ as the degeneracy in the distribution of the eigenvalues of
the reduced density matrix.

Now we will derive the entanglement gap of the system after partial projective measurement. The entanglement gaps are
defined as the difference between the logarithms of the eigenvalues of the reduced density matrices. We first define 
$Z^{cyl}_{\alpha}=e^{-c\frac{h}{12\alpha}}Z^{ann}_{\alpha}$. Then
we  note that one can write
\begin{eqnarray}\label{tr rho}
\tr \rho^{\alpha}=\frac{Z^{geom}_{\alpha}}{(Z^{geom}_{1})^{\alpha}}e^{-\frac{ch}{12}(\frac{1}{\alpha}-\alpha)}\Big{(}
(\frac{q_1^{-c/24}}{Z^{cyl}_{1}})^{\alpha}+\sum_jn_j (\frac{q_1^{\Delta_j-c/24}}{Z^{cyl}_{1}})^{\alpha}\Big{)}.
\end{eqnarray}
Note that for all the limiting cases that we studied so far
the $\frac{Z^{geom}_{\alpha}}{(Z^{geom}_{1})^{\alpha}}e^{-\frac{ch}{12}(\frac{1}{\alpha}-\alpha)}$ is approximately one, 
see \cite{Rajabpour2016}. Then it is easy 
to see that one can identify the following quantities as the eigenvalues of the reduced density matrix:
\begin{eqnarray}\label{tr rhof}
\lambda_j\approx \frac{q_1^{-c/24+\Delta_j+N}}{Z_1^{cyl}}, 
\end{eqnarray}
where the integer $N$ appears because the sum in (\ref{tr rho}) contains also the descendants of the operator with the conformal weight $\Delta_j$.
Finally we can write
\begin{eqnarray}\label{entanglement gap}
\delta\lambda=\ln\lambda_j-\ln\lambda_0\approx(\Delta_j+N)\ln q=-\frac{2\pi^2(\Delta_j+N)}{h}. 
\end{eqnarray}
The above formula is valid for all the cases that we studied in the previous sections. One just needs to use an appropriate $h$ to derive the entanglement
gap in the particular situation. When $s_1=s_2\ll l$ the above formula gives back the result of \cite{CalabreseLefevre,Peschel2,Cardy2015}.
Note that the smaller the $h$ the bigger the gap gets, consequently one expects huge entanglement gap when the two parts 
are disconnected and far from each other. 


\section{Massive field theories}

In this section, we make a list of predictions regarding post measurement entanglement entropy
in massive systems. Most of the upcoming statements are already appeared in \cite{Rajabpour2016} and they were based on numerical calculations on 
the massive Klein-Gordon field theory. It is quite well-known, see \cite{Calabrese2009,Doyon2007,Doyon2009,Casini2009a}, that in the $1+1$ dimensional
massive field theories the entanglement entropy of a subsystem  saturates with the size of the subsystem and is given by
\begin{eqnarray}\label{massive field theory}
S_{\alpha}=-\kappa \frac{c}{12}(1+\frac{1}{\alpha})\ln am+\beta(\kappa),\hspace{1cm}a\ll m^{-1}\ll l,
\end{eqnarray}
where $\kappa$ is the number of contact points between the subsystem and the rest of the system and $l$ is the size of the subsystem. Finally $\beta(\kappa)$
is a model-dependent universal constant \cite{Doyon2009}.
For  results regarding the 
non-critical spin chains see \cite{Korepin}. Note that one can interpret $\xi=m^{-1}$ as the correlation length of the system. The above equation
is an example of
the area-law in the $1+1$ dimension. It has been argued that one way to understand the area-law is based on the short-range correlations
present in the system which has significant contributions just around the contact points of the two regions. Note that the above formula is 
independent of the boundary conditions, in other words, it is valid for also periodic and open systems as far as $\xi$
is much smaller than the length of the system. Based on the above line of thinking
it was argued in \cite{Rajabpour2016} that the above equation should be valid also in the presence of the measurement region
as far as $a\ll m^{-1}\ll l,s$, where $s$ is the length of the measurement region.
This was simply because since projective measurement in part of the system in the massive field theories does not change 
the value of the correlations in the other parts of the system one naturally expects that the only effect of the measurement region be producing
a boundary condition in that part of the system which can just affect the value of $\kappa$ and nothing more. Of course, a priory
it is not guaranteed that the coefficient of the logarithm should be the central charge and indeed we think that
this might be the case just when we perform our measurement in the conformal basis. An exact derivation of the above formula should
be in principle possible by following the arguments based on the form factors of twist operators as it is done for the
 non-measurement case in \cite{Doyon2007,Doyon2009}. It is worth mentioning that if the measurement region is not much bigger than the correlation 
 length $\xi$ we expect 
 \begin{eqnarray}\label{massive field theory 2}
S_{\alpha}=\kappa \frac{c}{12}(1+\frac{1}{\alpha})\ln \frac{\xi(\xi+s)}{as}+\beta(\kappa),\hspace{1cm}a\ll \xi\ll l.
\end{eqnarray}
In the limit of $\xi\ll s$  we are back again to the equation (\ref{massive field theory}). Note that when $s\ll \xi$ we have
just the case without any projective measurement.
The equation (\ref{massive field theory})
make sense just when $\kappa$ is not zero. If the two regions after the projective measurement are completely
decoupled one naturally expect an exponential decay of the entanglement entropy with respect to the distance of the two regions \cite{Rajabpour2016}.
In other words,
\begin{eqnarray}\label{massive field theory disconnected}
S_{\alpha}\asymp_{_{s_{min}\to\infty}} e^{-\gamma(\alpha)ms_{min}},
\end{eqnarray}
where $\gamma(\alpha)$ is a number and $s_{min}$ is the minimum distance between the two regions. In other words,
with the notation of the previous section $s_{min}=\mbox{min}(s_1,s_2)$.

The massive theories are also studied in the presence of the temperature. In the presence of a weak temperature
the R\'enyi entropy follows the following formula \cite{Herzog2013}:
\begin{eqnarray}\label{massive field theory finite temperature}
S(T)-S(0)\sim e^{-\frac{m}{T}}, 
\end{eqnarray}
where $S(0)$ is the R\'enyi entropy of the bipartite system in the zero temperature limit. Because of the short-range nature
of the correlations in the massive systems, it is expected that
the above result is true also in the presence of the measurement region. We will support the above guess later with some numerical
calculations performed on the non-critical Ising model. We summarize this section as follows: because of the short-range nature
of the correlations in the massive systems as far as one does the measurements in the conformal basis we
expect that all the results regarding the non-measurement case be valid also for the post measurement entanglement entropy.
We conjecture that the conclusion is valid independent of the dimensionality of the system.

\section{Post-measurement entanglement entropy in the free fermions}

In this section, we present an efficient numerical method to calculate the entanglement entropy after partial measurement
on the number of fermions on some  of the sites. A similar method was already used in \cite{Rajabpour2015b} to calculate the same quantity 
for the XX-model. The method was inspired by the papers \cite{Peschela}. 
To extend the work of \cite{Rajabpour2015b} we use the results of 
\cite{Barthel2008} as the starting point.
The most general free fermion Hamiltonian is 
\begin{eqnarray}\label{free fermion Hamiltonian}
H=\sum_{ij}\Big{[}c_i^{\dagger}A_{ij}c_j+\frac{1}{2}(c_i^{\dagger}B_{ij}c_j^{\dagger}+h.c.))\Big{]}.	
\end{eqnarray}
We first write the reduced density matrix of a block of fermions $D$  by using block Green matrices. Following \cite{Barthel2008} 
we first define the   operators
\begin{eqnarray}\label{New operators}
a_i=c_i^{\dagger}+c_i,\hspace{1cm}b_i=c_i^{\dagger}-c_i.	
\end{eqnarray}
Then the block Green matrix is defined as
\begin{eqnarray}\label{Green matrix}
G_{ij}=\tr[\rho_Db_ia_j].	
\end{eqnarray}
To calculate the reduced density matrix after partial measurement we need to first define fermionic coherent states. They can be defined as follows:

\begin{eqnarray}\label{fermionic coherent states1}
 |\boldsymbol{\xi}>= |\xi_1,\xi_2,...,\xi_N\rangle= e^{-\sum_{i=1}^N\xi_ic_i^{\dagger}}|0\rangle,
\end{eqnarray}
where $\xi_i$'s are the Grassmann numbers following the  properties: $\xi_n\xi_m+\xi_m\xi_n=0$ and $\xi_n^2=\xi_m^2=0$. Then it is easy to show that
\begin{eqnarray}\label{fermionic coherent states2}
c_i |\boldsymbol{\xi}\rangle= -\xi_i  |\boldsymbol{\xi}\rangle.
\end{eqnarray}
With the same method one can also define another kind of fermionic coherent state as 
\begin{eqnarray}\label{fermionic coherent states3}
 |\boldsymbol{\eta}\rangle= |\eta_1,\eta_2,...,\eta_N\rangle= e^{-\sum_{i=1}^N\eta_ic_i}|1\rangle,
\end{eqnarray}
where $\eta_i$'s are the Grassmann numbers. Then it is easy to show that
\begin{eqnarray}\label{fermionic coherent states4}
c_i^{\dagger} |\boldsymbol{\eta}\rangle= -\eta_i  |\boldsymbol{\eta}\rangle.
\end{eqnarray}
Using the coherent states (\ref{fermionic coherent states1}) the reduced density matrix has the following form \cite{Barthel2008}
\begin{eqnarray}\label{reduced density matrix xi}
<\boldsymbol{\xi}|\rho_D|\boldsymbol{\xi'}>=\det\frac{1}{2}(1-G)e^{-\frac{1}{2}(\boldsymbol{\xi}^*-\boldsymbol{\xi}')^T
F(\boldsymbol{\xi}^*+\boldsymbol{\xi}')},
\end{eqnarray}
where $F=(G+1)(G-1)^{-1}$. If we use (\ref{fermionic coherent states3}) the same reduced density matrix can be written as
\begin{eqnarray}\label{reduced density matrix eta}
\langle\boldsymbol{\eta}|\rho_D|\boldsymbol{\eta'}\rangle=
\det\frac{1}{2}(1+G)e^{-\frac{1}{2}(\boldsymbol{\eta}^*-\boldsymbol{\eta}')^T
F^{-1}(\boldsymbol{\eta}^*+\boldsymbol{\eta}')},
\end{eqnarray}
where $F^{-1}$ is the inverse of the matrix $F$. After diagonalization of the reduced density matrix the R\'enyi
entanglement entropy has the following form \cite{Peschela,Barthel2008,Casini2009a}:
\begin{eqnarray}\label{entanglement entropy}
S_{\alpha}=\frac{1}{1-\alpha}\tr\ln[\Big{(}\frac{1-\sqrt{G^T.G}}{2}\Big{)}^{\alpha}+\Big{(}\frac{1+\sqrt{G^T.G}}{2}\Big{)}^{\alpha}],
\end{eqnarray}
where $G=(F-1)^{-1}(F+1)$. The reason that we prefer to have the form of the entanglement entropy with respect to the
$F$ matrix will be clear soon. Consider now the reduced density matrix of the subsystem $B$ after partial measurement
of the occupation number of the region $A$. This can be calculated in few different but equivalent ways as follows \cite{Rajabpour2015b}: for simplicity consider $1+1$ dimensional system
with the measurement performed on a string of sites (region $A$) with the outcome $|n_1,n_2,...,n_s\rangle$ with $n_j=0,1$ and we are 
interested in the entanglement entropy
of the region $B$ with respect to the rest. To calculate $S_B$ we first calculate $\rho_{A\cup B}$ for the pre-measurement state. To calculate $S_B$
 we need $\rho_{ B}=<n_1,n_2,...,n_s|\rho_{A\cup B}|n_1,n_2,...,n_s\rangle$. The right-hand side can be calculated  using the two equations
 (\ref{reduced density matrix xi}) and (\ref{reduced density matrix eta}). For example, consider that the outcome
 of the measurement on site $j$ is $|0_j\rangle$; then $\rho_{ B}$ can be calculated by  using the equation (\ref{reduced density matrix xi}) and
 putting $\xi_j$ equal to zero. This means that now one can think about a new reduced density matrix 
 \begin{eqnarray}\label{reduced density matrix xi2}
\langle\boldsymbol{\xi},0_j|\rho_{AB}|0_j,\boldsymbol{\xi'}\rangle
=\langle\boldsymbol{\xi}|\tilde{\rho}_B|\boldsymbol{\xi'}\rangle\sim e^{-\frac{1}{2}(\boldsymbol{\xi}^*-\boldsymbol{\xi}')^T
\tilde{F}(\boldsymbol{\xi}^*+\boldsymbol{\xi}')},
\end{eqnarray}
  with the matrix 
 $\tilde{F}_{ln}$ being a subblock of the matrix $F$ with  $l,n\in B$. Putting the new $\tilde{F}$ matrix
 in the equation (\ref{entanglement entropy}) one can find the entanglement entropy of the subsystem $B$ with this condition that
  the site $j$ is empty. Now consider that the outcome
 of the measurement on the site $k$ is $|1_k\rangle$; in this case, one  needs to use the equation (\ref{reduced density matrix eta})
 instead of the equation (\ref{reduced density matrix xi}) and follow the same procedure. 
   For an arbitrary
 outcome $|n_1,n_2,...,n_s\rangle$ one just needs to use the equations (\ref{reduced density matrix xi}) and 
 (\ref{reduced density matrix eta}) as follows: first we put   $\xi_j=0$ for all the empty sites $\{j\}$. Now we have a new Gaussian
 reduced density matrix with $F=\tilde{F}$. After going to the $\eta$
 representation by calculating $(\tilde{F})^{-1}$ we put   $\eta_k=0$ for all the filled sites $\{k\}$. The new reduced density matrix in the
 $\eta$ representation has the form $e^{-\frac{1}{2}(\boldsymbol{\eta}^*-\boldsymbol{\eta}')^T
F_f^{-1}(\boldsymbol{\eta}^*+\boldsymbol{\eta}')}$ with $F_f^{-1}$ being a subblock of the matrix $(\tilde{F})^{-1}$. Finally, we put $F_f$
in the equation (\ref{entanglement entropy}) to calculate the entanglement entropy. 
 Note that the order of using the two
 equations does not change the final outcome as it is expected. In principle, the above procedure
 works in any dimension with an arbitrary outcome for the occupation number measurement. It is worth mentioning that 
 one can totally avoid using (\ref{reduced density matrix eta}) by just starting with (\ref{reduced density matrix xi})  and putting
  $\xi_j=0$ for the $j$'s that correspond to zero fermions. Then for those sites that we have a fermion we just need to Grassmann integrate 
  over the corresponding sites. Note that the Grassmann integration over particular $\xi_k$
  is like putting a fermion in that site. This is simply because we have
 \begin{eqnarray}\label{Grassmann integral}
\int  |\boldsymbol{\xi}\rangle d\xi_k= -|\xi_1,\xi_2,...1_k,..,\xi_N\rangle,
\end{eqnarray} 
 We can now summarize the algorithm for the latter method as follows:  we first calculate $\rho_{A\cup B}$
  with the corresponding $F=F_0$, then we put $\xi_j=0$ whenever the corresponding sites are empty. Now we have
  a new Gaussian reduced density matrix with $F=F_1$. Finally, we perform Grassmann integral of the last
  reduced density matrix over all the $\xi_k$'s
  with the occupied $k$'s. The final reduced density matrix is still Gaussian but with $F=F_2$. Putting this matrix in (\ref{entanglement entropy})
  one can easily calculate the entanglement entropy. 
 In the next sections, we will
 use the above procedure to calculate the post measurement entanglement entropy in the quantum XY chain in the 
 $\sigma^z$ basis.

\section{XY spin chain}

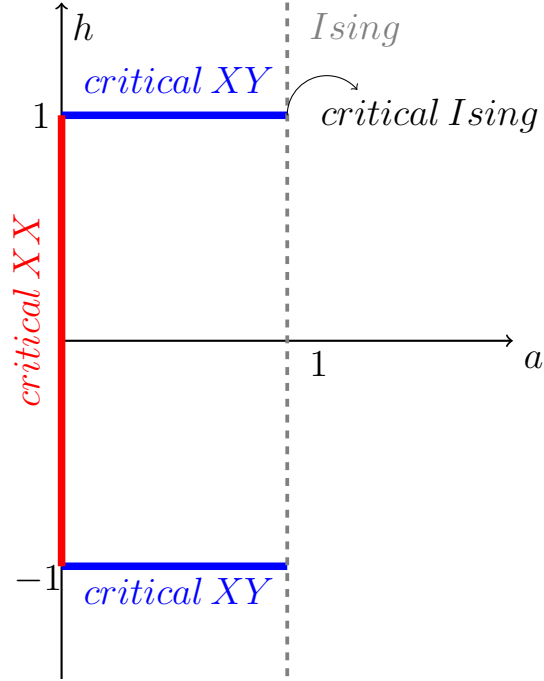
\begin{figure} [htb] 
\centering
\begin{tikzpicture}[scale=3]
\draw [thick, ->] (0,0) -- (2.,0);
\draw [thick] (0,-1) -- (0,-1.5);
\draw [thick, ->] (0,0) -- (0,1.5);
\draw [line width=0.1cm,blue] (0,1) -- (1,1);
\draw [line width=0.1cm,blue] (0,-1) -- (1,-1);
\draw [dashed,line width=0.05cm,gray] (1,1.5) -- (1,-1.5);
\draw [line width=0.1cm,red] (0,1) -- (0,-1);
\node [below right] at (2,0) {\large $a$};
\node [below right] at (0,1.5) {\large$h$};
\node [below right] at (-0.18,1.10) {\large$1$};
\node [below right] at (-0.26,-0.95) {\large$-1$};
\node [below right] at (1.05,0) {\large$1$};
\draw [->] (1,1) arc (180:40:5pt);
\node [above right,rotate=90, red] at (-.05,-0.34){\large $critical\, XX$};
\node [above right, blue] at (.05,1.05){\large $critical\, XY$};
\node [below right, blue] at (.05,-1.0){\large $critical\, XY$};
\node [below right, gray] at (1.05,1.5){\large$Ising$};
\node [right, black] at (1.10,1){\large$critical\,Ising$};
\end{tikzpicture}
\caption{(Color online)Different critical regions in the quantum $XY$ chain. The critical $XX$ 
chain has central charge $c=1$ and critical $XY$ chain has $c=\frac{1}{2}$. } 
\label{fig:XYpahse space}
\end{figure}

In this section we summarize all the necessary formulas and facts regarding the XY-chain. The necessary ingredients for our numerical calculations are 
the $G$ matrices and the  configurations that lead to the conformal boundary conditions. 
The Hamiltonian of the XY-chain is  as follows:
\begin{eqnarray}\label{Hamiltonian Ising}
H=-\sum_{j=1}^L\Big{[}(\frac{1+a}{2})\sigma_j^x\sigma_{j+1}^x+(\frac{1-a}{2})\sigma_j^y\sigma_{j+1}^y+h\sigma_j^z\Big{]}.
\end{eqnarray}
After using  Jordan-Wigner transformation, i.e. 
$c_j=\prod_{_{m<j}}\sigma_m^z\frac{\sigma_j^x-i\sigma_j^y}{2}$ and $\mathcal{N}=\prod_{_{m=1}}^{^{L}}\sigma_m^z=\pm1$ 
with $c_{L+1}^{\dagger}=0$ and $c_{L+1}^{\dagger}=\mathcal{N}c_{1}^{\dagger}$ for open and periodic boundary conditions respectively
 the Hamiltonian will have the following form:
 \begin{equation}\label{Ising Hamiltonian free fermion}
H=\sum_{j=1}^{L-1} \Big{[}(c_j^{\dagger}c_{j+1}+ac_j^{\dagger}c_{j+1}^{\dagger}+h.c.)-h(2c_j^{\dagger}c_j-1)\Big{]}+
\mathcal{N}(c_L^{\dagger}c_1+ac_L^{\dagger}c_1^{\dagger}+h.c.).
\end{equation}
Note that since $[H,\mathcal{N}]=0$ one needs to consider the two sectors independently and find the ground state 
of the spin chain as the ground state of the sector $\mathcal{N}=1$ or the first excited state of the sector $\mathcal{N}=1$.
Here we always concentrate on the cases that the ground state of the spin chain is in the sector $\mathcal{N}=1$.  The above Hamiltonian has a very rich phase diagram with different critical regions \cite{Mccoy1971}. 
In figure ~(\ref{fig:XYpahse space}) we show different critical regions of 
the system.

Because of the Jordan-Wigner transformation if the $\sigma^z$ is up(down) at site $j$ one can interpret it as having (lacking) a fermion at the same site.
Because of the Jordan-Wigner transformation if the $\sigma^z$ is up(down) at site $j$ one can interpret it as having (lacking) a fermion at the same site.
This correspondence helps us to calculate the entanglement entropy in the XY chain after projective measurement in the $\sigma^z$ basis by using the 
results of the previous section.
In the next subsections, we will summarize the formulas regarding the $G$ matrix for the XY chain, see for example \cite{Fagotti}.
We also comment on the configurations that lead to conformal boundary conditions \cite{Najafi}.

\subsection{Critical transverse field Ising chain}\label{sec:Critical transverse field Ising chain1}

In this section, we list some of the known facts about Ising model. Here, we first list the correlation matrices
necessary to calculate the post measurement entanglement entropy and then we present the results known about the conformal configurations and the
conformal field theory of the Ising model.

\subsubsection{Correlation matrices:}

When the size of the total system is finite $L$, depending on the form of the boundary conditions, periodic or open; $G$ at the Ising critical point
has the following forms:
  \begin{eqnarray}\label{G Ising PBC }
G^P_{ij}&=&-\frac{1}{L\sin(\frac{\pi(i-j+1/2)}{L})},\\
\label{G Ising OBC }
G^O_{ij}&=&-\frac{1}{2L+1}\Big{(}\frac{1}{\sin(\frac{\pi(i-j+1/2)}{2L+1})}+\frac{1}{\sin(\frac{\pi(i+j+1/2)}{2L+1})}\Big{)}.
\end{eqnarray}
Notice that for $L\to\infty$ the first equation reduces to the one corresponding to the infinite chain
 and the second equation  gives the result for the semi-infinite chain. The critical XY line in the figure ~3
 is in the same universality class as the Ising critical point and has the central charge $c=\frac{1}{2}$. The Green matrix for the infinite
 system is given by
 \begin{eqnarray}\label{Green matrix XY chain universality}
G_{ij}=\int_0^{\pi}\frac{d\phi}{\pi}\frac{(\cos\phi-1)\cos[(i-j)\phi]-a\sin\phi\sin\big[(i-j)\phi]}{\sqrt{(1-\cos\phi)^2+a^2\sin^2\phi}}.
\end{eqnarray}
The above Green matrix is useful to check the universality of the results.

\subsubsection{Boundary conformal field theory of the Ising model:}

There are two different conformal boundary conditions compatible with the  CFT of the Ising model, free and fixed boundary conditions \cite{Cardy1989}.
Here, free and fixed refers to the state of the spin in the $\sigma^x$ direction. These two boundary conditions can produce four different
partition functions: 1) fixed with spins in the same direction on both boundaries  "Fi1-Fi1"
2) fixed with spins in the opposite direction "Fi1-Fi2" 3) free on one boundary and fixed on the other one "Fr-Fi" and 4) free on both boundaries "Fr-Fr".
The corresponding partition functions on the cylinder with the length $\frac{h}{\alpha}$ and the circumference $2\pi$ can be written 
with respect to characters as follows
\begin{eqnarray}\label{partition functions Ising}
Z_{_{Fi1-Fi1}}=\chi_0(\tau)+\chi_{1/2}(\tau)+\sqrt{2}\chi_{1/16}(\tau),\\
Z_{_{Fi1-Fi2}}=\chi_0(\tau)+\chi_{1/2}(\tau)-\sqrt{2}\chi_{1/16}(\tau),\\
Z_{_{Fr-Fr}}=\chi_0(\tau)+\chi_{1/2}(\tau),\\
Z_{_{Fr-Fi}}=\chi_0(\tau)-\chi_{1/2}(\tau),
\end{eqnarray}
where the characters are defined as follows:
\begin{eqnarray}\label{characters}
\chi_0(\tau)=\frac{1}{2\sqrt{\eta(\tau)}}\Big{(}\sqrt{\Theta_3(\tilde{q}_{\alpha}^{1/2})}+\sqrt{\Theta_4(\tilde{q}_{\alpha}^{1/2})}\Big{)}
=\tilde{q}_{\alpha}^{-1/48}(1+\tilde{q}_{\alpha}^2+\tilde{q}_{\alpha}^3+...),\hspace{1cm}\\
\chi_{1/16}(\tau)=\frac{1}{2\sqrt{\eta(\tau)}}\sqrt{\Theta_2(\tilde{q}_{\alpha}^{1/2})}
=\tilde{q}_{\alpha}^{-1/48+1/16}(1+\tilde{q}_{\alpha}+\tilde{q}_{\alpha}^2+2\tilde{q}_{\alpha}^3+...),\\
\chi_{1/2}(\tau)=\frac{1}{2\sqrt{\eta(\tau)}}\Big{(}\sqrt{\Theta_3(\tilde{q}_{\alpha}^{1/2})}-\sqrt{\Theta_4(\tilde{q}_{\alpha}^{1/2})}\Big{)}
=\tilde{q}_{\alpha}^{-1/48+1/2}(1+\tilde{q}_{\alpha}+\tilde{q}_{\alpha}^2+...).
\end{eqnarray}
where $\Theta_i$'s are the Jacobi theta functions and  $\tilde{q}_{\alpha}=e^{\pi i\tau}$ with $\tau=i\frac{h}{\pi\alpha}$ is as before. Finally
$\eta$ is the Dedekind function with the following definition
\begin{eqnarray}\label{Dedekind eta}
\eta(q)=q^{\frac{1}{24}}\prod_{n=1}^{\infty}(1-q^n).
\end{eqnarray}
There are some comments in order: first of all, for the first two partition functions, the smallest
non-trivial scaling dimension is $\Delta_1=\frac{1}{16}$ which is the scaling dimension of the spin operator. However, for the last two
$\Delta_1=\frac{1}{2}$ which is the scaling dimension of the energy operator. Another interesting fact is that
\begin{eqnarray}\label{partition functions identities}
Z_{_{Fi1-Fi1}}+Z_{_{Fi1-Fi2}}=2Z_{_{Fr-Fr}}.
\end{eqnarray}
Which means that the partition function of the Ising model with the fixed boundaries, as far as we do not know the nature of the fixed boundary conditions,
is proportional to the partition function of the Ising model with the free boundaries. In the next subsection, we will comment
on the configurations that lead to the above boundary conditions. Finally, it is important to also comment on the parameter $b_0$
that appears in the study of Affleck-Ludwig term
for different boundaries. Based on the above formulas it is easy to identify
\begin{eqnarray}\label{b0}
b_0^{Fr}=1,\hspace{1cm}b_0^{Fi}=\frac{1}{\sqrt{2}},
\end{eqnarray}
for the free and fixed boundary conditions respectively.

\subsubsection{Conformal configurations:}

The conformal configurations for the  critical XY line (including the Ising point) in the $\sigma^z$ basis are already studied in \cite{Najafi} and we summarize
the results here. All the   configurations with the crystal structure are flowing to  conformal boundary conditions. This has
been shown by studying the formation probability of {\it{crystal}} configurations and comparing the results with the CFT predictions. 
Formation probability of a configuration is the probability of occurrence of that configuration in the spin chain.
We list here the most interesting examples of the crystal configurations:

\begin{description}
 \item  \item[a] $(|\uparrow,\uparrow,\uparrow,\uparrow,...>)$ 

\item  \item[b] $(|\downarrow,\downarrow,\downarrow,\downarrow,...>)$

 \item  \item[c]  $(|\downarrow,\uparrow,\downarrow,\uparrow,...>)$ 
\end{description}
Definition of more complicated crystal configurations is quite straightforward. We can define some of  them by labeling the configuration by a number $x$
which is the ratio of the number of down spins to the total number of  spins in a base of a crystal configuration. For example,
we have $x_a=0$, $x_b=1$ and $x_c=\frac{1}{2}$. Note that there are infinite different crystal configurations with
the same $x$. For example, the configuration $(|\downarrow,\downarrow,\uparrow,\uparrow,\downarrow,\downarrow,...>)$ is also $x=\frac{1}{2}$.
We call this configuration, which can be derived from the configuration \textbf{c} by doubling every spin, the configuration $(\frac{1}{2},2)$.
We can now define a class of crystal configurations $(x,k)$, where $x$ is defined as before and $k$ is the number of neighboring down
spins in a base of the crystal configuration with this condition that in the base of the crystal all the up (down) spins are neighbors. With the above definition  $(1,1)$, $(\frac{1}{2},1)$ are the configurations 
 \textbf{b} and \textbf{c} respectively. Exceptionally, for later convinience, we take the configuration $(1,0)$ as the configuration \textbf{a}. Note that although the above configurations do not exhaust all the possible crystal
configurations they are quite enough for our purpose.
It is expected that all of the above crystal configurations flow to conformal boundary conditions \cite{Najafi}. It is worth mentioning that
although all of these configurations flow to conformal boundary conditions it is not a priory clear that they flow to what kind of conformal boundary
conditions. For example, in the case of the Ising model we have two possible different conformal boundary conditions, free and fixed \cite{Cardy2004,Cardy1989}.
It was argued in \cite{Stephan2013} that all the spins up configuration should flow to free boundary condition. 
 In the case of the free-free boundary conditions
on the two slits 
the smallest scaling dimension present in the partition function of the annulus is $\Delta=\frac{1}{2}$ which is the scaling dimension of
the energy operator \cite{Cardy1989}. Of course, this fact is important when we discuss disconnected cases.
Numerical calculations of the formation probabilities performed in the presence of a boundary
show that all the configurations $(x,2k)$ also flow to free boundary conditions \cite{Najafi}. However, 
the configurations $(x,2k+1)$, including the configurations \textbf{b} and \textbf{c}, flow to fixed boundary conditions. 
 The above considerations suggest that all of our CFT results should 
 be valid for all the crystal configurations as far as the system is infinite or we have periodic boundary condition. We do not expect 
 the validity of our results for the configurations $(x,2k+1)$ when the system has an open boundary condition. This is simply because
 since the natural boundary of the Ising chain that we are considering has a free boundary condition
 if the configuration induces a fixed boundary condition on the slit one needs to consider also the effect of the boundary changing operator.
For the configurations $(x,2k)$ the presented CFT results should be valid also in the presence of the open boundary condition.
  We will numerically show
 that the above conclusions are indeed the case when we study the critical transverse field
 Ising chain.

Using the numerical calculations 
in \cite{Najafi} it was argued that not only the crystal configurations but also some configurations that 
although not perfectly crystal  but very close to that can also flow to a boundary conformal field theory. This fact will be important 
in our later discussion regarding the localizble entanglement \cite{Najafi-nonselective}.
Finally, it is worth mentioning that all of the above results are valid when we are making the measurement in the $\sigma^z$ basis. The
situation changes
completely if one makes a measurement in the $\sigma^x$ basis.

 \subsection{XX critical line}
 In this subsection, we list all the relevant results regarding the correlation matrices and the 
 conformal configurations of the XX model. We will also list the formulas regarding the CFT of the XX chain.
 
 \subsubsection{Correlation matrices:}

The critical XX chain $a=0$ has a very different structure than the critical Ising chain. It has $U(1)$ symmetry which guaranties
the conservation of the total number of up spins, in other words, the number of fermions.  Since in this model $< c_i^{\dagger}c_j^{\dagger}>=< c_ic_j>=0$
one can write $G_{ij}=2C_{ij}-\delta_{ij}$, where  $C_{ij}=< c_i^{\dagger}c_j>$. 
For the periodic boundary condition provided $\frac{L}{2\pi }\arccos(-h)\notin N$ the form of the  $C$ matrix is \cite{Fagotti}
 \begin{eqnarray}\label{correlation matrix XX PBC}
C_{ij}=\frac{1}{L}\sum_{k=1}^L e^{\frac{2\pi i k(j-i)}{L}}\theta(h+\cos\frac{2\pi k}{L});
\end{eqnarray}
where $\theta(x)=\frac{1+\mbox{sgn(x)}}{2}$. When the ground state is 
non-degenerate \footnote{the ground state is degenerate, for example, when $\frac{L}{2\pi }\arccos(-h)\in N$, see \cite{Fagotti}. }
and the magnetic field is non-zero we have

\begin{eqnarray}\label{G XX PBC }
C^P_{ij}=\frac{n_f}{\pi}\delta_{ij}+(1-\delta_{ij})\frac{\sin(n_f(i-j))}{L\sin(\frac{\pi(i-j)}{L})},
\end{eqnarray}
where $n_f=\frac{\pi}{L}\Big{(}2 \lceil\frac{L}{2\pi}\arccos(-h) \rceil-1\Big{)}$ is the Fermi
momentum  and 
$\lceil x\rceil $ is the closest integer larger than $x$.

For the open boundary condition provided $h+\cos\frac{\pi k}{L+1}\neq 0$ the form of the  $C$ matrix is:
\begin{eqnarray}\label{correlation matrix XX OBC}
C_{ij}=\frac{2}{L+1}\sum_{k=1}^L \sin\frac{\pi k i}{L+1} \sin\frac{\pi k j}{L+1}      \theta(h+\cos\frac{\pi k}{L+1});
\end{eqnarray}
where $\theta(x)$ is defined as above. For a non-zero magnetic field when the ground state is non-degenerate, i.e. $h+\cos\frac{\pi k}{L+1}\neq 0$ we have
\begin{eqnarray}\label{G-XX-OBC}
\hspace{-2cm}C^O_{ij}=\Big{(}\frac{1}{2}-(\frac{L}{2(L+1)}-\frac{n'_f}{\pi})\Big{)}\delta_{ij}\nonumber\\
+(1-\delta_{ij})\frac{1}{2(L+1)}
\Big{(}\frac{\sin(n'_f(i-j))}{\sin(\frac{\pi(i-j)}{2L+2})}-\frac{\sin(n'_f(i+j))}{\sin(\frac{\pi(i+j)}{2L+2})}\Big{)},
\end{eqnarray}
where $n'_f=\frac{\pi}{2(L+1)}\Big{(}1+2 	\lfloor\frac{(L+1)}{\pi}\arccos(-h)) 	\rfloor\Big{)}$ and  
$\lfloor x\rfloor$ is the closest integer smaller than $x$. The form of the above correlation matrix is consistent with
the Dirichlet boundary condition.
Note that  one can get the results for the infinite and the semi-infinite cases by just sending $L$
to infinity.

\subsubsection{Boundary conformal field theory of the XX chain:}

It is quite well-known that the continuum limit of the XX chain can be described by a compactified free bosonic theory,
\begin{eqnarray}\label{free boson}
S=\frac{1}{2}\int dx_1 dx_2 [(\partial_1\phi)^2+(\partial_2\phi)^2];
\end{eqnarray}
where $\phi\equiv \phi+2\pi r$ with $r=\frac{1}{2\sqrt{\pi}}$. There are two possible conformal boundary conditions,
Dirichlet and Neumann. Since in this work we do not face Neumann boundary condition we will just focus on the Dirichlet boundary condition. 
 The partition function of the free compactified bosonic theory
on the finite cylinder
has the following form
\begin{eqnarray}\label{partition function Dirichlet}
Z_{DD}(q)=\frac{1}{\eta(q)}\sum_{n\in \mathbb{Z}} q^{\frac{1}{2}(n+\delta)^2},
\end{eqnarray}
where $\delta=\frac{\phi_1-\phi_2}{\sqrt{\pi}}$ with $\phi_1$ and $\phi_2$ being the value of the field $\phi$ on the two boundaries.
 The above equation
means that 
for positive $\delta$ the smallest scaling dimension is
\begin{eqnarray}\label{scaling dimension dirichlet1}
\Delta_{1}=\left\{
\begin{array}{c l} 
\frac{1}{2} & \delta=0,\hspace{1cm}\\
    \frac{\delta^2}{2} & 0<\delta\leq \frac{1}{2},\hspace{1cm}\\
        \frac{(\delta-1)^2}{2}& \frac{1}{2}\leq\delta<1.\\        
\end{array}\right.
\end{eqnarray}
The above scaling dimensions will frequently appear in our later numerical calculations.

\subsubsection{Conformal configurations:}

It has been already shown that
the all spins up and all spins down configurations, i.e. \textbf{a} and \textbf{b}, do not lead to conformal boundary conditions, see for example
\cite{Stephan2013,Najafi}. This is possible because  the XX chain has a $U(1)$ symmetry which keeps the number of fermions fixed. To have all the spins
up one needs to inject fermions which are in contrast with the $U(1)$ symmetry. However, the antiferromagnetic configuration, i.e. \textbf{c}, 
leads to a conformal
boundary condition if one works with the half-filling case, see \cite{Najafi}. It was shown in \cite{Najafi} that for an infinite  system
with the Fermi momentum $n_f$ just the configurations with $x=\frac{n_f}{\pi}$ flow to conformal boundary conditions. 
In addition based on the numerical results of \cite{RajabCasimir} one can conjecture that
the corresponding boundary conditions are all Dirichlet boundary conditions.

We argued in the above that all the configurations $(\frac{n_f}{\pi},k)$ flow to Dirichlet boundary conditions but a priory 
it is not clear what is the value of $\phi$ on the boundary for different configurations. If one takes similar configurations
on the two slits one is left with $\delta=0$ and consequently the smallest scaling dimension in the spectrum is $\Delta=\frac{1}{2}$.
However, if the configurations on the two slits are different one expect to find non-zero $\delta$ which means a different spectrum
for the system. Our CFT results suggest that the post measurement entanglement entropy changes like a power-law with an exponent which depends on
the smallest scaling dimension present in the system. This means that one can find $\delta$ corresponding to Dirichlet boundary conditions
by studying the post measurement entanglement entropy. Note that since different values of $\delta$ can lead to the same 
$\Delta_1$ the value of $\delta$ can not be fixed uniquely. In principle, we  have $\delta=\sqrt{2\Delta_1}$
or $\delta=1-\sqrt{2\Delta_1}$. The two different $\delta$'s although lead to the same smallest scaling dimension they have different
partition functions. To fix the total spectrum of the system with a Dirichlet boundary condition one needs to also extract the second smallest scaling dimension.
In this work, we will concentrate on the smallest scaling dimension and leave the corrections to future studies.
The conclusion of the above argument is that the  post measurement entanglement entropy provides a  method to 
characterize the conformal boundary conditions. We will study in the next sections many different configurations based on the above idea. It is worth mentioning that
one can also extract similar results using the formation probabilities, see \cite{Najafi}.

The Dirichlet-Dirichlet partition function that we wrote in the above can be also expressed in the $\tilde{q}$ representation as follows
\begin{eqnarray}\label{partition function Dirichlet tilde q}
Z_{DD}(\tilde{q})=\frac{(b_0^{DD})^2}{\eta(\tilde{q})}\sum_{n\in \mathbb{Z}} \tilde{q}^{\frac{n^2}{4}}e^{2\pi n\delta i},
\end{eqnarray}
where $\eta(\tilde{q})=\sqrt{\frac{\pi}{h}}\eta(q)$ and $b_0^{DD}=1$. The above results indicate that in this case the boundary
entropy $S^{AL}$ independent of the configuration is zero.

Finally, it is worth mentioning that all of the above results are valid if we make the projective measurement in the $\sigma^z$ basis.
When the measurement is done in the $\sigma^x$-basis it is expected that the boundary  flows to a conformal Neumann boundary condition. Consequently,
one needs to work with either $Z_{NN}$ or $Z_{DN}$, where $N$ and $D$ stands for the Neumann and the Dirichlet. In these cases, first of
all, the spectrum of the system is different and in addition the Affleck-Ludwig term is not zero anymore. We leave more through analysis of
the $\sigma^x$ basis for a future work.

\section{Entanglement entropy after selective measurements in the  critical Ising chain}

In this section, we will check the validity of the post measurement entanglement entropy
formulas derived in the section ~3 for the Ising chain. In other words we will check the validity of the formulas:  (\ref{rajab2015})
, (\ref{power-law decay1}), (\ref{SB for PBC}), (\ref{power-law decay 2}), (\ref{power-law decay 3}) and (\ref{SB for OBC}). 
The  formulas (\ref{rajab2015}), (\ref{SB for PBC}) and (\ref{SB for OBC}) are
the post measurement entanglement entropy of two connected regions and the other three are the ones related to the disconnected regions, see Figures (\ref{fig:setups1}) and (\ref{fig:setups2}).
We perform the measurement  in the $\sigma^z$
basis so that we can use the results of the section ~5.  From now on it is useful to fix some notations regarding the exponents appearing in the disconnected
cases. First of all we define the setups leading to the equations (\ref{power-law decay1}), (\ref{power-law decay 2}), (\ref{power-law decay 3})
as follows:

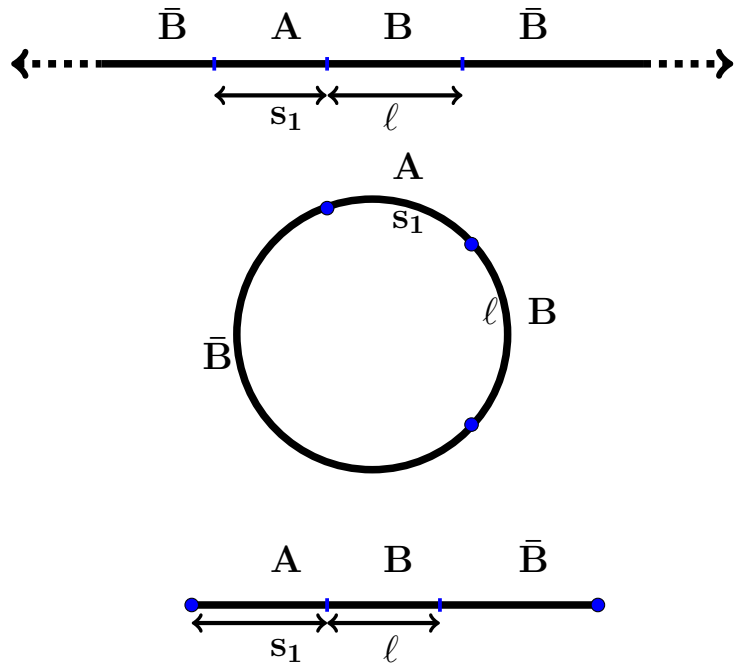
\begin{figure} [htb] 
\centering
\begin{tikzpicture}[scale=0.6]

\draw [line width=0.1cm ,black] (0,0) -- (12,0);
\draw [<- ,line width=0.1cm,dashed,black] (-2,0) -- (0,0);
\draw [-> ,line width=0.1cm,dashed,black] (12,0) -- (14,0);
\draw [line width=0.05cm ,blue] (2.5,0.15) -- (2.5,-0.15);
\draw [line width=0.05cm ,blue] (5,0.15) -- (5,-0.15);
\draw [line width=0.05cm ,blue] (8,0.15) -- (8,-0.15);
\node [above right, black] at (1.,0.3){\large$\bf{\bar{B}}$};
\node [above right, black] at (3.5,0.3){\large$\bf{A}$};
\node [above right, black] at (6,0.3){\large$\bf{B}$};
\node [above right, black] at (9,0.3){\large$\bf{\bar{B}}$};
\draw [ ultra thick , <-> ] (2.5,-0.7) -- (4.99,-0.7);
\draw [ ultra thick , <-> ] (5.01,-0.7) -- (8,-0.7);
\node [above right, black] at (3.5,-1.7){\large$\bf{s_{1}}$};
\node [above right, black] at (6.,-1.7){\large$\bf{\ell\enspace}$};

\draw [line width=0.1cm,black ] (6,-6) circle [radius=3];
\draw [fill=blue] (5.,-3.2) circle [radius=0.15];
\draw [fill=blue] (8.2,-4.) circle [radius=0.15];
\draw [fill=blue] (8.2,-8) circle [radius=0.15];
\node [above right, black] at (6.2,-2.8){\large$\bf{A}$};
\node [above right, black] at (9.2,-6){\large$\bf{B}$};
\node [above right, black] at (2.,-7){\large$\bf{\bar{B}}$};
\node [above right, black] at (6.2,-4){\large$\bf{s_{1}}$};
\node [above right, black] at (8.2,-6){\large$\bf{\ell\enspace}$};

\draw [line width=0.1cm ,black] (2,-12) -- (11,-12);
\draw [fill=blue] (2,-12) circle [radius=0.15];
\draw [fill=blue] (11,-12) circle [radius=0.15];
\draw [line width=0.05cm ,blue] (5,-12.15) -- (5,-11.85);
\draw [line width=0.05cm ,blue] (7.5,-12.15) -- (7.5,-11.85);
\draw [ ultra thick , <-> ] (2.,-12.4) -- (5,-12.4);
\draw [ ultra thick , <-> ] (5.,-12.4) -- (7.5,-12.4);
\node [above right, black] at (3.5,-11.5){\large$\bf{A}$};
\node [above right, black] at (6,-11.5){\large$\bf{B}$};
\node [above right, black] at (9,-11.5){\large$\bf{\bar{B}}$};
\node [above right, black] at (3.5,-13.5){\large$\bf{s_{1}}$};
\node [above right, black] at (6,-13.5){\large$\bf{\ell\enspace}$};

\end{tikzpicture}
\caption{Color online)  different setups for the post measurement entanglement entropy in the connected cases. } 
\label{fig:setups1}
\end{figure}



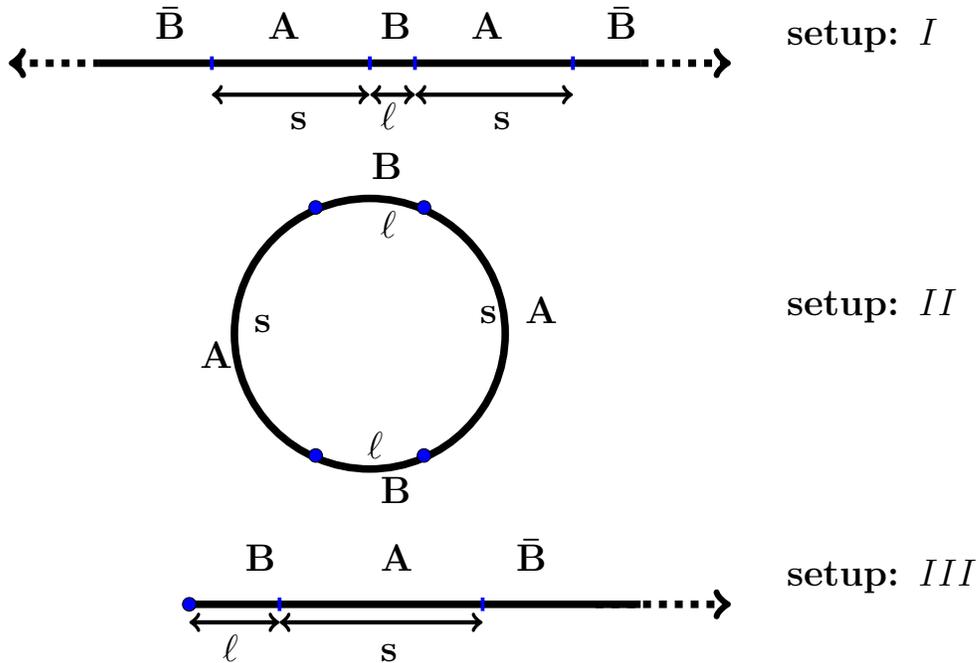
\begin{figure} [hthp!] 
\centering
\begin{tikzpicture}[scale=0.6]

\draw [line width=0.1cm ,black] (0,0) -- (12,0);
\draw [<- ,line width=0.1cm,dashed,black] (-2,0) -- (0,0);
\draw [-> ,line width=0.1cm,dashed,black] (12,0) -- (14,0);
\draw [line width=0.05cm ,blue] (2.5,0.15) -- (2.5,-0.15);
\draw [line width=0.05cm ,blue] (6,0.15) -- (6,-0.15);
\draw [line width=0.05cm ,blue] (7,0.15) -- (7,-0.15);
\draw [line width=0.05cm ,blue] (10.5,0.15) -- (10.5,-0.15);
\node [above right, black] at (1.,0.3){\large$\bf{\bar{B}}$};
\node [above right, black] at (3.5,0.3){\large$\bf{A}$};
\node [above right, black] at (6,0.3){\large$\bf{B}$};
\node [above right, black] at (8,0.3){\large$\bf{A}$};
\node [above right, black] at (11.,0.3){\large$\bf{\bar{B}}$};
\draw [ ultra thick , <-> ] (2.5,-0.7) -- (5.99,-0.7);
\draw [ ultra thick , <-> ] (6.01,-0.7) -- (7,-0.7);
\draw [ ultra thick , <-> ] (7.01,-0.7) -- (10.5,-0.7);
\node [above right, black] at (4.,-1.7){\large$\bf{s}$};
\node [above right, black] at (6.,-1.7){\large$\bf{\ell\enspace}$};
\node [above right, black] at (8.5,-1.7){\large$\bf{s}$};
\node [above right, black] at (15,0){\large \bf{setup: $I$}};

\draw [line width=0.1cm,black ] (6,-6) circle [radius=3];
\draw [fill=blue] (4.8,-3.2) circle [radius=0.15];
\draw [fill=blue] (7.2,-3.2) circle [radius=0.15];
\draw [fill=blue] (7.2,-8.7) circle [radius=0.15];
\draw [fill=blue] (4.8,-8.7) circle [radius=0.15];
\node [above right, black] at (5.8,-2.8){\large$\bf{B}$};
\node [above right, black] at (9.2,-6){\large$\bf{A}$};
\node [above right, black] at (6.,-10){\large$\bf{B}$};
\node [above right, black] at (2.,-7){\large$\bf{A}$};
\node [above right, black] at (6.,-4.1){\large$\bf{\ell\enspace}$};
\node [above right, black] at (8.2,-6){\large$\bf{s}$};
\node [above right, black] at (5.7,-9){\large$\bf{\ell\enspace}$};
\node [above right, black] at (3.2,-6.2){\large$\bf{s}$};
\node [above right, black] at (15,-6){\large \bf{setup: $II$}};

\draw [line width=0.1cm ,black] (2,-12) -- (12,-12);
\draw [fill=blue] (2,-12) circle [radius=0.15];
\draw [-> ,line width=0.1cm,dashed,black] (11,-12) -- (14,-12);
\draw [line width=0.05cm ,blue] (4,-12.15) -- (4,-11.85);
\draw [line width=0.05cm ,blue] (8.5,-12.15) -- (8.5,-11.85);
\draw [ ultra thick , <-> ] (2.,-12.4) -- (4,-12.4);
\draw [ ultra thick , <-> ] (4.,-12.4) -- (8.5,-12.4);
\node [above right, black] at (3.,-11.5){\large$\bf{B}$};
\node [above right, black] at (6,-11.5){\large$\bf{A}$};
\node [above right, black] at (9,-11.5){\large$\bf{\bar{B}}$};
\node [above right, black] at (2.5,-13.5){\large$\bf{\ell\enspace}$};
\node [above right, black] at (6,-13.5){\large$\bf{s}$};
\node [above right, black] at (15,-12){\large \bf{setup: $III$}};
\end{tikzpicture}
\caption{Color online)  Three setups regarding post measurement entanglement
entropy for disconnected cases.} 
\label{fig:setups2}
\end{figure}


\begin{description}
 \item  \item[Setup I:]The total system is infinite and the measurement region $A$ is made of two  
 large disconnected regions with each of them with the length $s$
 around the domain $B$ with length $l$: The post measurement entanglement entropy, with this condition that the result of the measurement
 on the two regions are $C_1$ and $C_2$, (up to a logarithm for $\alpha=1$) decays as
  \begin{eqnarray}\label{exponential setup I}
S_{\alpha}\asymp \Big{(}\frac{l}{8s}\Big{)}^{\Delta_{I}^{\{C_1,C_2\}}(\alpha)},
\end{eqnarray}
where
 \begin{eqnarray}\label{exponent setup I}
\Delta_{I}^{\{C_1,C_2\}}(\alpha)=\left\{
\begin{array}{c l}      
    2\alpha\Delta_1^{\{C_1,C_2\}}, & \alpha<1,\hspace{1cm}\\
        2\Delta_1^{\{C_1,C_2\}} &\alpha\geq1,\\        
\end{array}\right.
\end{eqnarray}
where $\Delta_1^{\{C_1,C_2\}}$ is the smallest scaling dimension present in the spectrum of the system. Note that this exponent can
be dependent on the configuration.
\item  \item[Setup II:] The system is periodic with the finite size $L$. The  measurement region $A$ is made of two equal 
 large disconnected regions in a way that the regions $B$ and $\bar{B}$ have the same size $l$. The post measurement entanglement entropy
 (up to a logarithm for $\alpha=1$) changes as
  \begin{eqnarray}\label{exponential setup II}
S_{\alpha}\asymp \Big{(}\frac{\pi l}{4L}\Big{)}^{\Delta_{P}^{\{C_1,C_2\}}(\alpha)},
\end{eqnarray}
where
 \begin{eqnarray}\label{exponent setup II2}
\Delta_{P}^{\{C_1,C_2\}}(\alpha)=\left\{
\begin{array}{c l}      
    4\alpha\Delta_1^{\{C_1,C_2\}}, & \alpha<1,\hspace{1cm}\\
        4\Delta_1^{\{C_1,C_2\}} &\alpha\geq1,\\        
\end{array}\right.
\end{eqnarray}
where as before $\Delta_1^{\{C_1,C_2\}}$ is the smallest scaling dimension present in the spectrum of the system.

 \item  \item[Setup III:] The system is semi-infinite and the measurement region $A$ is made of one connected large 
 domain with the size $s$ and the configuration $C$. The simply connected domain  $B$ with the size $l$ starts from the origin. The post measurement entanglement entropy
 (up to a logarithm for $\alpha=1$) changes as
  \begin{eqnarray}\label{exponential setup III}
S_{\alpha}\asymp \Big{(}\frac{ l}{4s}\Big{)}^{\Delta_{O}^{\{C\}}(\alpha)},
\end{eqnarray}
where
 \begin{eqnarray}\label{exponent setup II}
\Delta_{O}^{\{C\}}(\alpha)=\left\{
\begin{array}{c l}      
    4\alpha\Delta_1^{\{C\}}, & \alpha<1,\hspace{1cm}\\
        4\Delta_1^{\{C\}} &\alpha\geq1,\\        
\end{array}\right.
\end{eqnarray}
where $\Delta_1^{\{C\}}$ is again the smallest scaling dimension present in the spectrum of the system. Note that we will follow the same
notation also for the XX model.

\end{description}

\subsection{Connected regions}\label{Critical transverse field Ising chain}

In this subsection, we check the validity of the formulas  (\ref{rajab2015}), (\ref{SB for PBC}) and (\ref{SB for OBC})
for the critical transverse field Ising chain. We will just focus on the leading term in the corresponding formulas.

\begin{figure}   [hthp!]
\centering
\includegraphics[width=0.4\textwidth,angle =-90]{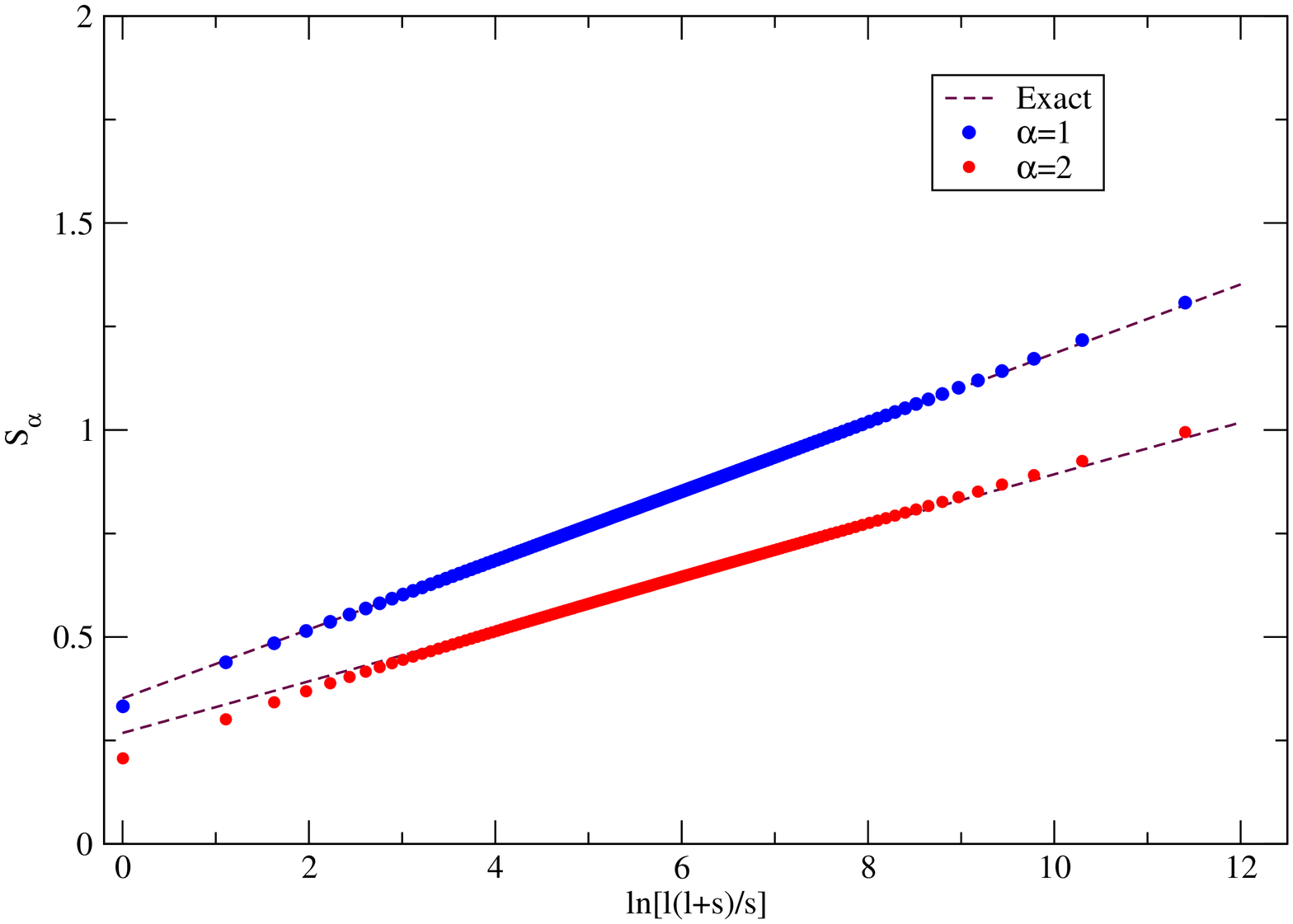}
\includegraphics[width=0.4\textwidth,angle =-90]{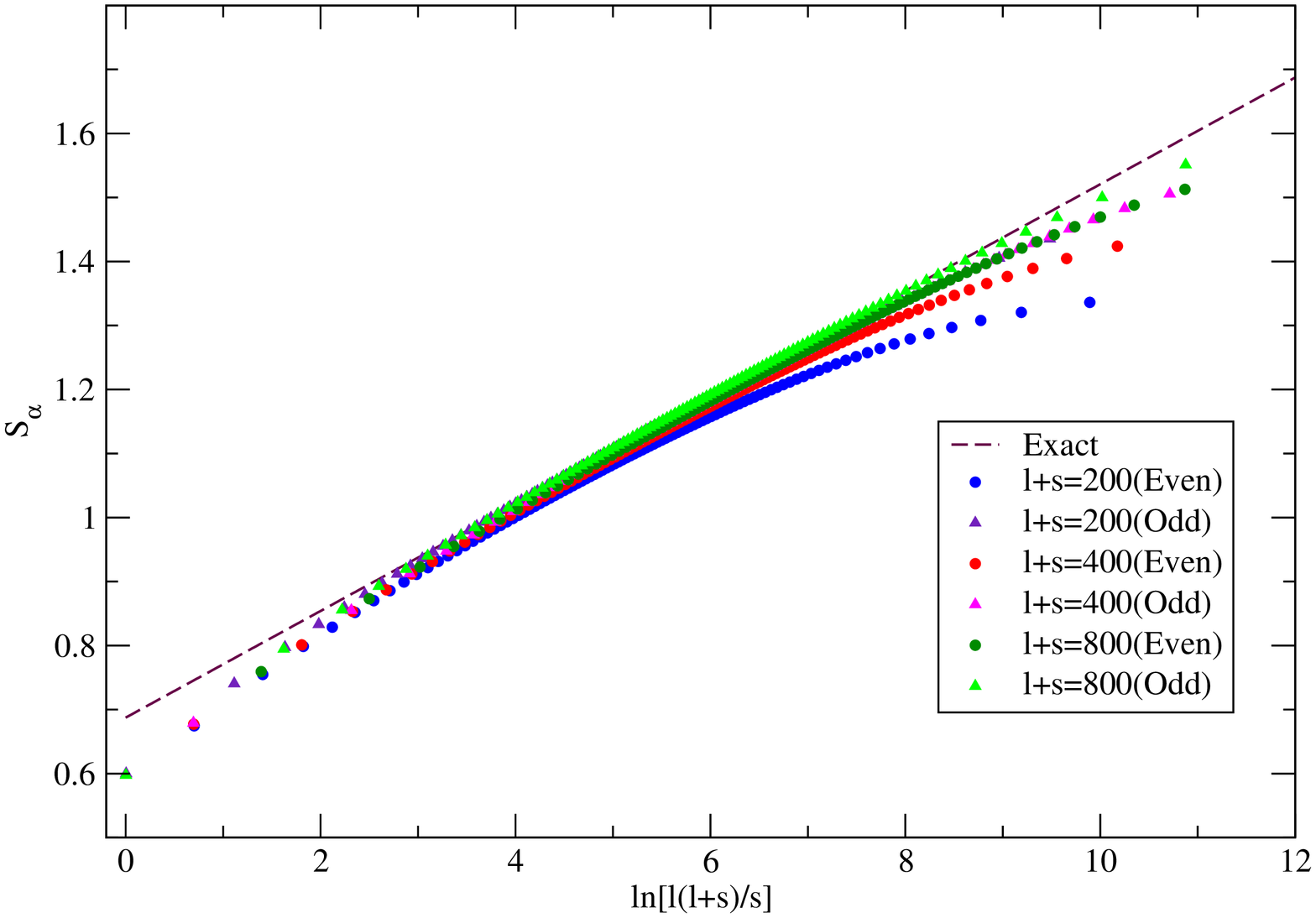}
\includegraphics[width=0.4\textwidth,angle =-90]{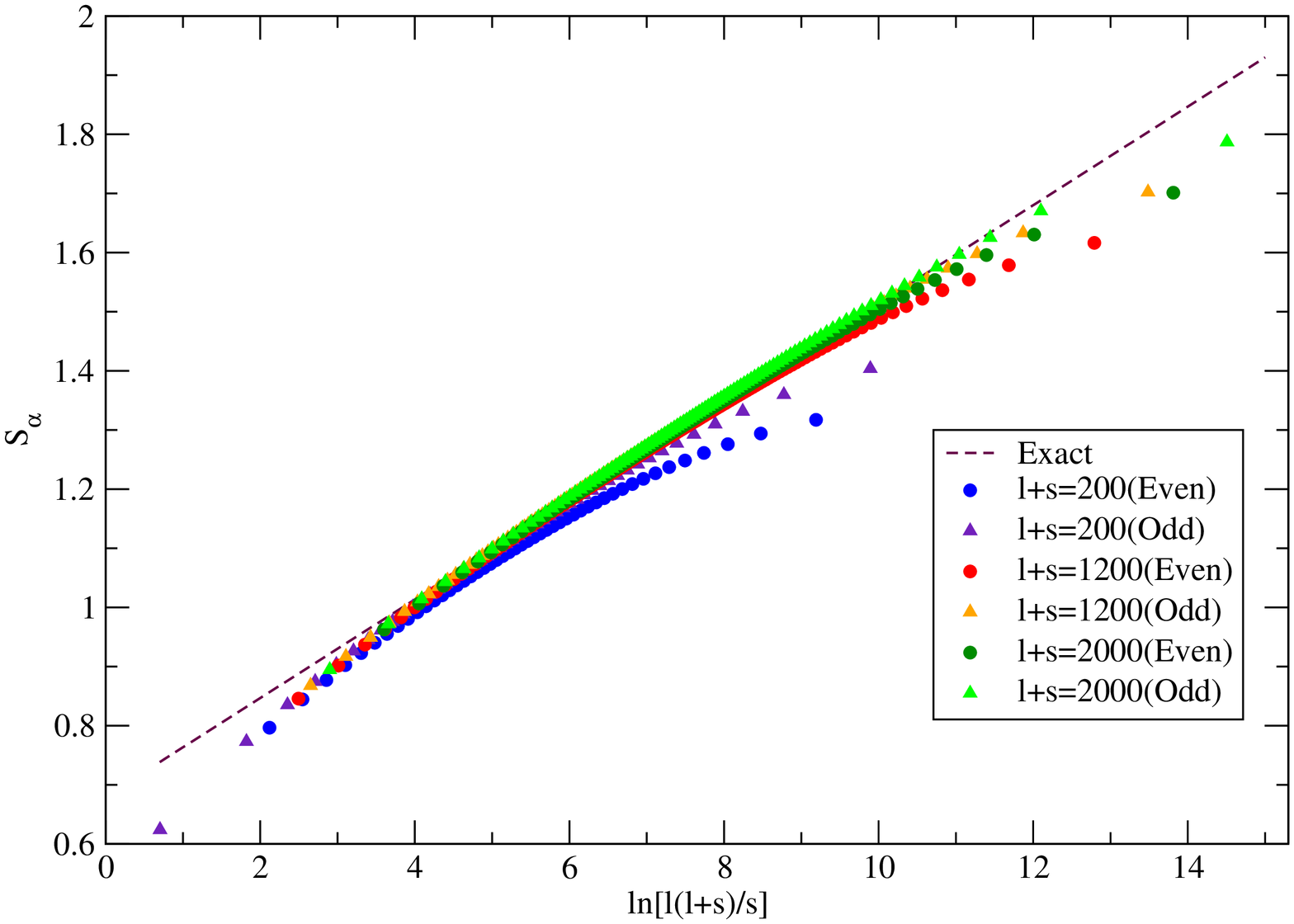}
\caption{(Color online) Post measurement entanglement entropy for the infinite transverse field Ising model. Top: post measurement entanglement entropy of the configuration \textbf{a} with $\alpha=1$ and $\alpha=2$. In the numerics we fixed
$l+s=300$. Middle: post measurement entanglement entropy of the configuration \textbf{b} with $\alpha=1$ for different values of $l+s$.  Bottom:
post measurement entanglement entropy of the configuration \textbf{c} with $\alpha=1$ for different values of $l+s$. In the inset the even and odd means that
$l$ and $s$ are both even or both odd numbers.
In all the figures the dashed lines are the CFT predictions.} 
\label{fig:Ising-connected}
\end{figure}
\begin{figure} [hthp!] 
\center
\includegraphics[width=0.45\textwidth,angle =-90]{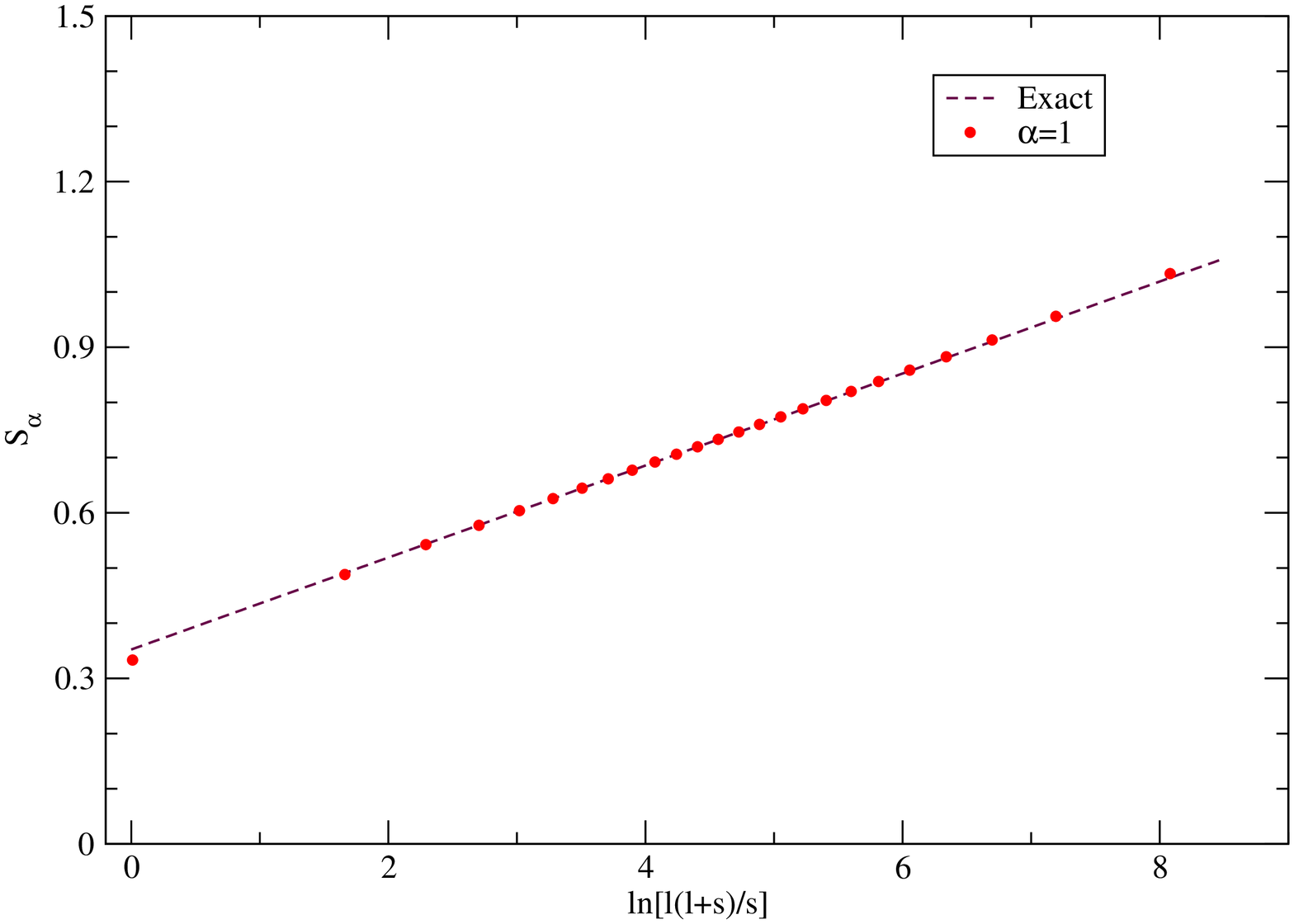}
\caption{(Color online) 
Post measurement entanglement entropy (with $\alpha=1$) of the configuration \textbf{a} for a point on the critical XY line with $a=\frac{1}{2}$  and fixed $l+s=100$.
The dashed line is the CFT prediction.} 
\label{fig:Ising-connected-universality}
\end{figure}

We first check the formula  (\ref{rajab2015}) valid for the infinite system by fixing the spins in the subsystem
$A$ in the up direction. The results for $\alpha=1$ and $\alpha=2$  shown in the Figure ~\ref{fig:Ising-connected} are in good agreement with the formula
(\ref{rajab2015}). We repeated the same calculation for the case when all the spins are down, see Figure ~\ref{fig:Ising-connected}. Here 
we realized that for small
subsystem sizes we have two branches for the two possible parities of the number of fermions in the subsystem. However,
the two branches start to get closer to each other by taking larger and larger subsystem sizes. There is fairly
a big deviation from the CFT result when $l$ is very small or when $s$ is very small. We do not know the exact reason
for this effect. One possibility is the presence of the boundary changing operators or it might be the lattice effect coming from the extrapolation length. We observed similar effect also for the case when the result of the projective measurement is the
antiferromagnetic configuration, see Figure ~\ref{fig:Ising-connected}. However, the effect disappears when we consider the configuration
$(\frac{1}{2},2)$. We checked the universality of our results by 
calculating the post measurement entanglement entropy on
the infinite critical XY line for the configuration \textbf{a}. The result is shown in the  Figure ~\ref{fig:Ising-connected-universality} 
is consistent with the formula (\ref{rajab2015}) which confirms
the universality of our results. Note that we observed the above behavior for also  other crystal configurations mentioned in the previous
section. We expect that the CFT results are valid for all the crystal configurations.  

\begin{figure} [hthp!] 
\centering
 \includegraphics[width=0.5\textwidth,angle =-90]{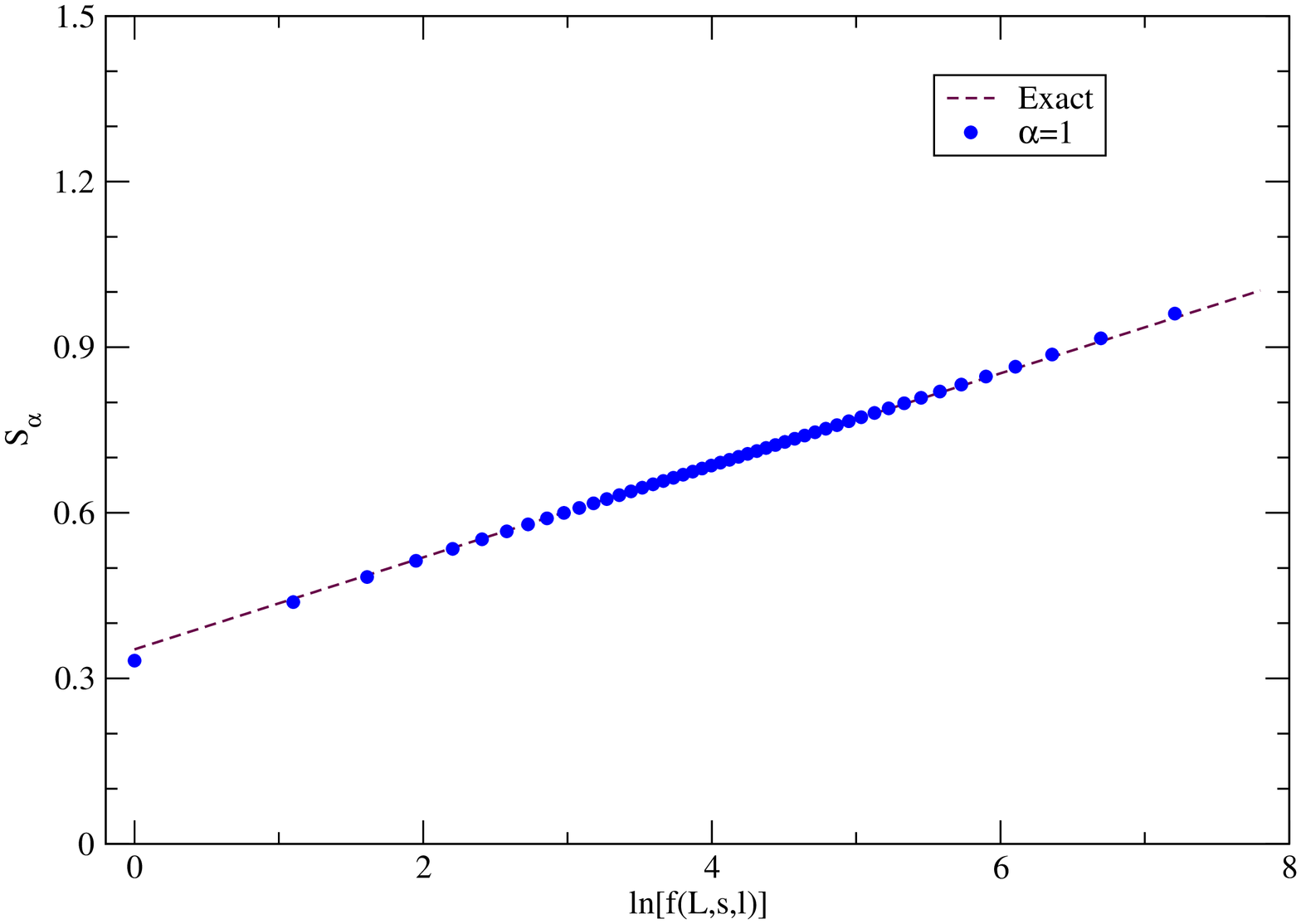}
\caption{(Color online) post measurement entanglement entropy in the periodic transverse field Ising model for the configuration \textbf{a} 
with respect to $\ln f (L,s,l)$, where 
$f (L,s,l)=\frac{L}{\pi}\frac{\sin\frac{\pi}{L}(l+s)\sin\frac{\pi}{L}l}{a\sin\frac{\pi}{L}s}$. In the numerics we fixed
$L=200$ and $l+s=100$.   The dashed line is the CFT prediction (\ref{SB for PBC}).}
\label{Ising-connected-PBC}
\end{figure}

\begin{figure} [hthp!] 
\centering 
\includegraphics[width=0.5\textwidth,angle =-90]{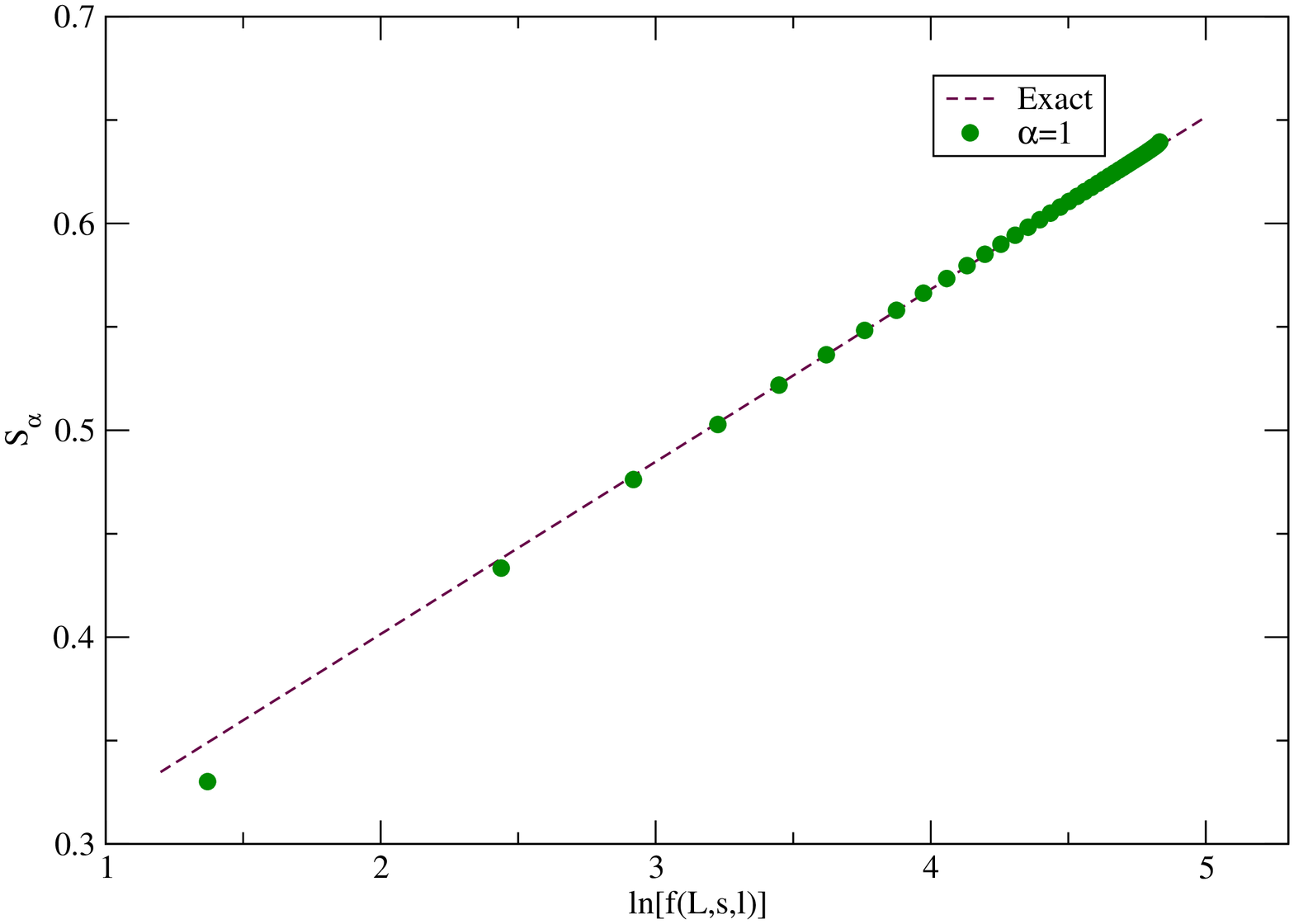}
\caption{(Color online) post measurement entanglement entropy for the  transverse field Ising model with open boundary conditions. The corresponding configuration is
\textbf{a} and the post measurement entanglement entropy is depicted
with respect to $\ln f (L,s,l)$, where 
$f (L,s,l)=\frac{2L}{\pi}\frac{\cos\frac{\pi s}{L}-\cos\pi\frac{l+s}{L}}{a\cos^2\frac{\pi s}{2L}}\cot\frac{\pi(l+s)}{2L}$ for the OBC. In the numerics we fixed
$L=200$ and $l+s=100$. The dashed line is the CFT prediction (\ref{SB for OBC}).} 
\label{Ising-connected-OBC}
\end{figure}

We then checked the formula 
(\ref{SB for PBC}), valid for the periodic systems, for the case when the result of the measurement is the configuration \textbf{a}. 
The numerical results are shown in the Figure ~\ref{Ising-connected-PBC}   are consistent with the CFT formulas.
Similar results are also valid for the configurations \textbf{b} and \textbf{c}. The conclusion is that the formula  (\ref{SB for PBC})
is valid for all the crystal configurations. Finally, we studied the open boundary condition in the presence
of different configurations. Our numerical results  for the configurations $(x,2k)$ are consistent with the formula  (\ref{SB for OBC}).
In  the Figure ~\ref{Ising-connected-OBC},  we depicted the result for the configuration \textbf{a}. We obtained similar result also for the configuration
$(\frac{1}{2},2)$. However,  the results for the configurations $(x,2k+1)$ do not follow the formula (\ref{SB for OBC}). This might be, as we discussed
before, because of the presence of the boundary changing operators. It will be  interesting to study the effect of boundary changing operators 
on our CFT calculations. We leave more through analyses of the configurations $(x,2k+1)$ for a future work.

\newpage

\subsection{Affleck-Ludwig boundary entropy}

To study the Affleck-Ludwig boundary entropy we first calculated the entanglement entropy of a sub-region without
projective measurement and fit the data to
\begin{eqnarray}\label{SB for no-measurement PBC 2}
S_{\alpha}=\frac{c}{6}(1+\frac{1}{\alpha})\ln \Big{(}\frac{L}{\pi}\sin\frac{\pi l}{L}\Big{)}+a_{\alpha},
\end{eqnarray}
and determined $a_{\alpha}$.
Then we did the same calculation in the presence of the measurement region and fit the data to 
\begin{eqnarray}\label{SB for PBC affleck-ludwig 2}
S_{\alpha}=\frac{c}{12}(1+\frac{1}{\alpha})\ln \Big{(}\frac{4L}{\pi}\frac{\sin\frac{\pi}{L}(l+s_1)\sin\frac{\pi}{L}l}{s_2\sin\frac{\pi}{L}s_1}\Big{)}+
b_{\alpha}
\end{eqnarray}
and determined $b_{\alpha}$. Finally the Affleck-Ludwig boundary entropy is given by
\begin{eqnarray}\label{Affleck-ludwig numericas}
S^{AL}=\ln b_0=b_{\alpha}-\frac{a_{\alpha}}{2}. 
\end{eqnarray}
We did this calculation for the configurations
$(x,2k)$ and for $b_0$ found a value incredibly close to one, for example, we derived
\begin{eqnarray}\label{b free}
b^{\textbf{a}}_{0}=0.996\hspace{1cm}b^{(\frac{1}{2},2)}_{0}=1.009.
\end{eqnarray}
The above results are consistent with the free nature of the configurations $(x,2k)$. Then we repeated the same calculations for
the configurations $(x,2k+1)$. Here for $S^{AL}$ we found a value very close to $\frac{\ln 2}{2}$. This is not exactly compatible
with what we expect for the fixed boundary condition which we have $S^{AL}=-\frac{\ln 2}{2}$. The extra  $\ln 2$ factor can be understood
as follows: Although all the configurations $(x,2k+1)$ flow to fixed boundary conditions a priory it is not known that they are 
flowing to the up (down) fixed boundary conditions (here with up (down) we mean in the euclidean version when we consider
the $\sigma^x$ basis). This ambiguity contributes a factor of two to the partition function and a factor of $\ln 2$ to the entanglement
entropy. Taking to the account this factor we find the desired boundary entropy. For example, our numerical calculations show
\begin{eqnarray}\label{b fixed}
b^{\textbf{b}}_{0}=0.698\approx \frac{1}{\sqrt{2}}\hspace{1cm}b^{(\frac{1}{2},1)}_{0}=0.696\approx \frac{1}{\sqrt{2}}.
\end{eqnarray}
The above reasoning will appear again in the next subsection when we discuss the disconnected cases.

\subsection{Disconnected regions}

In this sub-section, we calculate the entanglement entropy of two regions that are disconnected after 
projective measurement. 
In other words, we verify the validity of the equations (\ref{power-law decay1}), 
 (\ref{power-law decay 2}) and (\ref{power-law decay 3}) for the critical Ising chain. As we discussed before
 it is expected that most of the crystal configurations flow to free or fixed boundary conditions.
For the Ising model with the free-free boundary conditions the operator with the 
 smallest scaling dimension is the energy operator with $\Delta_1=\frac{1}{2}$. However,
 for the fixed-fixed boundary condition, it is the spin operator with $\Delta_1=\frac{1}{16}$.
 We will show in the next subsections that working in the $\sigma^z$ basis we can just detect 
 the first scaling dimension, $\Delta_1=\frac{1}{2}$.
 
 \subsubsection{Infinite chain:}
 
\begin{figure} [hthp!] 
\centering
 \includegraphics[width=0.5\textwidth,angle =-90]{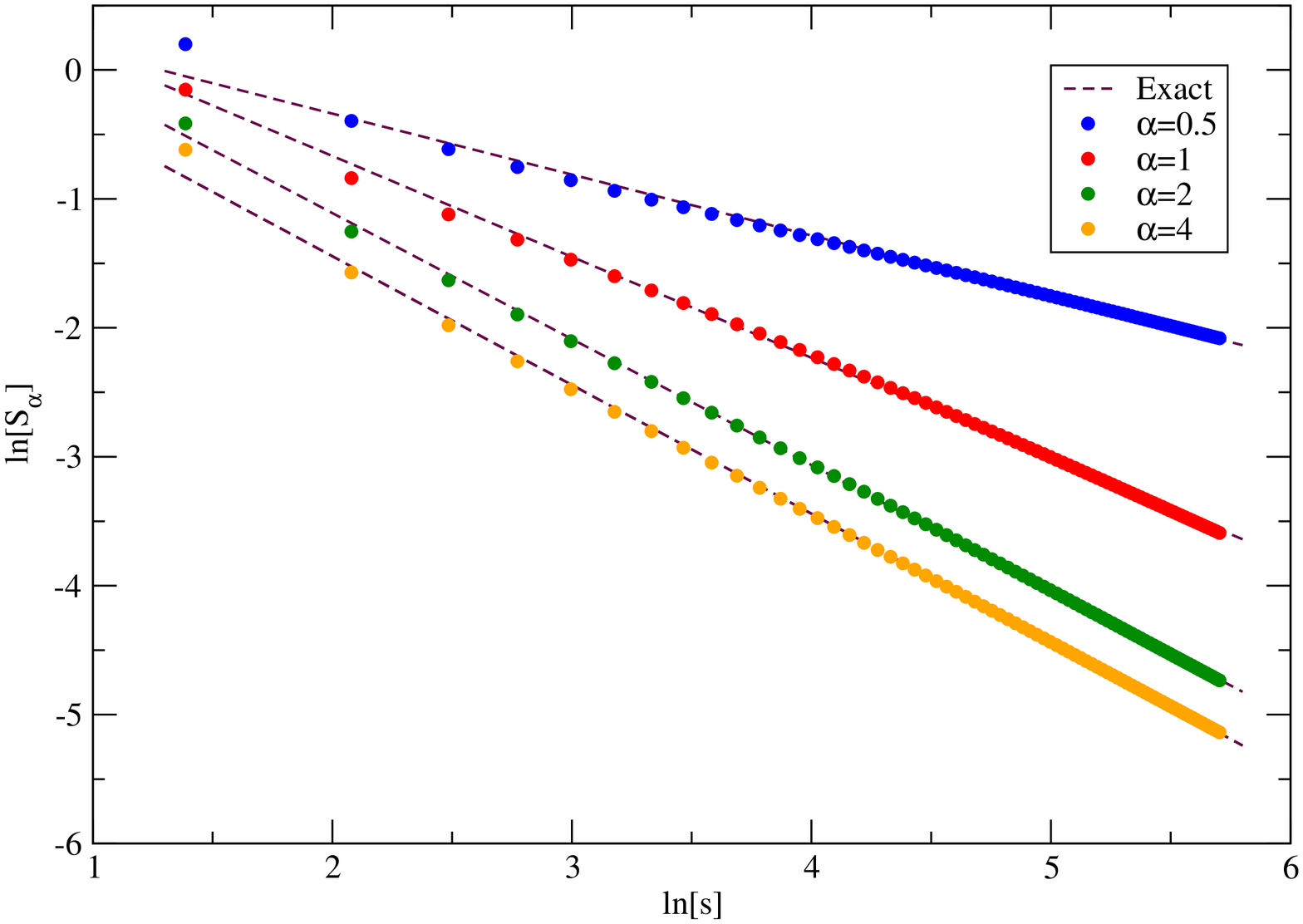}
\includegraphics[width=0.5\textwidth,angle =-90]{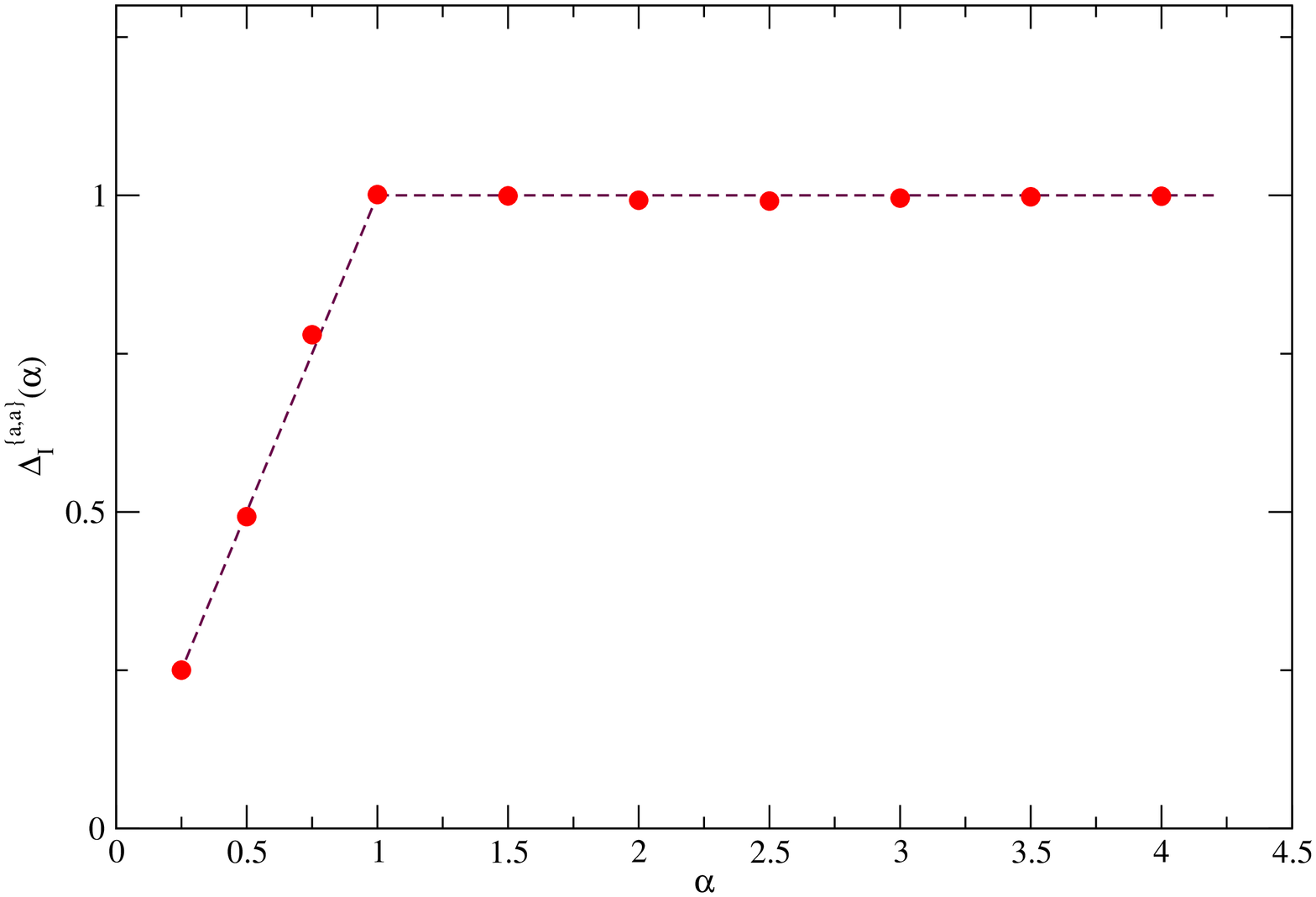}
\caption{(Color online) Post measurement entanglement entropy in the infinite transverse field Ising model for the setup I with the configuration \textbf{a}. Top: log-log plot of the post measurement entanglement entropy of the configuration 
\textbf{a} 
when the two subsystems are disconnected. We took $l=10$ and $s$ goes from $10$ to $300$. Bottom: The exponent of the power-law
 $\Delta_{I}^{\{\textbf{a},\textbf{a}\}}(\alpha)$
with respect to $\alpha$. The dashed line is the formula (\ref{exponent1Ising}).}
\label{fig:Ising-disconnected-infinite-up}
\end{figure}

  Putting all the pieces of the above argument together for the setup I we expect
 \begin{eqnarray}\label{exponent1Ising}
\Delta_{I}^{\{C_1,C_2\}}(\alpha)=\left\{
\begin{array}{c l}      
    \alpha, & \alpha<1,\hspace{1cm}\\
        1 &\alpha\geq1,\\        
\end{array}\right.
\end{eqnarray}
In the Figure ~\ref{fig:Ising-disconnected-infinite-up},  we first showed that the power-law behavior is valid for the Ising model
when we consider the configuration \textbf{a} on both regions.
Then we showed the validity of the equation (\ref{exponent1Ising}) for $\Delta_{I}^{\{\textbf{a},\textbf{a}\}}(\alpha)$. 

\begin{figure} [hthp!] 
\centering
\includegraphics[width=0.45\textwidth,angle =-90]{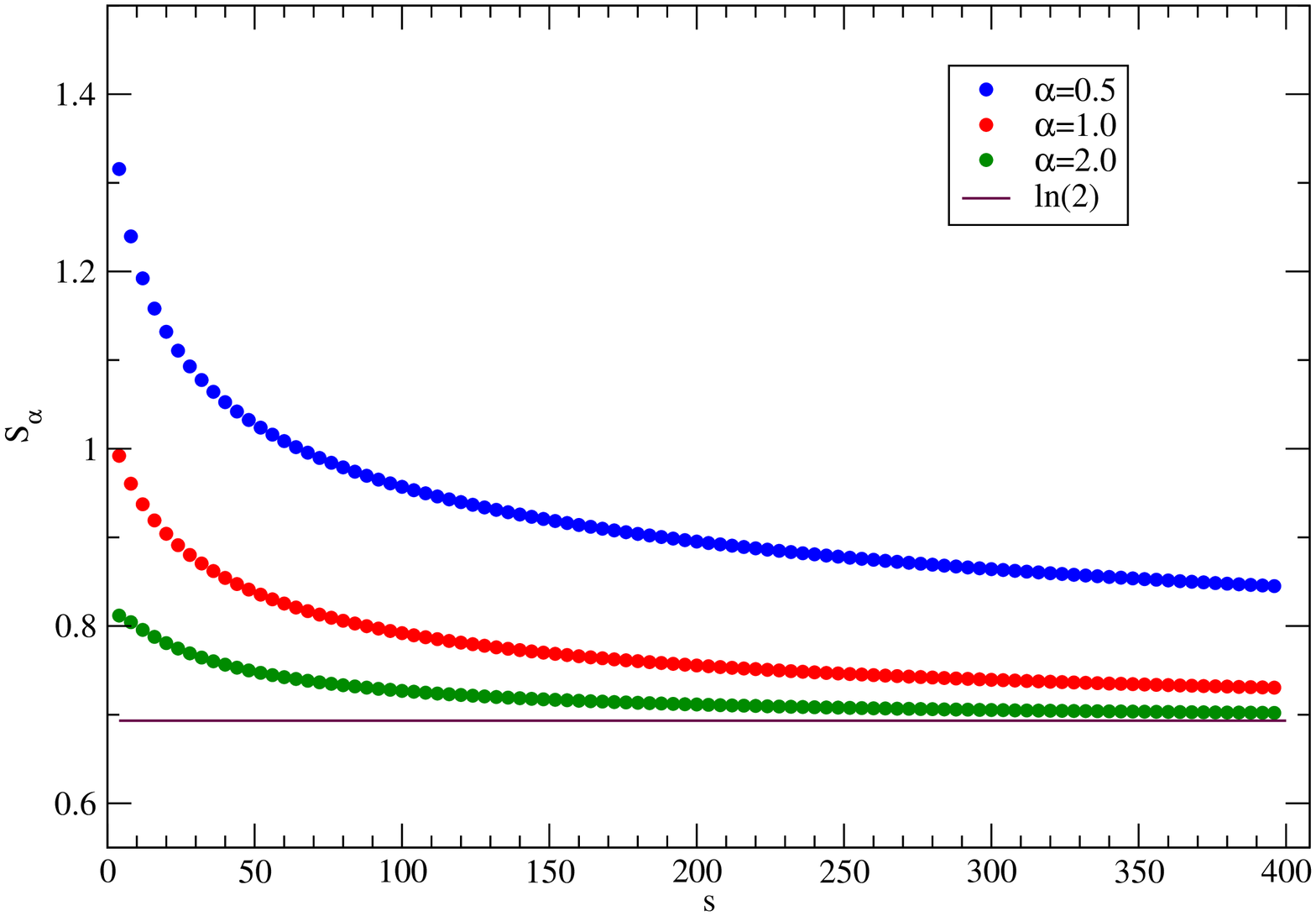}
\includegraphics[width=0.45\textwidth,angle =-90]{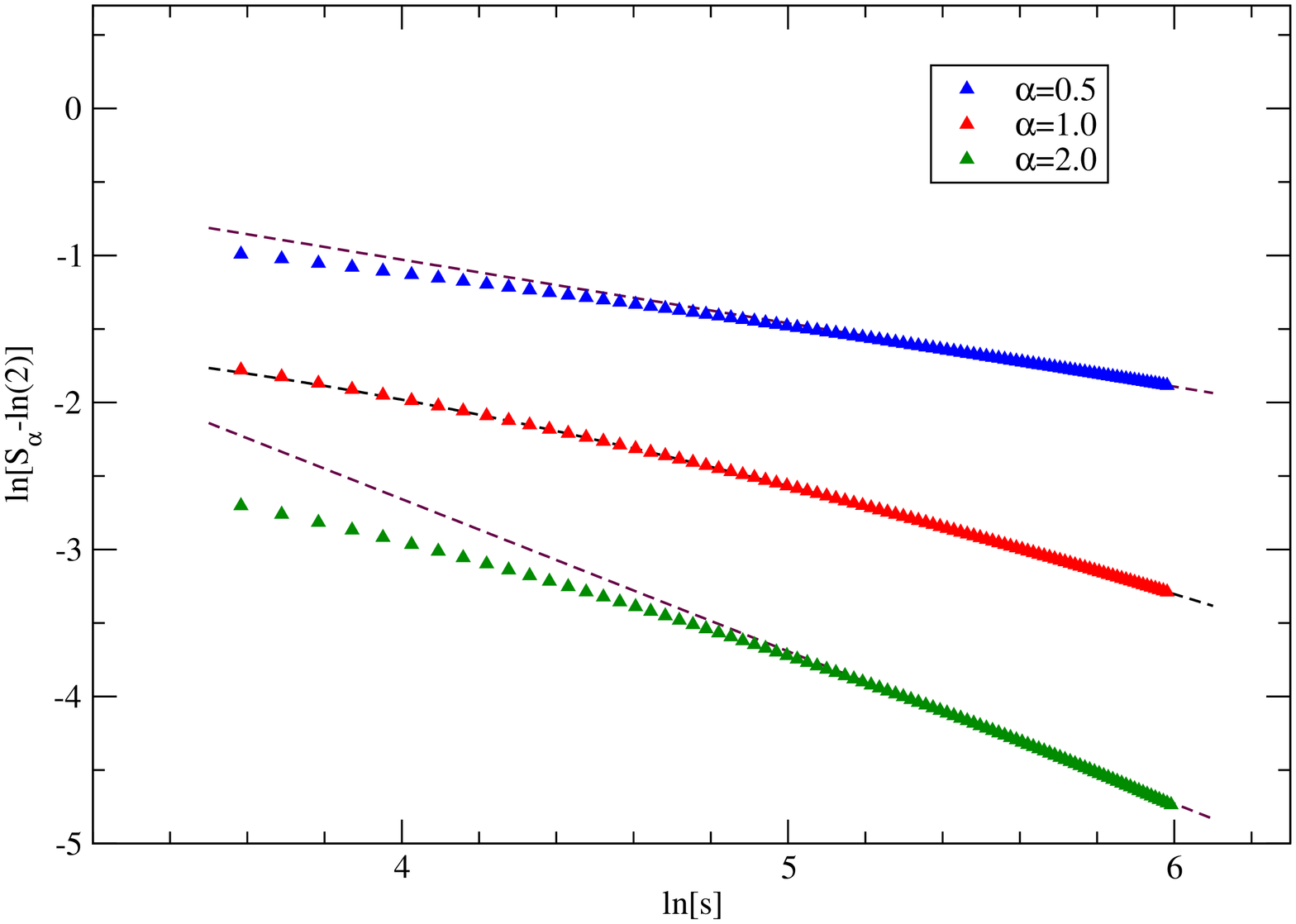}
\includegraphics[width=0.45\textwidth,angle =-90]{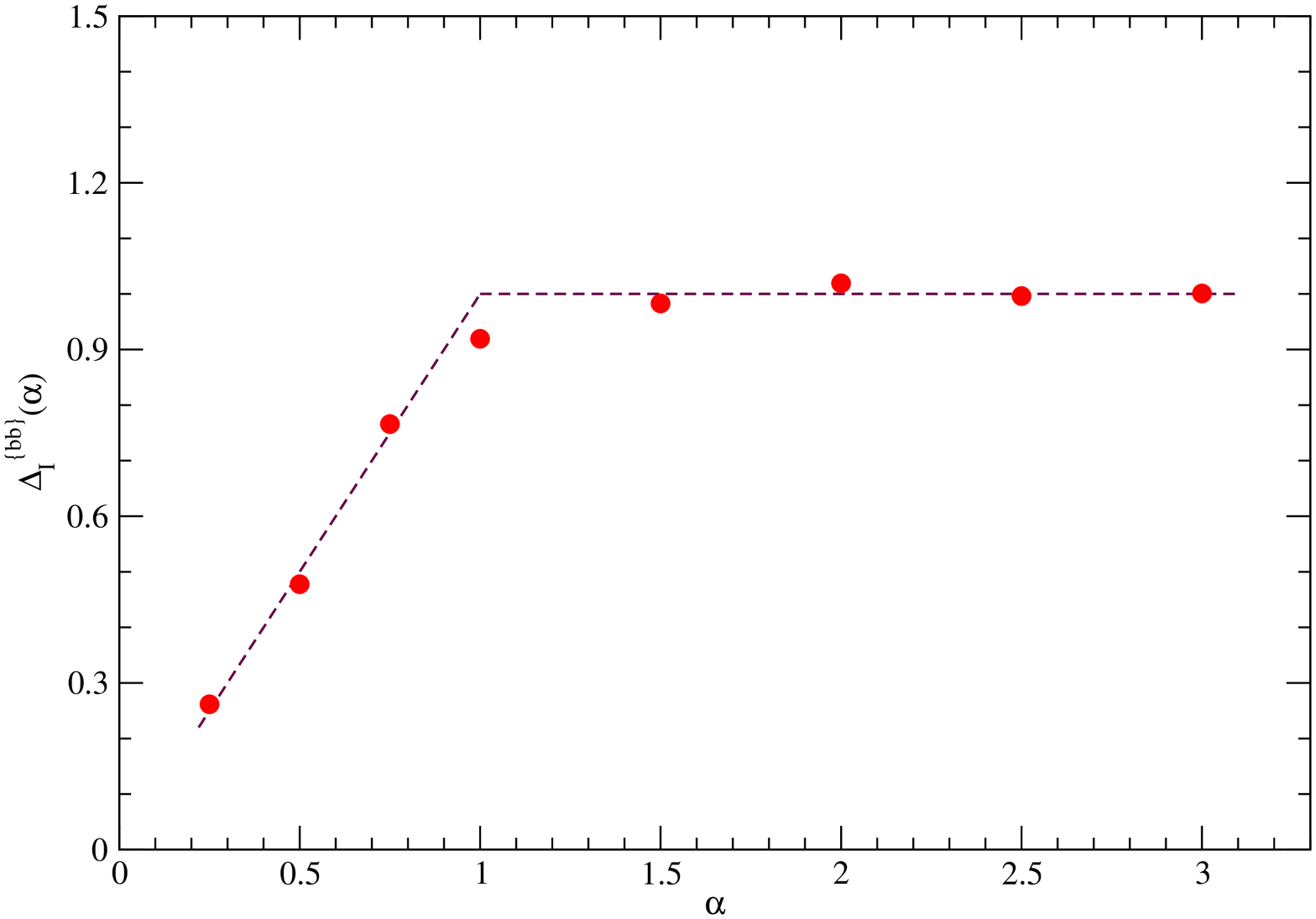}
\caption{(Color online) Post measurement entanglement entropy for the infinite transverse field Ising model in the setup I with the corresponding configuration \textbf{b}
for different values of $\alpha$.a) $S_{\alpha}$ with respect to the size of the measurement region. b) log-log plot of the post measurement entanglement entropy.
 Bottom: The exponent of the power-law
 $\Delta_{I}^{\{\textbf{b},\textbf{b}\}}(\alpha)$
with respect to $\alpha$. The dashed line is the formula (\ref{exponent1Ising}). We took $l=30$ and $s$ goes from $10$ to $400$. } 
\label{fig:Ising-disconnected-infinite-down}
\vspace{-20pt}
\end{figure}

To check that our results are configuration independent or not we also calculated the post measurement entanglement
entropy for the set-up I when the outcome of the measurement is the configurations $(x,2k)$. We found that 
the power-law behavior with the exponent (\ref{exponent1Ising}) is  valid also in these cases. However, for all the configurations
$(x,2k+1)$ 
surprisingly we found a very different behavior. The R\'enyi entanglement entropy \footnote{Note that depending on the $l$
for some values of $\alpha$ the R\'enyi entanglement entropy increases with $s$ and then saturates to $\log 2$} decreases
with respect to $s$ and then saturates for a value which is very close to $\ln 2$, see Figure \ref{fig:Ising-disconnected-infinite-down}. 
This behavior is totally counterintuitive because
 we expect that the R\'enyi entanglement entropy always decreases to zero by increasing the size of the measurement region.
 The above strange behavior can possibly be understood as follows: as we discussed in the previous section 
 although these configurations flow to fixed boundary conditions a priory we do not know that they flow to $Z_{_{Fi1-Fi1}}$ 
 or $Z_{_{Fi1-Fi2}}$. This means that the total partition function for these configurations on the cylinder
 is 

\begin{eqnarray}\label{partition functions identities 2}
Z=Z_{_{Fi1-Fi1}}+Z_{_{Fi1-Fi2}}=2Z_{_{Fr-Fr}}.
\end{eqnarray}
The factor $2$ in the above formula is independent of $s$ and produces a $\ln 2$ in the calculations of the $S_{\alpha}$
which survives even when $s$ goes to infinity. Another interesting feature of the above formula is that now instead of
the partition function of fixed-fixed on the cylinder we have the partition function of free-free. If the above argument is correct we expect that
the R\'enyi entropy approaches to the $\ln 2$ like a power-law with an exponent which is the same as (\ref{exponent1Ising}). In other words
 \begin{eqnarray}\label{exponential setup I unusual}
S_{\alpha}\asymp \ln 2+\beta(\alpha)\Big{(}\frac{l}{8s}\Big{)}^{\Delta_{I}^{\{C_1,C_2\}}(\alpha)},
\end{eqnarray}
where $C_1,C_2\in (x,2k+1)$. Our numerical results depicted in the Figure \ref{fig:Ising-disconnected-infinite-down} are consistent with the above picture. The conclusion is that
although the configurations $(x,2k+1)$ flow to fixed boundary conditions all of the exponents are the ones that come from
the free boundary conditions. In the discrete level we realized that for the large $s$ the post measurement $G$ is in a way that the eigenvalues of the  matrix $\tilde{G}^{T}.\tilde{G}$
are all close to one except one eigenvalue which is approximately zero. Then having the equation (\ref{entanglement entropy}) it is obvious that one expects
$S_{\alpha}=\ln 2$ for the large $s$. It will be interesting to prove this fact by exact calculations starting with the configurations $(x,2k+1)$.

Finally, we also studied the case with $C_1\in (x,2k)$ and $C_2\in (x,2k+1)$. Based on the previous arguments
this example should be related to the free-fixed partition function. We expect that the entanglement entropy
follows the equation (\ref{exponential setup I unusual}). Although not shown here our numerical calculations confirmed our expectations.
The conclusion is that as far as one of the configurations is from the set $(x,2k+1)$  the entanglement entropy follows the equation 
(\ref{exponential setup I unusual}).

 \subsubsection{Periodic chain:}
 
Following similar argument as above one can write for  the setup II 
 \begin{eqnarray}\label{exponent2Ising}
\Delta_{P}^{\{C_1,C_2\}}(\alpha)=\left\{
\begin{array}{c l}      
    2\alpha, & \alpha<1,\hspace{1cm}\\
        2 &\alpha\geq1,\\        
\end{array}\right.
\end{eqnarray}

\begin{figure} [hthp!] 
\centering
 \includegraphics[width=0.5\textwidth,angle =-90]{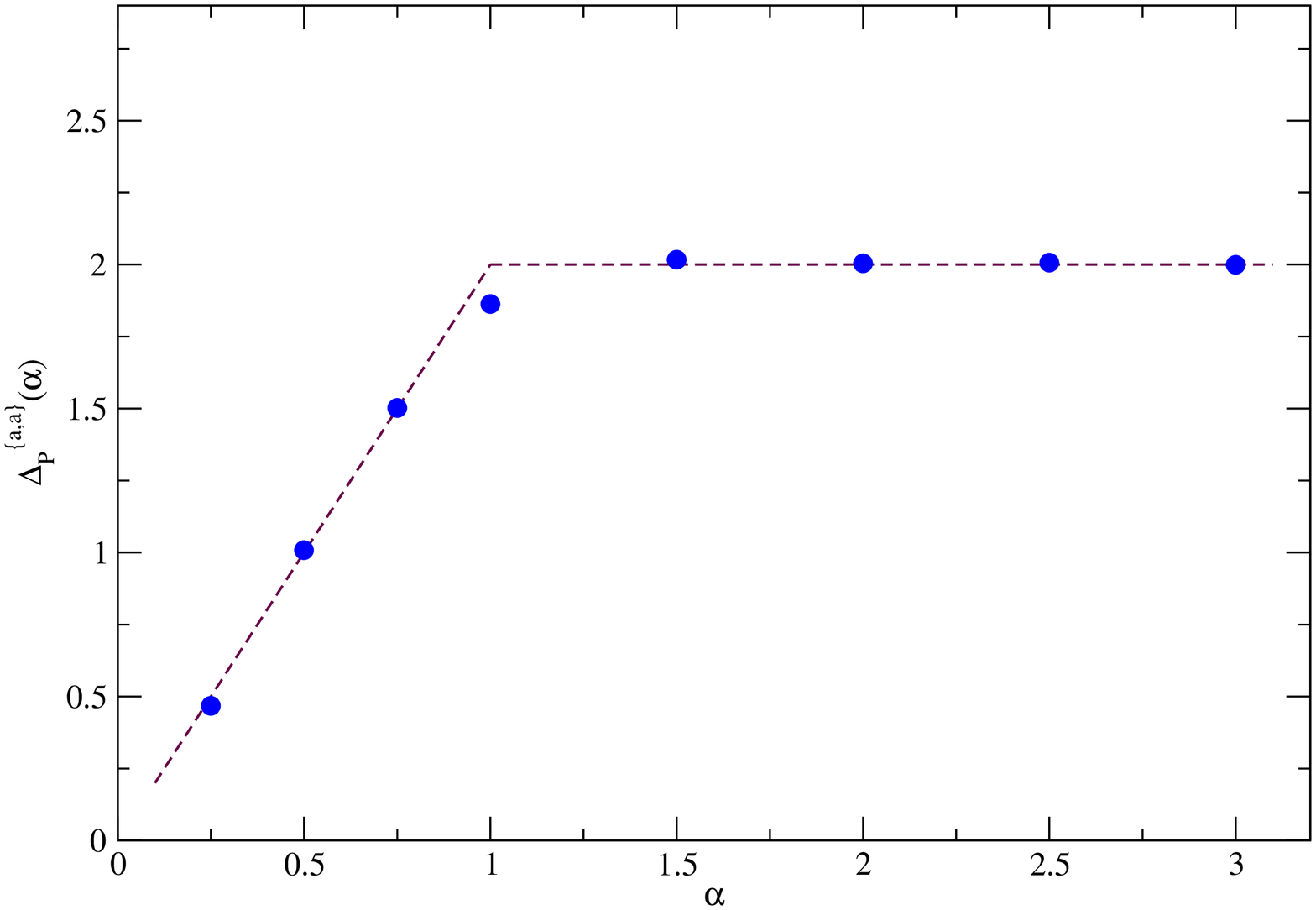}
\caption{(Color online) The exponent of the power-law
 $\Delta_{P}^{\{\textbf{a},\textbf{a}\}}(\alpha)$
with respect to $\alpha$. We took $L=400$
 and $l$ goes from $4$ to $50$. 
The dashed line is the formula (\ref{exponent2Ising}).} 
\label{Ising-disconnected-PBC-up}
\end{figure}

In the Figure ~\ref{Ising-disconnected-PBC-up},  we checked the
validity of the equation (\ref{exponent2Ising})  for the $\Delta_{P}^{\{\textbf{a},\textbf{a}\}}(\alpha)$ 
 in the finite periodic  system. 
The results are consistent with the CFT predictions. Note that the above result should be correct for 
all the crystal configurations discussed in this paper. However, one needs to be careful that for $C_1=C_2=(x,2k+1)$
we expect a factor of two in the partition functions which leads us  to have
\begin{eqnarray}\label{exponential setup II unusual}
S_{\alpha}\asymp \ln 2+\beta(\alpha)\Big{(}\frac{\pi l}{4L}\Big{)}^{\Delta_{P}^{\{C_1,C_2\}}(\alpha)}.
\end{eqnarray}

   \subsubsection{Semi-infinite chain:}
   
\begin{figure} [hthp!] 
\centering
 \includegraphics[width=0.5\textwidth,angle =-90]{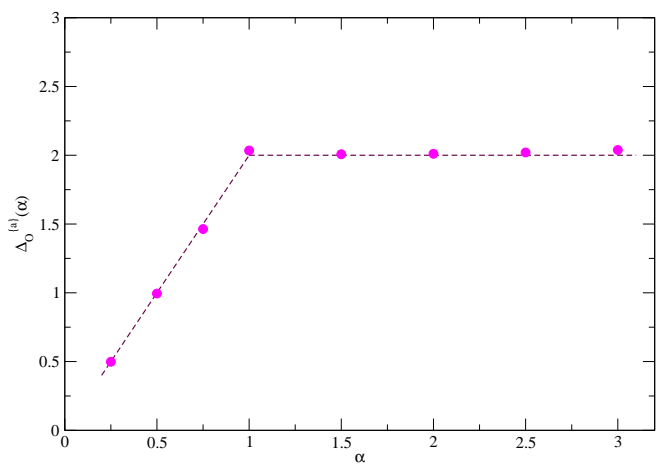}
\caption{(Color online) The exponent of the power-law
 $\Delta_{O}^{\{\textbf{a}\}}(\alpha)$
with respect to $\alpha$. 
We took  $l+s=200$ and the fit is done for the interval $l\in(1,100)$. 
The dashed line is the formula (\ref{exponent3Ising}).} 
\label{Ising-disconnected-semi-infinite-up}
\end{figure}

In the setup III for the configuration $(x,2k)$ we expect
 \begin{eqnarray}\label{exponent3Ising}
\Delta_{O}^{\{C\}}(\alpha)=\left\{
\begin{array}{c l}      
    2\alpha, & \alpha<1,\hspace{1cm}\\
        2 &\alpha\geq1,\\        
\end{array}\right.
\end{eqnarray}
 In the Figure  ~\ref{Ising-disconnected-semi-infinite-up}, we checked the validity of
 (\ref{exponent3Ising}) with $\Delta_{O}^{\{\textbf{a}\}}(\alpha)$. The results are consistent with our CFT calculations. Note that again we expect that the
 von Neumann entropy saturates to $\ln 2$ for the configurations $(x,2k+1)$.

\section{Entanglement entropy after selective measurements in the  critical XX chain}

In this section, we will check the validity of the post measurement entanglement entropy
formulas derived in the section ~3 for the critical XX chain. In other words we will check the validity of the formulas:  (\ref{rajab2015})
, (\ref{power-law decay1}), (\ref{SB for PBC}), (\ref{power-law decay 2}), (\ref{power-law decay 3}) and (\ref{SB for OBC}). 
The  formulas (\ref{rajab2015}), (\ref{SB for PBC}) and (\ref{SB for OBC}) are
the post measurement entanglement entropy of two connected regions and the other three are the ones related to the disconnected regions.
We perform the measurement  in the $\sigma^z$
basis so that we can use the results of the section ~5. 
For the critical XX chain as we mentioned in the previous section the configurations \textbf{a} and \textbf{b}
are not conformal configurations, however, the configurations $(\frac{n_f}{\pi},k)$ lead to  conformal boundaries. We mostly focus here on these
configurations and
check the CFT results. It is worth mentioning that although it is expected that the configurations $(x,k)$ with $x\neq \frac{n_f}{\pi}$ are not conformal
it was shown numerically \cite{Rajabpour2015b} that if $l$ is sufficiently large with respect to $s$ the CFT results still can be used. For example, for the configuration
\textbf{a} the CFT results are valid for $l>\frac{\pi}{n_f}s$. Of course, the range of the validity of the CFT results is bigger for those cases
that $x$ is closer to $\frac{n_f}{\pi}$. We will comment more about this fact in the upcoming subsections.

\subsection{Connected regions}
In this subsection, we first study the entanglement entropy in the presence of the
configurations $(x,k)$ with $x= \frac{n_f}{\pi}$ which we call them conformal configurations. Then
we comment about the effect of the non-conformal configurations, i.e. $(x,k)$ with $x\neq \frac{n_f}{\pi}$.

\subsubsection{conformal configurations}
\begin{figure} [hthp!] 
\centering
 \includegraphics[width=0.5\textwidth,angle =-90]{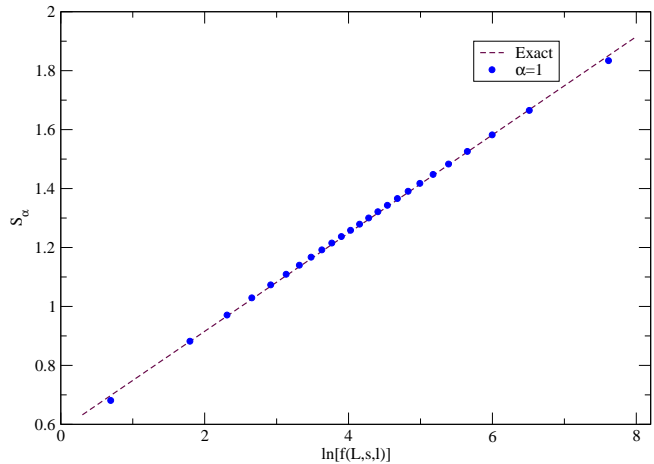}
\caption{(Color online) Post measurement entanglement entropy for the periodic XX model with the configuration \textbf{c} 
with respect to $\ln f (L,s,l)$, where 
$f (L,s,l)=\frac{L}{\pi}\frac{\sin\frac{\pi}{L}(l+s)\sin\frac{\pi}{L}l}{a\sin\frac{\pi}{L}s}$ for the PBC. In the numerics we fixed
$L=200$ and $l+s=100$. In  the figure  the dashed line is the CFT prediction (\ref{SB for PBC}) with $c=1$.}
\label{fig:XX-connected-PBC}
\end{figure}

\begin{figure} [hthp!] 
\centering
\includegraphics[width=0.5\textwidth,angle =-90]{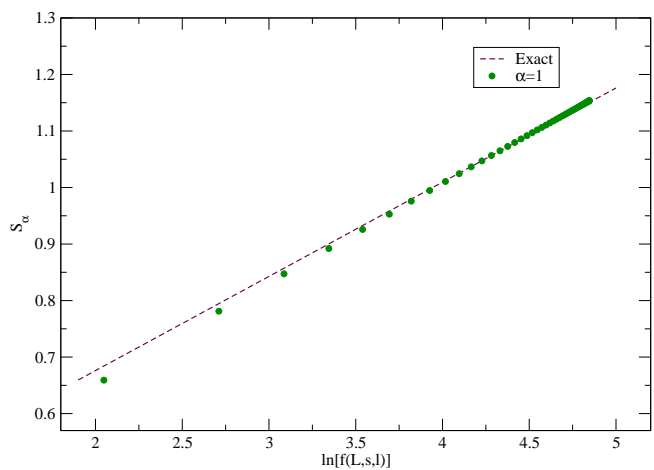}
\caption{(Color online) Post measurement entanglement entropy for the  XX model with open boundary conditions. The corresponding configuration is \textbf{c}
and post measurement entanglement entropy  is depicted 
with respect to $\ln f (L,s,l)$, where 
$f (L,s,l)=\frac{2L}{\pi}\frac{\cos\frac{\pi s}{L}-\cos\pi\frac{l+s}{L}}{a\cos^2\frac{\pi s}{2L}}\cot\frac{\pi(l+s)}{2L}$ for the OBC. In the numerics we fixed
$L=200$ and $l+s=100$. In  the figure  the dashed line is the CFT prediction  (\ref{SB for OBC})  with $c=1$.} 
\label{fig:XX-connected-OBC}
\end{figure}

The formula (\ref{rajab2015}) has been already checked for the XX chain when the outcome of the measurement is
an antiferromagnetic configuration \cite{Rajabpour2015b}. We calculated numerically the post measurement entanglement entropy
of two connected 
regions when the corresponding configuration is \textbf{c}, for the finite periodic and   open chains.
The numerical results depicted in the Figures ~\ref{fig:XX-connected-PBC} and ~\ref{fig:XX-connected-OBC} 
show a reasonable compatibility with 
the CFT formulas   (\ref{SB for PBC}) and (\ref{SB for OBC}). We have obtained similar results for also the configurations $(\frac{n_f}{\pi},k)$ in the case of
infinite and periodic boundary conditions. For the open chain when $k>1$ one needs to take into account also boundary changing operators. We leave more through analysis of this point to a future work. Final 
conclusion is that the CFT results are valid for all the conformal configurations $(\frac{n_f}{\pi},k)$.

\subsubsection{non-conformal configurations:} 

\begin{figure} [hthp!] 
\centering
\includegraphics[width=0.5\textwidth,angle =-90]{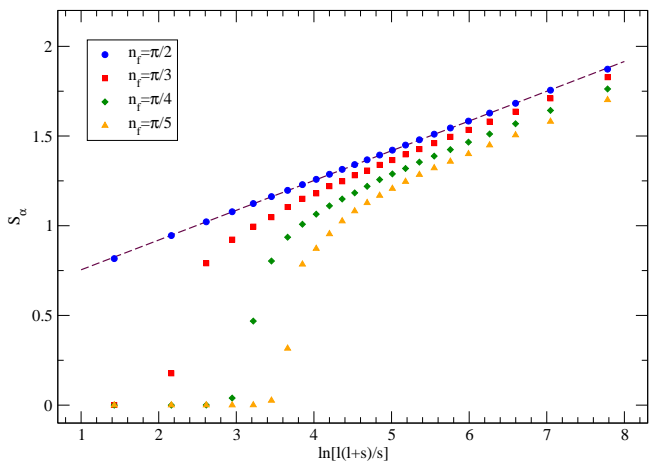}
\caption{(Color online) Post-measurement entanglement entropy in the XX-chain for an infinite chain in the presence of the configuration
\textbf{c} for different values of the fillings $n_f$.  } 
\label{fig:pmcxx_nf}
\end{figure}

As we mentioned before all the configurations $(x,k)$ with $x\neq\frac{n_f}{\pi}$ are not conformal configurations, however, it is expected
that for large $\frac{l}{s}$, in other words small measurement region, the CFT results be valid. This has been already shown in
(\cite{Rajabpour2015b}) for the configuration \textbf{a} with different $n_f$'s. Here we examined similar phenomena for the
configuration \textbf{c}. This configuration is conformal just for $n_f=\frac{\pi}{2}$ and not for other fillings. In the Figure
(\ref{fig:pmcxx_nf}), we change the filling but with fixed configuration calculated the post measurement entanglement entropy.
Numerical result show that for this configuration as far as $\frac{l}{s}>1-2\frac{n_f}{\pi}$
the CFT results are valid. We expect similar behavior also for the other configurations. At the moment it is not clear how one
can predict the regime of the validity of the CFT results. However, it is not difficult to see that whenever we need to inject fermions to the subsystem
in contrast to the filling factor of the system one leads to the non-conformal configurations. The more fermions we inject the  bigger $l$
we need to have results consistent with the CFT. In the regime that the CFT results are not valid, we see an exponential decay
of the entanglement entropy.

\subsection{Affleck-Ludwig boundary entropy and the g-theorm}

In this subsection, we make some comments regarding the Affleck-Ludwig boundary entropy and the g-theorem.
We  calculated
the Affleck-Ludwig term for the conformal configurations as we did for the transverse field Ising model. We followed the same procedure
and basically used the equation (\ref{Affleck-ludwig numericas}).
Our numerical results performed for $n_f=\frac{\pi}{2}$ show that
\begin{eqnarray}\label{b XX}
b^{(\frac{1}{2},1)}_{0}=1.00,\hspace{1cm}b^{(\frac{1}{2},2)}_{0}=1.00.
\end{eqnarray}
The above results are perfectly consistent with what we expect for the Dirichlet boundary conditions which we have $b_0=1$.

\begin{figure} [hthp!] 
\centering
\includegraphics[width=0.5\textwidth,angle =-90]{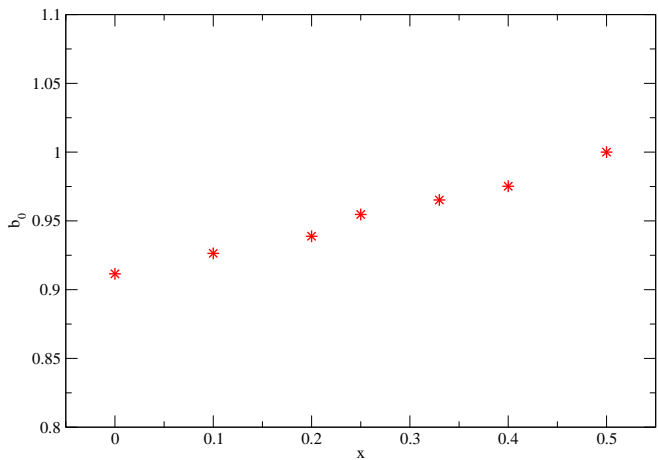}
\caption{(Color online) $b_0$ for different configurations $(x,1)$  for the half-filling case $n_f=\frac{\pi}{2}$.  } 
\label{fig:ALxxnf}
\end{figure}

After finding the $b_0$ for the conformal configurations we calculated the same quantity for the
non-conformal configurations. As we mentioned before all the configurations $(x,k)$ with $x\neq\frac{n_f}{\pi}$
are not conformal so in principle, they are a good laboratory to verify the entropic version of the g-theorem.
For this reason, we followed the same procedure as above but this time, we just used the regime that the CFT results are valid. 
The results shown in the figure (\ref{fig:ALxxnf}) show that for $n_f=\frac{\pi}{2}$ the $b_0$ for the configurations
$(x,k)$ start to decrease by decreasing $x$ from $\frac{1}{2}$ which is the conformal Dirichlet point to the 
non-conformal point at $x=0$. This is  compatible with the g-theorem which states that the 
$b_0$ decreases to the infrared. It is worth mentioning that in principle for the XX chain we have
two boundary fixed points, Dirichlet with $b_0=1$ and Neumann with $b_0=\frac{1}{2}$. Every other boundary conditions
should be between these two values. As it is clear from the Figure (\ref{fig:ALxxnf}) our results are in complete agreement with the above arguments.

\subsection{Disconnected regions}

In this subsection, we study the post measurement entanglement entropy in the XX chain
by using the configurations $(\frac{n_f}{\pi},k)$. We will show that based on the chosen configuration and the boundary condition the smallest
scaling dimension in the spectrum of the system changes. Because of this subtlety we  study the 
infinite (setup I),  the periodic (setup II)
and the open (setup III) chains separately.

\subsubsection{Infinite  chain}

\begin{figure} [hthp!] 
\centering
\includegraphics[width=0.45\textwidth,angle =-90]{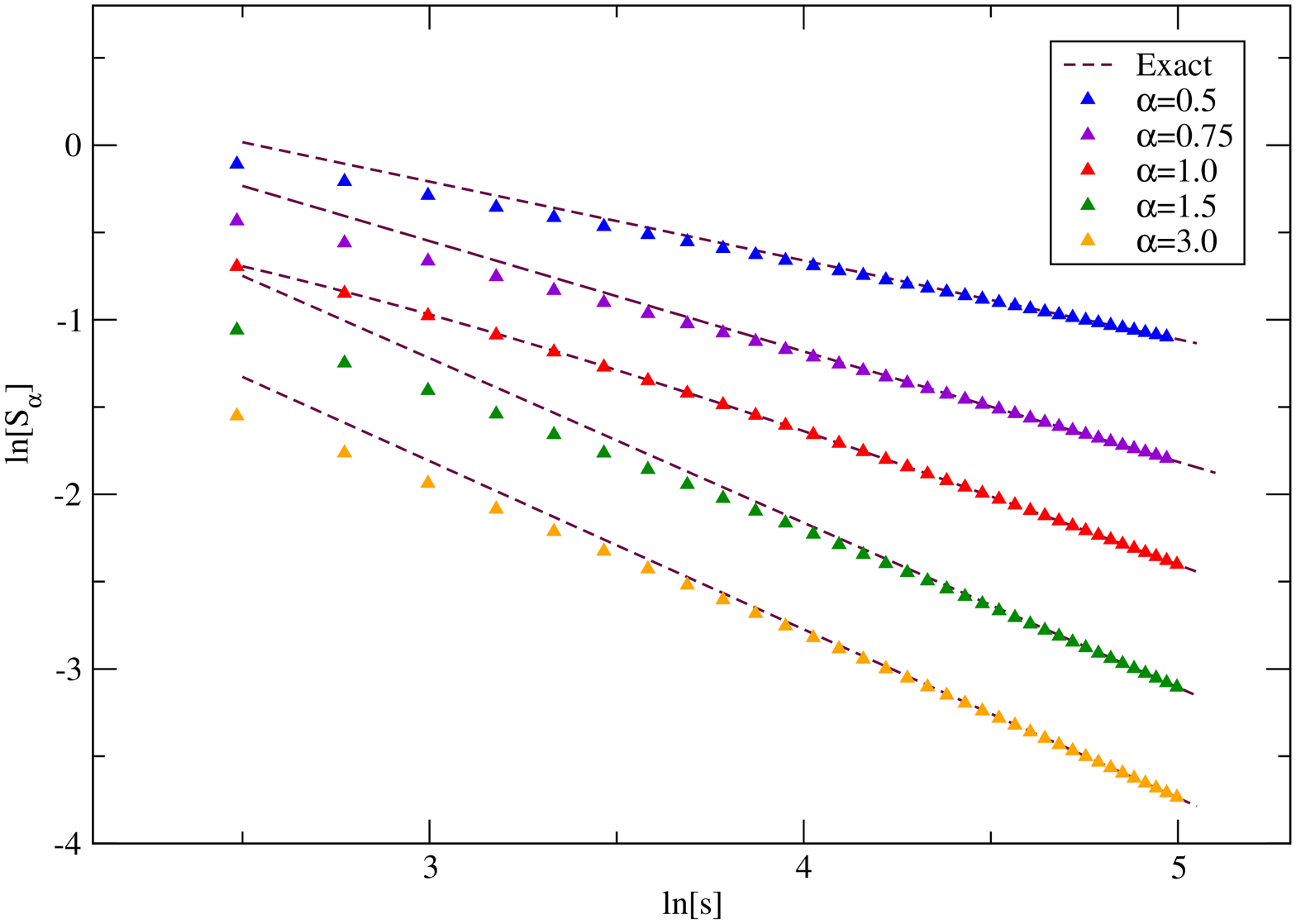}
\includegraphics[width=0.45\textwidth,angle =-90]{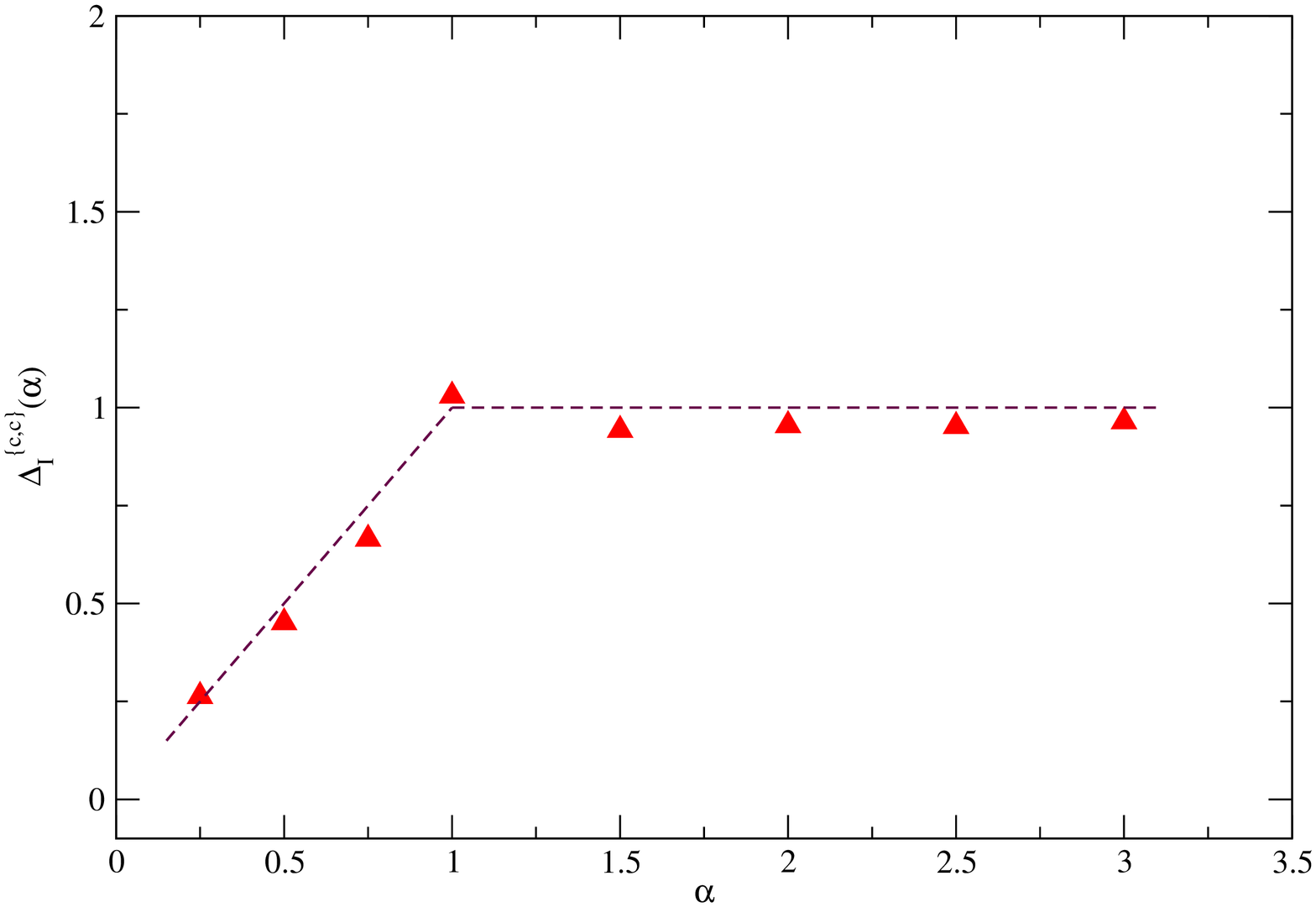}
\caption{(Color online) Post-measurement entanglement entropy of the disconnected regions in the half filling XX model  for
the setup I with the configuration \textbf{c}. Top: log-log plot of $S_{\alpha}$ with respect to $s$ for different $\alpha$'s. The dashed lines
are the CFT predictions. Bottom: The exponent $\Delta_{I}^{\{\textbf{c},\textbf{c}\}}(\alpha)$ for different  $\alpha$'s is extracted by taking $l=10$
and fitting the data to a straight line in the region $s\in(100,160)$.} 
\label{fig:XX-infinite--disconnected}
\end{figure}

\begin{figure} [hthp!] 
\centering
\includegraphics[width=0.45\textwidth,angle =-90]{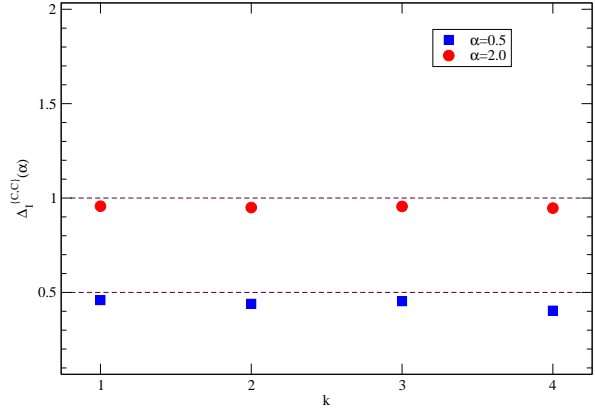}
\caption{(Color online) The exponent $\Delta_{I}^{\{C,C\}}(\alpha)$ for   $\alpha=\frac{1}{2}$ and $2$ for different
configurations $C=(\frac{1}{2},k)$ with $k=1,2,3$ and $4$. We took the half filling case $n_f=\frac{\pi}{2}$. The exponents are extracted by taking $l=10$
and fitting the data to a straight line in the region $s\in(200,250)$. The dashed lines are the CFT predictions for the Dirichlet boundary conditions.
The large deviation for $k=4$ is most likely the  finite size effect.} 
\label{fig:XX-infinite--disconnected-half filling- different configurations}
\end{figure}

As we mentioned in the section ~6 if we take equal configurations on the two slits   the operator with the smallest
 scaling dimension has  $\Delta_1=\frac{1}{2}$ \cite{Alcaraz1989,Bilstein}. Consequently for the setup I if the result of the projective measurement
 is a conformal configuration, for example, the configuration \textbf{c} for $n_f=\frac{\pi}{2}$,  we have
 \begin{eqnarray}
\label{exponent1XX}
\Delta_{I}^{\{C,C\}}(\alpha)=\left\{
\begin{array}{c l}      
    \alpha, & \alpha<1,\hspace{1cm}\\
        1 &\alpha\geq1.       
\end{array}\right.
\end{eqnarray}
where $C$ stands here for $(\frac{n_f}{\pi},k)$. In the Figure \ref{fig:XX-infinite--disconnected}, we checked the validity of 
the equation (\ref{exponent1XX}) for the configuration \textbf{c}. 
Our numerical results are consistent
with the CFT predictions.
To check that the above result for $n_f=\frac{\pi}{2}$ is independent of the conformal configuration we also calculated
the entanglement entropy for the configurations $(\frac{1}{2},k)$ with $k=2,3$ and $4$. The results shown in the 
Figure \ref{fig:XX-infinite--disconnected-half filling- different configurations}
demonstrate that the smallest scaling dimension in all of the above cases are the same. In other words all of the 
configurations $(\frac{1}{2},k)$ flow to a Dirichlet boundary condition. Note that based on the above results although 
one can  conclude that all of the boundary conditions are the Dirichlet boundary conditions it is not yet clear that they are all
the same Dirichlet boundaries. We will come back to this point in a few lines.

\begin{figure} [hthp!] 
\centering
\includegraphics[width=0.45\textwidth,angle =-90]{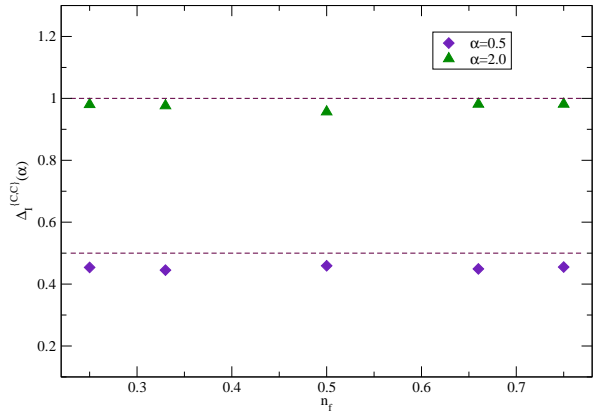}
\caption{(Color online) The exponent $\Delta_{I}^{\{C,C\}}(\alpha)$ for   $\alpha=\frac{1}{2}$ and $2$ for different
fillings. Here the $C$ stands for the configurations $C=(\frac{n_f}{\pi},1)$. The exponents are extracted by taking $l=10$
and fitting the logarithm of the  data to a straight line in the region $s\in(200,250)$. The dashed lines are the CFT predictions for the Dirichlet boundary conditions.} 
\label{fig:XX-infinite--disconnected-half filling- different fillings}
\end{figure}

To study the effect of the Fermi momentum $n_f$ we also studied the entanglement entropy in the presence of
the configurations $(\frac{n_f}{\pi},k)$. The results shown in the Figure \ref{fig:XX-infinite--disconnected-half filling- different fillings} 
demonstrate that the smallest scaling dimension
present in the spectrum is the same as before. In other words as far as we take similar configurations on the two slits
the smallest scaling dimension is $\Delta_1=\frac{1}{2}$. 

\begin{figure} [hthp!] 
\centering
\includegraphics[width=0.45\textwidth,angle =-90]{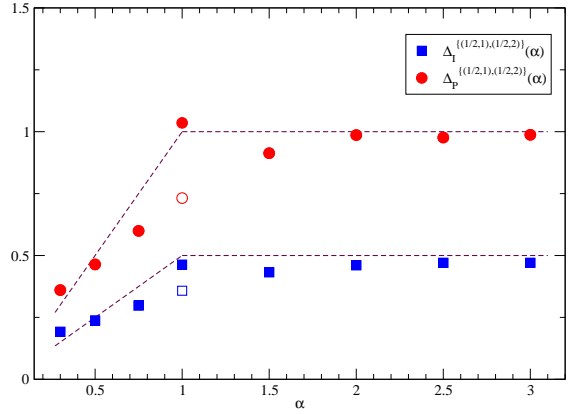}
\caption{(Color online) The exponents $\Delta_I^{\{(\frac{1}{2},1),(\frac{1}{2},2)\}}(\alpha)$ and $\Delta_P^{\{(\frac{1}{2},1),(\frac{1}{2},2)\}}(\alpha)$
for different R\'enyi entropies.  The exponents are extracted by taking $l=10$
and fitting the logarithm of the data to a straight line in the region $s\in(200,250)$. For $\alpha=1$ the empty
circle and square are the results without considering the logarithmic correction. However the filled ones are the correct ones after
considering also the logarithm corrections. The dashed lines are the CFT predictions for the Dirichlet boundary conditions with
$\Delta_1^{\{(\frac{1}{2},1),(\frac{1}{2},2)\}}=\frac{1}{4}$.} 
\label{fig:XX-disconnected-different configurations}
\end{figure}

As we mentioned before although all of the above configurations flow to Dirichlet boundary condition it is yet unclear
what is the value of $\phi$ on the boundary for the different configurations. To have an idea about this quantity one can simply study
the post measurement entanglement entropy when there are different configurations on the two slits. For example, one can put the configuration
$(\frac{1}{2},1)$ on the slit one and the configuration $(\frac{1}{2},2)$ on the slit two and then calculate
the exponent of the power-law decay $\Delta(\alpha)$ of the entanglement. If the exponent is the same as before one can conclude that
most probably both of the configurations flow to the same Dirichlet boundary condition
but if the exponent is different one can simply write
 \begin{eqnarray}\label{exponent and Dirichlet value}
\Delta_{I}^{\{C_1,C_2\}}(\alpha)=\left\{
\begin{array}{c l}      
    2\alpha\Delta_1^{\{C_1,C_2\}}, & \alpha<1,\hspace{1cm}\\
        2\Delta_1^{\{C_1,C_2\}} &\alpha\geq1,       
\end{array}\right.
\end{eqnarray}
where $\Delta_1^{\{C_1,C_2\}}$ is the same as (\ref{scaling dimension dirichlet1}). This can give an idea about the nature of the corresponding
Dirichlet boundary condition. Having the above ideas in mind one can calculate the $\Delta_1^{\{C_1,C_2\}}$ by taking different conformal configurations.
In the Figure \ref{fig:XX-disconnected-different configurations}, we have depicted the results 
for the configurations $C_1=(\frac{1}{2},1)$ and $C_2=(\frac{1}{2},2)$ which
shows that indeed the two configurations apparently flow to two different Dirichlet boundary conditions. 
The $\Delta_1^{\{(\frac{1}{2},1),(\frac{1}{2},2)\}}$ in this case is around
$\frac{1}{4}$. We will show later that  this number is consistent with the calculations of the open boundary conditions.  For 
$\delta^{12}:=\delta^{{\{(\frac{1}{2},1),(\frac{1}{2},2)\}}}=\frac{\phi^{\{(\frac{1}{2},1)\}}-\phi^{\{(\frac{1}{2},2)\}}}{\sqrt{\pi}}$ at this level we have 
two possibilities $\delta^{12}=\frac{1}{\sqrt{2}}$ or $\delta^{12}=1-\frac{1}{\sqrt{2}}$. 
We have repeated the calculations for also other configurations and realized that the $\Delta_1^{\{C_1,C_2\}}$ changes by changing the configurations.
This numerical exercise means that although all the different configurations flow to the Dirichlet boundary conditions they are not equal.
We leave  more through analyzes of this point for a future study.

\subsubsection{Periodic chain}

\begin{figure} [hthp!] 
\centering
 \includegraphics[width=0.5\textwidth,angle =-90]{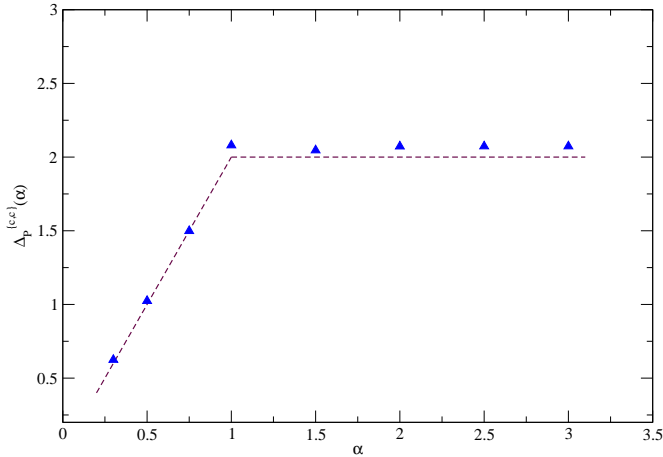}
\caption{(Color online) The exponent $\Delta_{P}^{\{\textbf{c},\textbf{c}\}}(\alpha)$ for the XX model in the setup II. We took $L=302$
 and $l$ goes from $4$ to $24$. 
The dashed line is the formula (\ref{exponent2XX}).} 
\label{fig:deltaxx_per1}
\end{figure}

We also studied the post measurement entanglement entropy for the periodic boundary condition.
In the XX model as we discussed before if we take the same conformal configurations on both lines  the operator with the smallest
 scaling dimension has  $\Delta_1=\frac{1}{2}$.  Consequently for the setup II   we have

\begin{eqnarray}
\label{exponent2XX}
\Delta_{P}^{\{C,C\}}(\alpha)=\left\{
\begin{array}{c l}      
    2\alpha, & \alpha<1,\hspace{1cm}\\
        2 &\alpha\geq1.        
\end{array}\right.
\end{eqnarray}
The numerical calculations are similar to the one done for the Ising model, however, one should be careful that because of the
presence of the zero mode the $\det (1+G)$ or the $\det (1-G)$ or both of them are zero. To overcome this issue first of all we take $h$ and $L$ 
in a way that $n_f=\frac{\pi}{2}$. Then
we change $G_{ii}$ with a small amount $\epsilon$ and then do the calculations. To find the most efficient $\epsilon$
we took smaller and smaller values up to time that the results were reasonably stable. In our calculations, we took effectively $\epsilon=10^{-6}$.
The results shown in the Figure \ref{fig:deltaxx_per1} are consistent with the CFT prediction (\ref{exponent2XX}).
In the more general case of different configurations on the two slits we have
 \begin{eqnarray}\label{exponent and Dirichlet value periodic}
\Delta_{P}^{\{C_1,C_2\}}(\alpha)=\left\{
\begin{array}{c l}      
    4\alpha\Delta_1^{\{C_1,C_2\}}, & \alpha<1,\hspace{1cm}\\
        4\Delta_1^{\{C_1,C_2\}} &\alpha\geq1,       
\end{array}\right.
\end{eqnarray}
where the $\Delta_1^{\{C_1,C_2\}}$'s are the same as the last subsection. The numerical results  presented in the 
Figure \ref{fig:XX-disconnected-different configurations} are consistent with CFT computations.

\subsubsection{Semi-infinite  chain}

\begin{figure} [hthp!] 
\centering
 \includegraphics[width=0.5\textwidth,angle =-90]{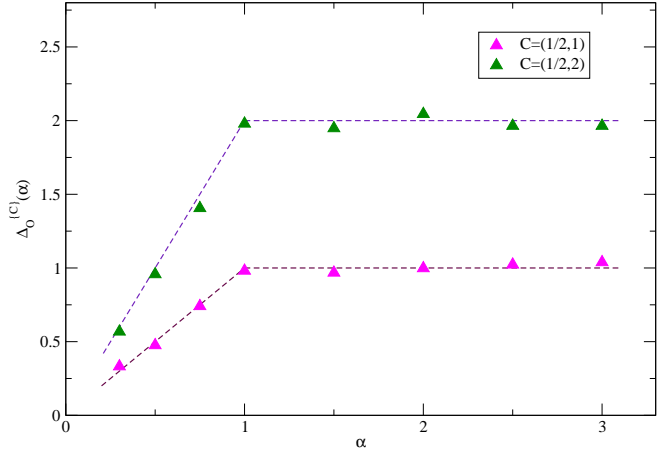}
\caption{(Color online) The exponent $\Delta_{O}^{\{C\}}(\alpha)$ for the XX model in the  setup III with 
the configurations  $C=(\frac{1}{2},1)$ and $C=(\frac{1}{2},2)$. We took $l+s=300$
 and $l$ goes from $4$ to $24$. 
The dashed line is the formula (\ref{exponent and Dirichlet value semi}).} 
\label{fig:deltaxx_si}
\end{figure}

Finally, we repeated the calculations for the semi-infinite system. Note that we assumed Dirichlet boundary condition
for the actual boundary of the system. However this Dirichlet boundary condition can be different
from the one induced by the projective measurement. Based on the CFT calculations the entanglement entropy of the 
two disconnected systems, i. e. setup III, should decay like a power-law with an exponent coming from the formula 
 \begin{eqnarray}\label{exponent and Dirichlet value semi}
\Delta_{O}^{\{C\}}(\alpha)=\left\{
\begin{array}{c l}      
    4\alpha\Delta_1^{\{C\}}, & \alpha<1,\hspace{1cm}\\
        4\Delta_1^{\{C\}} &\alpha\geq1,       
\end{array}\right.
\end{eqnarray}
where the $\Delta_1^{\{C\}}$ is unknown a priory but can be determined by the numerical calculations for different configurations.
Our numerical results performed by using different configurations , i.e. $(\frac{1}{2},1)$ and $(\frac{1}{2},2)$ are shown in the Figure ~\ref{fig:deltaxx_si}. As it is clear
from the Figure the value of $\Delta_1^{\{C\}}$ is dependent on the configuration but  after fixing its value 
the other exponents can be derived using our CFT results. Based on the numerical results for the configuration
$(\frac{1}{2},2)$ the $\Delta_1^{\{(\frac{1}{2},2)\}}=\frac{1}{2}$ which in principle means that $\delta_2=0$. In other words  this 
configuration flows to a Dirichlet boundary condition which is exactly the same as the natural Dirichlet boundary condition
of the semi-infinite system at the origin. However, for the configuration
$(\frac{1}{2},1)$ the $\Delta_1^{\{(\frac{1}{2},1)\}}=\frac{1}{4}$. This value was expected from our earlier calculations based on the infinite
system with two slits, one slit with the configuration $(\frac{1}{2},1)$ and the other one with $(\frac{1}{2},2)$. Since the configuration 
$(\frac{1}{2},2)$ is exactly the same as the natural boundary we can simply find that 
$\Delta_{1}^{\{(\frac{1}{2},1)\}}=\Delta_1^{\{(\frac{1}{2},1),(\frac{1}{2},2)\}}=\frac{1}{4}$. This result
shows the consistency of our computation in a most revealing way.

\subsubsection{non-conformal configurations:} 

\begin{figure} [hthp!] 
\centering
\includegraphics[width=0.45\textwidth,angle =-90]{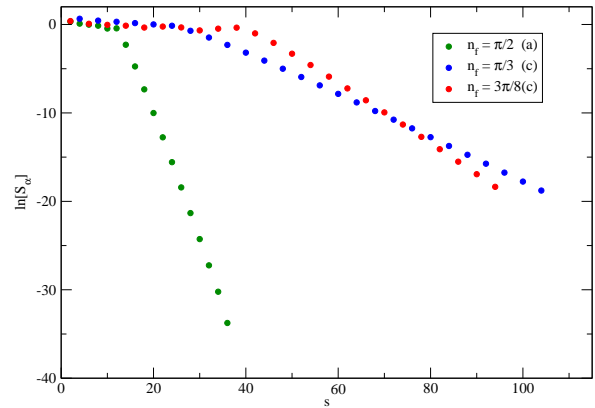}
\caption{(Color online) Post-measurement entanglement entropy of the disconnected regions with different fillings and
configurations in the XX model  for
the setup I. The letter inside the parenthesis is the corresponding configuration.} 
\label{fig:XX-infinite--disconnected-nonconformal}
\end{figure}

We also calculated the post measurement entanglement entropy when the result of the measurement is not a conformal configuration, for example,
\textbf{a} and \textbf{b}.
The numerical results performed in different conditions suggest that the entanglement entropy of the disconnected regions decays
exponentially for the large measurement regions, see Figure ~\ref{fig:XX-infinite--disconnected-nonconformal}. 
This  result which can have important consequences when we
discuss localizable entanglement could be expected from our discussion regarding the post measurement
entanglement entropy in the connected cases. Since here we are working in the large $s$ regime it is not expected that the CFT results be valid.
However, for small $s$ one might hope to see some agreement with the CFT formulas. Indeed as it is clear in the Figure 
~\ref{fig:XX-infinite--disconnected-nonconformal}
the entanglement entropy does not decay immediately after introducing the $s$. It just starts to decay exponentially when
$s$ is large enough with respect to the $l$.
Note that the exponential decay of
the post measurement entanglement entropy in this case is reminiscent of the the same quantity for the non-critical systems. This means that
for the large values of $s$ non-critical boundary conditions suppress the correlation functions between the subsystem and the rest of the system
strongly which effectively  mimic the behaviour of a massive system. This interpretation is consistent with what we argued during the discussion
regarding Affleck-Ludwig boundary entropy. 
The non-critical 
chains will be discussed in the upcoming section.  

\section{ Entanglement entropy after selective measurements in the non-critical Ising chain}

In this section, we study numerically the non-critical transverse field Ising chain. In particular, we study the formulas (\ref{massive field theory}), (\ref{massive field theory 2}) and
(\ref{massive field theory disconnected}). The elements of the Green matrix can be calculated using the following integral \cite{Lieb}
  \begin{eqnarray}\label{Green non-critical}
G_{ts}=\frac{1}{2\pi}\int_0^{2\pi}d\phi e^{-i(t-s)\phi}\frac{e^{-i\phi}-h}{\sqrt{(1-h e^{i\phi})(1-h e^{-i\phi})}}.
\end{eqnarray}
The above formula is valid for an infinite chain but we believe that all of our upcoming conclusions are equally valid for also finite systems. For the gapped Ising model
we have $m=|h-1|=\xi^{-1}$.

\begin{figure} [hthp!] 
\centering
 \includegraphics[width=0.5\textwidth,angle =-90]{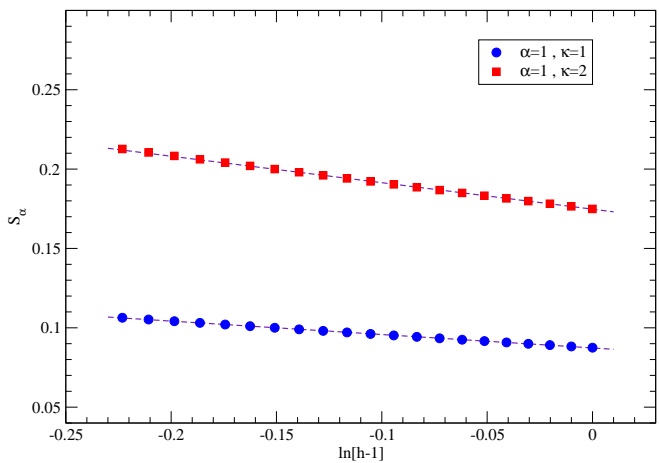}
\caption{(Color online) Post measurement von Neumann entanglement entropy in a non-critical transverse field Ising chain for two cases: the region $B$ after measurement has one $\kappa=1$ or two $\kappa=2$ contact points with $\bar{B}$.
The interval for $h$ is chosen in a way that $a<m^{-1}<l,s$. The dashed lines are the equation (\ref{massive field theory}).} 

\label{fig:massive1}
\end{figure}

\begin{figure} [hthp!] 
\centering
 \includegraphics[width=0.5\textwidth,angle =-90]{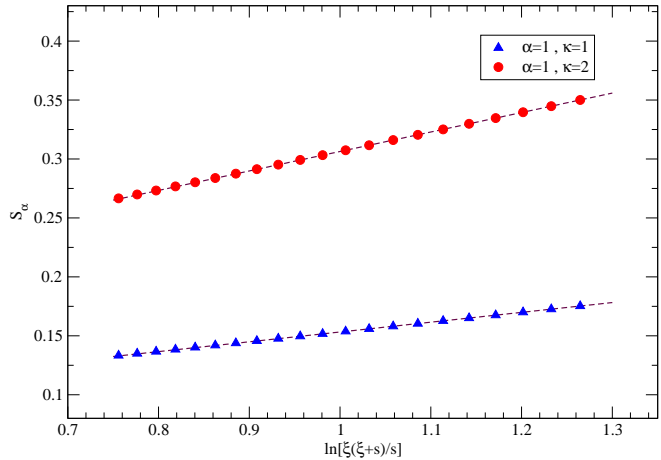}
\caption{(Color online)  Post measurement von Neumann entanglement entropy in a non-critical transverse field Ising chain for two cases: the region $B$ after measurement has one $\kappa=1$ or two $\kappa=2$ contact points with $\bar{B}$.
The interval for $h$ is chosen in a way that $s<m^{-1}<l$. The dashed lines are the equation (\ref{massive field theory 2})} 

\label{fig:massive_s1}
\end{figure}

\begin{figure} [hthp!] 
\centering
\includegraphics[width=0.5\textwidth,angle =-90]{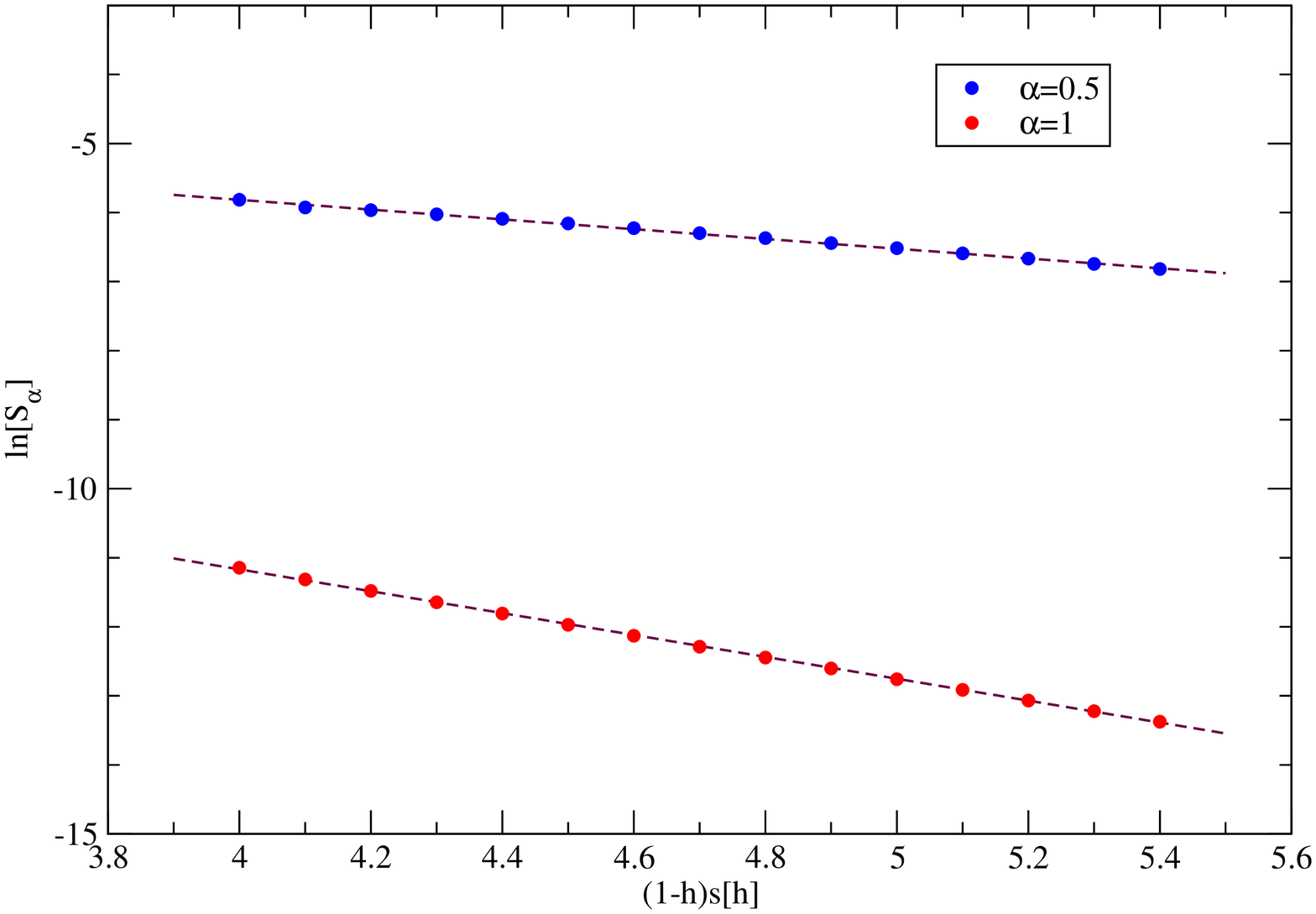}
\includegraphics[width=0.5\textwidth,angle =-90]{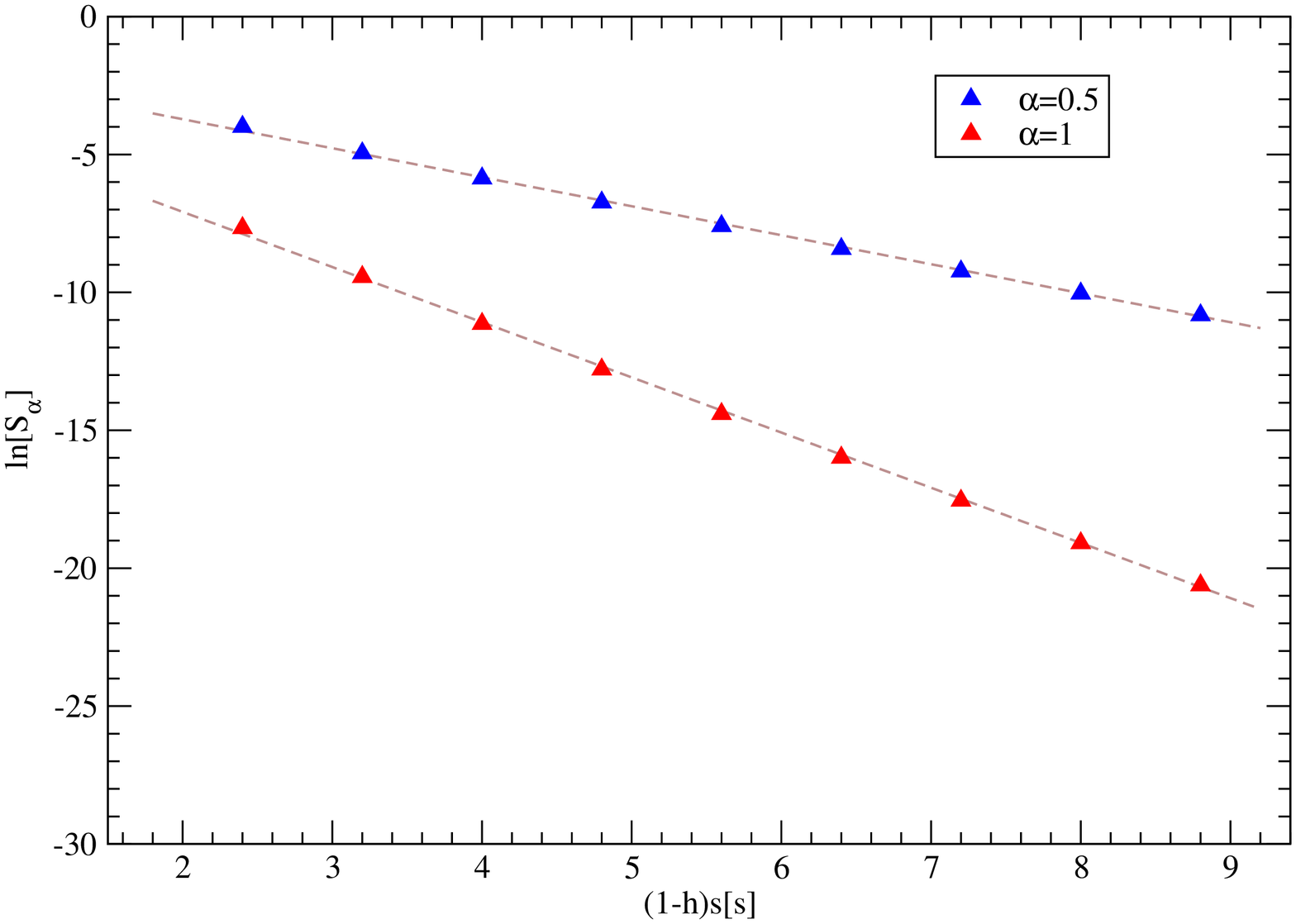}

\caption{(Color online) Post measurement von Neumann entanglement entropy in a non-critical transverse field Ising chain for disconnected regions (setup I): Up) $s$ is fixed
and $h$ changes. Down) $h$ is fixed and $s$ is changing. 
The intervals are chosen in a way that $a<m^{-1}<l,s$. The dashed lines are the equation (\ref{massive field theory disconnected}).} 

\label{fig:massive_dis}
\end{figure}

\begin{figure} [hthp!] 
\centering
\includegraphics[width=0.5\textwidth,angle =-90]{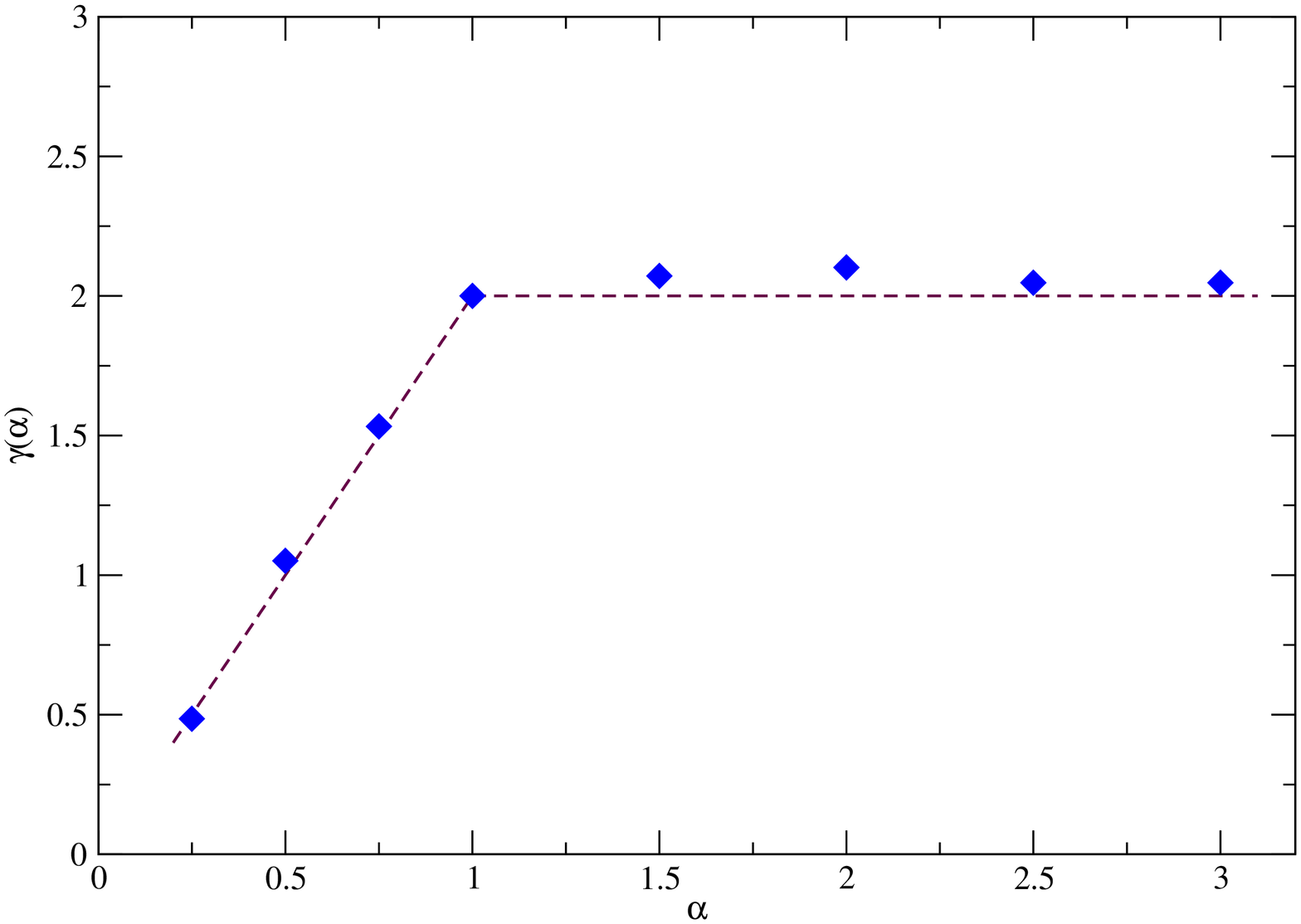}

\caption{(Color online) $\gamma(\alpha)$ vs $\alpha$. The dashed lines are the equation (\ref{gamma alpha}).} 

\label{fig:delta_mass_dis}
\end{figure}

We first study the post measurement entanglement entropy in the non-critical Ising chain for the connected cases, in other words,
we are interested to check the validity of the equations (\ref{massive field theory}) and (\ref{massive field theory 2}). The results of the numerical calculations are shown in the
Figures (\ref{fig:massive1}) and (\ref{fig:massive_s1}). The numerical calculations are in a reasonable agreement with the general predictions. Note that here we discussed just the 
post measurement entanglement entropy in the $\sigma^z$ basis. As we discussed before we do not expect the equations (\ref{massive field theory}) and (\ref{massive field theory 2})
be valid in generic bases. However, it is quite possible that if one stick to a domain which is far from the measurement region then again the equation (\ref{massive field theory})
be valid with $\kappa=2$. This is simply because any local measurement in part of a massive system affects very little the correlation functions far from the
measurement region.
Finally, we also studied the post measurement entanglement entropy of two decoupled regions. The results depicted in the Figure (\ref{fig:massive_dis}) shows that
the entanglement entropy decreases exponentially with respect to the size of the measurement region in complete agreement with the equation (\ref{massive field theory disconnected}). We also studied
$\gamma(\alpha)$ with respect to $\alpha$ and surprisingly found that it closely follows (see Figure \ref{fig:delta_mass_dis}):

 \begin{eqnarray}\label{gamma alpha}
\gamma(\alpha)=\left\{
\begin{array}{c l}      
    2\alpha, & \alpha<1,\hspace{1cm}\\
        2 &\alpha\geq1,\\        
\end{array}\right.
\end{eqnarray}
Although we do not expect the above formula be universal the general behaviour, linear increase and then saturation, might be a universal pattern for the massive systems. 

\section{Entanglement entropy after selective measurements in the finite temperature  XY chain}

In this section, we study numerically the effect of the temperature on the post measurement R\'enyi entropy of the critical
XY chain. In other words we would like to verify the equations 
(\ref{SB for finite temperature 1}) and (\ref{power-law decay finite temperature2}) for the critical XY chain.
The method of the calculation is exactly the same as before, one just needs to use the finite temperature Green matrix
in the formulas of the section ~5. The Green matrix of the finite temperature XY chain is given by
\begin{eqnarray}\label{Green matrix of finite temperature}
G_{ij}=\int_0^{2\pi}\frac{d\phi}{2\pi}\tanh \frac{\epsilon_{\phi}}{2T}e^{i\theta_{\phi}}e^{i\phi(i-j)}
\end{eqnarray}
where
\begin{eqnarray}\label{theta and epsilon}
e^{i\theta_{\phi}}=\frac{\cos \phi -h+ia\sin\phi}{\epsilon_{\phi}},\\
\epsilon_{\phi}=\sqrt{(\cos\phi-h)^2+a^2\sin^2\phi}.
\end{eqnarray}
In the next two subsections we will use the above Green matrix for the critical transverse field Ising model and the critical
XX chain and calculate the R\'enyi entropies.

\subsection{Transverse field Ising chain:}

In this subsection, we first study the post measurement R\'enyi entropy in the critical transverse field Ising chain and later
we focus on the non-critical case.
\subsubsection{Critical transverse field Ising chain:}

\begin{figure} [hthp!] 
\centering
 \includegraphics[width=0.5\textwidth,angle =-90]{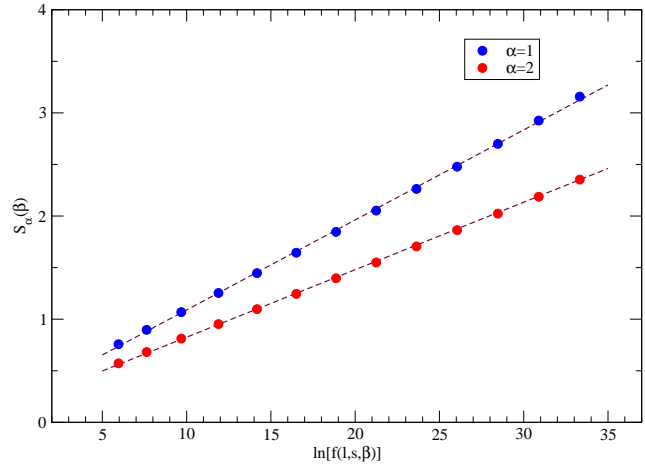}
\caption{(Color online) The finite temperature R\'enyi entropy for the critical transverse
field Ising chain for $\alpha=1$ and $2$. In the above 
$f(l,s,\beta)=
\frac{c}{12}(1+\frac{1}{\alpha})\ln \Big{(}\frac{\beta}{\pi}\frac{\sinh\frac{\pi}{\beta}(l+s_1)\sinh\frac{\pi}{\beta}l}{s_2\sinh\frac{\pi}{\beta}s_1}\Big{)}$
and  the dashed lines are the CFT results.} 

\label{fig:ft_ising1}
\end{figure}

To calculate the R\'enyi entropy of the finite temperature transverse field Ising point we first put $a=h=1$
in the equation (\ref{Green matrix of finite temperature}) then we fixed the configuration to \textbf{a}. The results for the infinite
connected case is demonstrated in the Figure ~\ref{fig:ft_ising1} which have a reasonable compatibility with our analytic
result (\ref{SB for finite temperature 1}).

\begin{figure} [hthp!] 
\centering
 \includegraphics[width=0.5\textwidth,angle =-90]{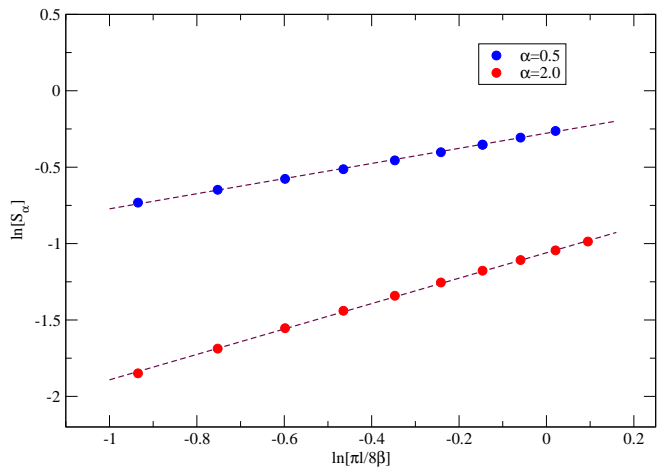}
\caption{(Color online) The finite temperature R\'enyi entropy for the critical transverse
field Ising chain for $\alpha=\frac{1}{2}$ and $2$ in the regime $\frac{\pi s}{\beta}\gg 1\gg \frac{\pi l}{8\beta}$. In the above 
the slop of the dashed lines are $0.83$ and $0.49$ for  $\alpha=2$ and $\frac{1}{2}$ respectively.} 

\label{fig:ft_pmd1}
\end{figure}

\begin{figure} [hthp!] 
\centering
 \includegraphics[width=0.5\textwidth,angle =-90]{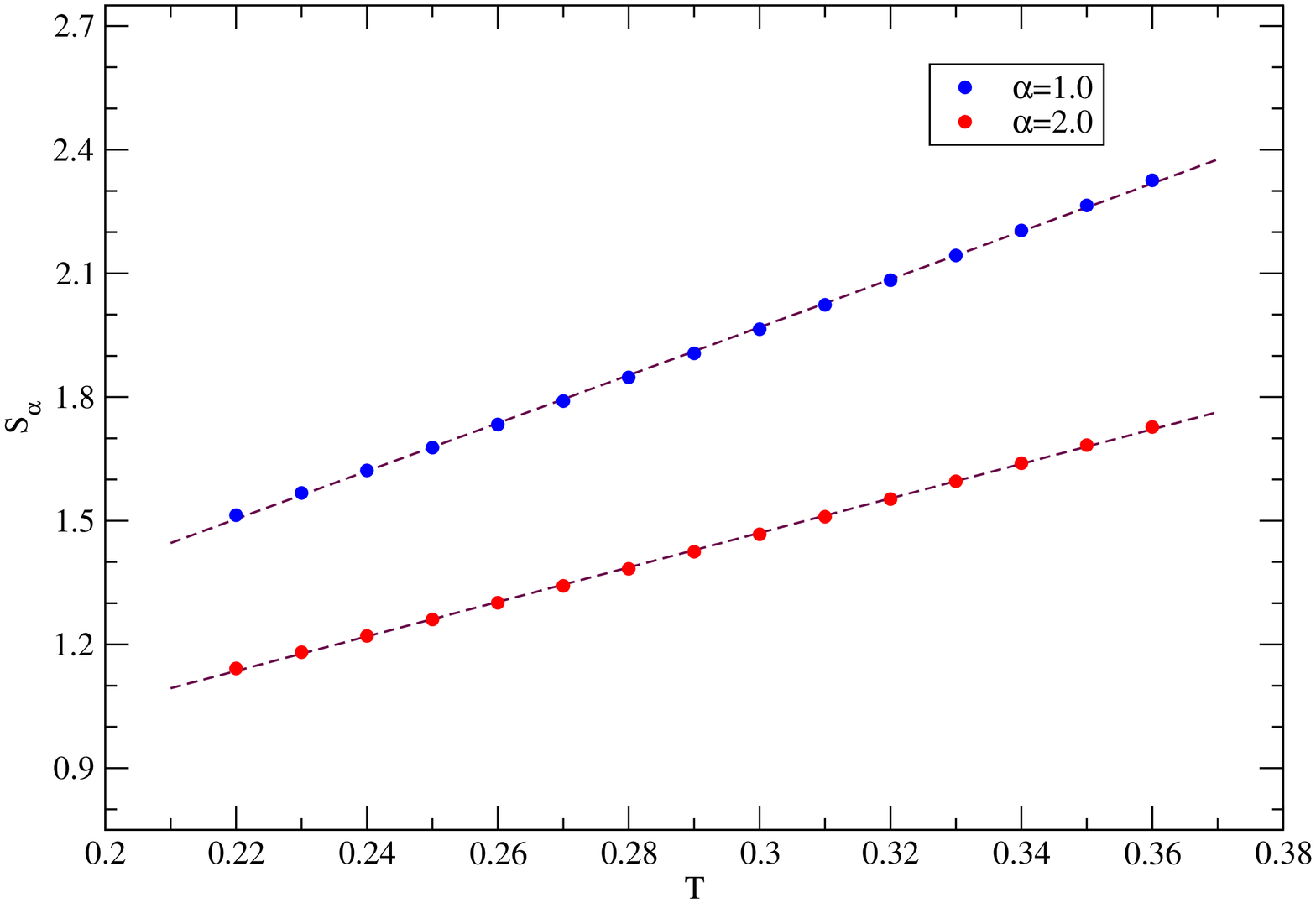}
 \includegraphics[width=0.5\textwidth,angle =-90]{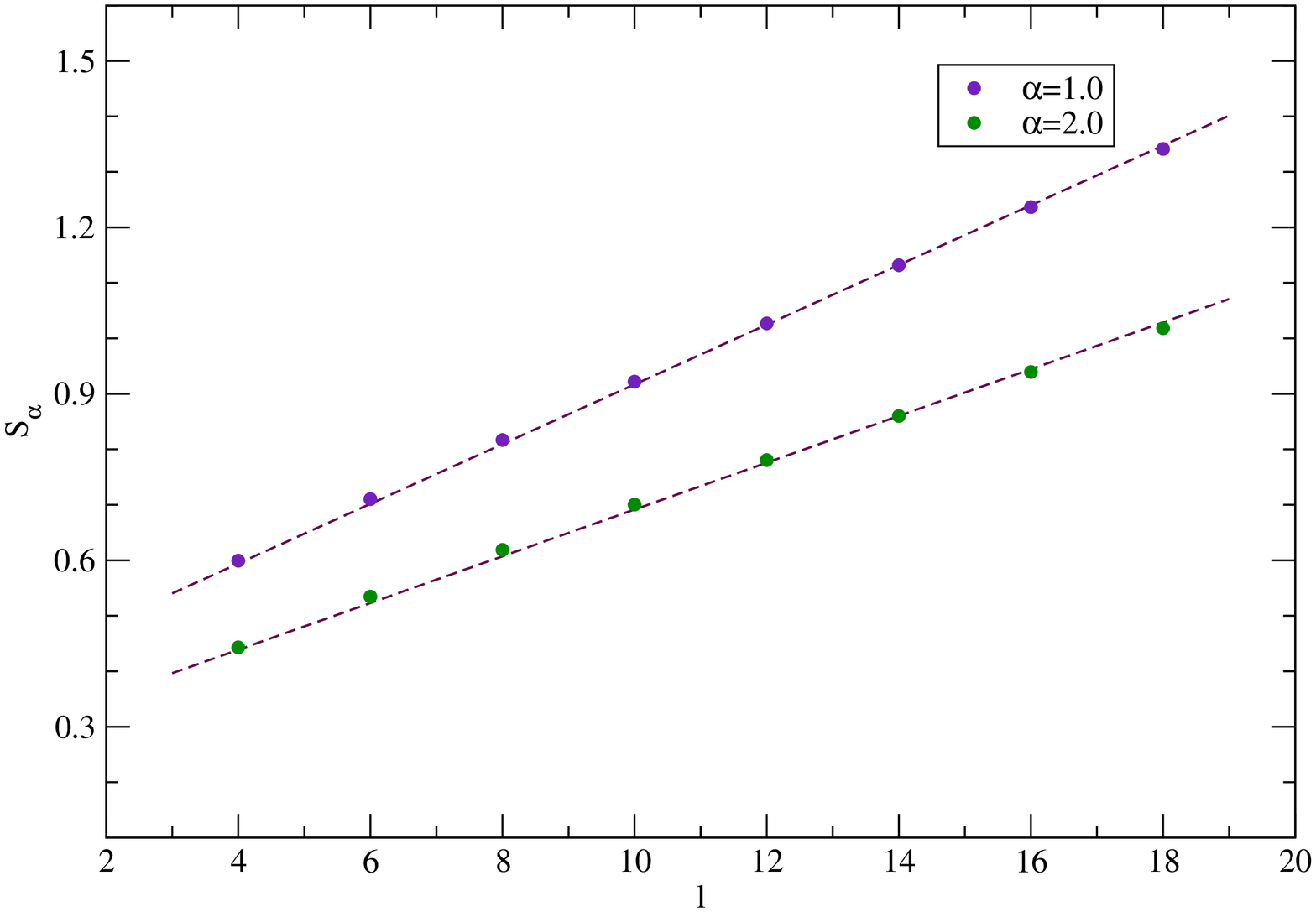}
\caption{(Color online) The high temperature R\'enyi entropy for the critical transverse
field Ising chain for $\alpha=1$ and $2$ in the non-connected setup I. Top) The R\'enyi entropy with respect to the temperature
with fixed $l=10$. Down) The R\'enyi entropy with respect to the length
with fixed temperature $T=0.1$. The ratio of the coefficient of the two lines is around $1.3$.} 

\label{fig:ft_pmchtl}
\end{figure}

We then extended our calculations to the non-connected cases especially we studied the regime
$\frac{\pi s}{\beta}\gg 1\gg \frac{\pi l}{8\beta}$ where the entropy increases like a power-law with respect to the measurement region.
The numerical results shown in the Figure (\ref{fig:ft_pmd1}) indeed confirm the power-law behaviour and the power of the exponent
is in a reasonable compatibility with the CFT formula (\ref{power-law decay finite temperature2}).
After confirming the CFT results for the small temperature regime  we studied the large temperature regime.
In this case, we expect a linear increase of the post measurement R\'enyi entropy with respect to the temperature
and the size of the region. The interesting setup to study in this regime is the setup I which we have two decoupled regions.
Here we expect to have the equation 
\begin{eqnarray}\label{SB for infinite temperature final}
S_{\alpha}(\beta)=
\frac{\pi c}{6}(1+\frac{1}{\alpha})\frac{l}{\beta}+....
\end{eqnarray}
The numerical results shown in the Figure (\ref{fig:ft_pmchtl})
show clearly the linear increase with respect to the temperature and also the volume law. Note that the coefficient
of the linear term is not a universal quantity but one expect 
\begin{eqnarray}\label{SB for infinite temperature final ratio}
\frac{S_{\alpha_1}(\beta)}{S_{\alpha_2}(\beta)}=\frac{\alpha_2}{\alpha_1}\frac{1+\alpha_1}{1+\alpha_2}
\end{eqnarray}
to be a universal quantity. Our numerical results are consistent with the above ratio.

\subsubsection{Non-critical transverse field Ising chain:}

\begin{figure} [hthp!] 
\centering
 \includegraphics[width=0.5\textwidth,angle =-90]{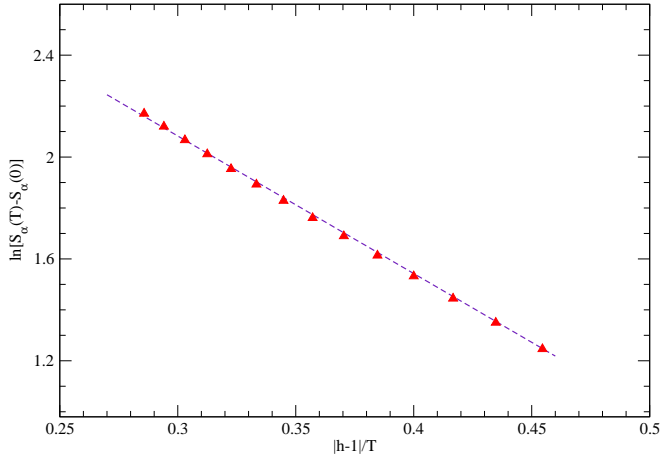}
\caption{(Color online) The high temperature R\'enyi entropy for the critical transverse
field Ising chain for $\alpha=1$. We took $h = 1.50$ and the sizes of the regions $A$ and $B$ are  $s=40$ and $l=20$ respectively.} 
\label{fig:massive_ft1}
\end{figure}

In this subsection, we study the von Neumann entropy in the finite temperature gapped transverse field Ising chain. Following the ideas of section ~4 we expect the entropy of a subsystem after projective measurement
decays exponentially with respect to the gap in the system. In other words, because of the Gibbs nature of the reduced density matrix one expect that the leading term of the entropy changes as \cite{Herzog2013}
\begin{eqnarray}\label{non-critical Ising finite temperature}
S(T)-S(0)\sim e^{-\frac{|h-1|}{T}}, 
\end{eqnarray}
In the Figure (\ref{fig:massive_ft1}) we verified the above equation for a connected case. We expect similar results for also non-connected cases.

\subsection{Critical XX chain:}

\begin{figure} [hthp!] 
\centering
 \includegraphics[width=0.5\textwidth,angle =-90]{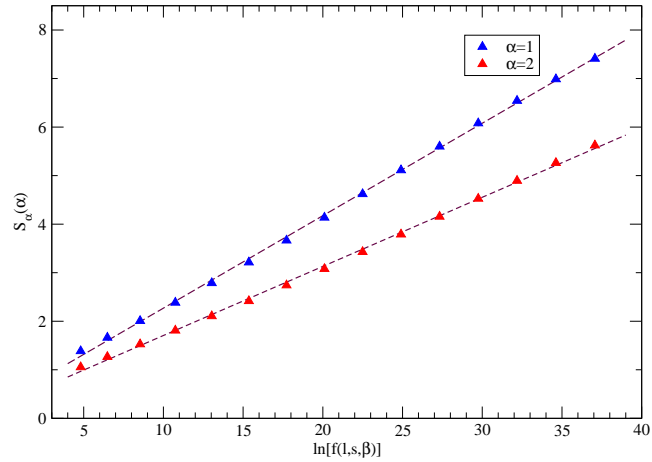}
\caption{(Color online) The finite temperature R\'enyi entropy for the critical XX chain for $\alpha=1$ and $2$. In the above 
$f(l,s,\beta)=
\frac{c}{12}(1+\frac{1}{\alpha})\ln \Big{(}\frac{\beta}{\pi}\frac{\sinh\frac{\pi}{\beta}(l+s_1)\sinh\frac{\pi}{\beta}l}{s_2\sinh\frac{\pi}{\beta}s_1}\Big{)}$
and  the dashed lines are the CFT results.} 

\label{fig:ft_xx1}
\end{figure}

In this section, we calculated the R\'enyi entropy of the finite temperature XX chain by first putting $a=h=0$
in the equation (\ref{Green matrix of finite temperature}). Then we fixed the configuration to \textbf{c}. The results for the infinite
connected case is demonstrated in the Figure ~(\ref{fig:ft_xx1}) which have a reasonable compatibility with our analytic
result (\ref{SB for finite temperature 1}).

\begin{figure} [hthp!] 
\centering
 \includegraphics[width=0.5\textwidth,angle =-90]{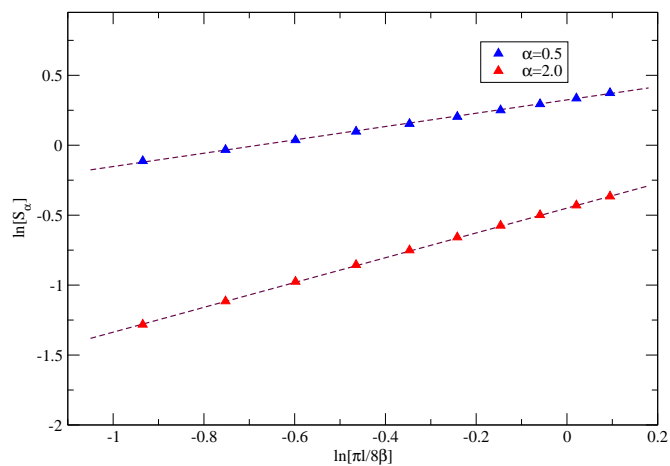}
\caption{(Color online) The finite temperature R\'enyi entropy for the critical XX chain for $\alpha=\frac{1}{2}$ and $2$ in the regime $\frac{\pi s}{\beta}\gg 1\gg \frac{\pi l}{8\beta}$. In the above 
the slop of the dashed lines are $0.88$ and $0.47$ for  $\alpha=2$ and $\frac{1}{2}$ respectively.} 

\label{fig:ft_pmdxx1}
\end{figure}

\begin{figure} [hthp!] 
\centering
 \includegraphics[width=0.5\textwidth,angle =-90]{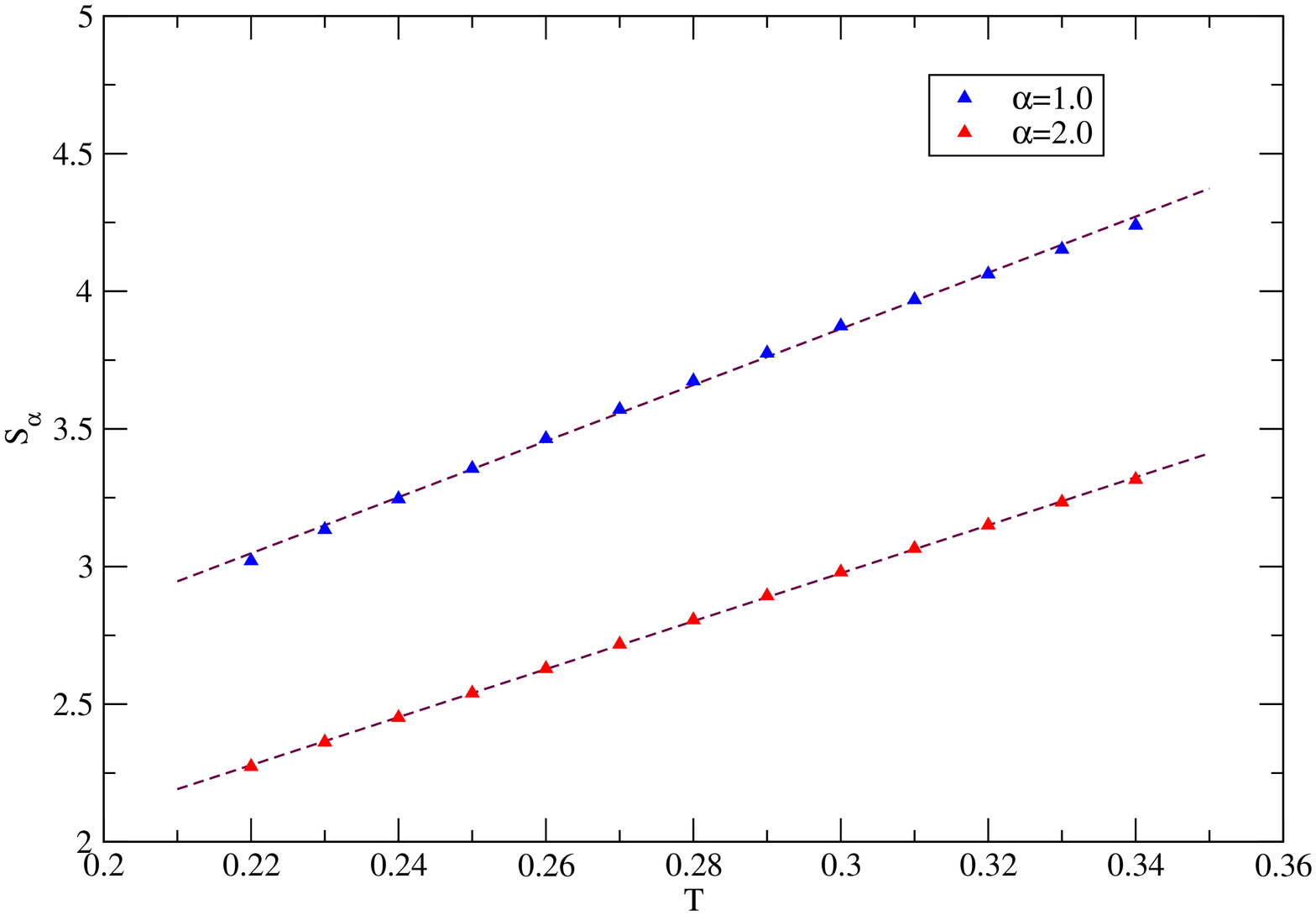}
 \includegraphics[width=0.5\textwidth,angle =-90]{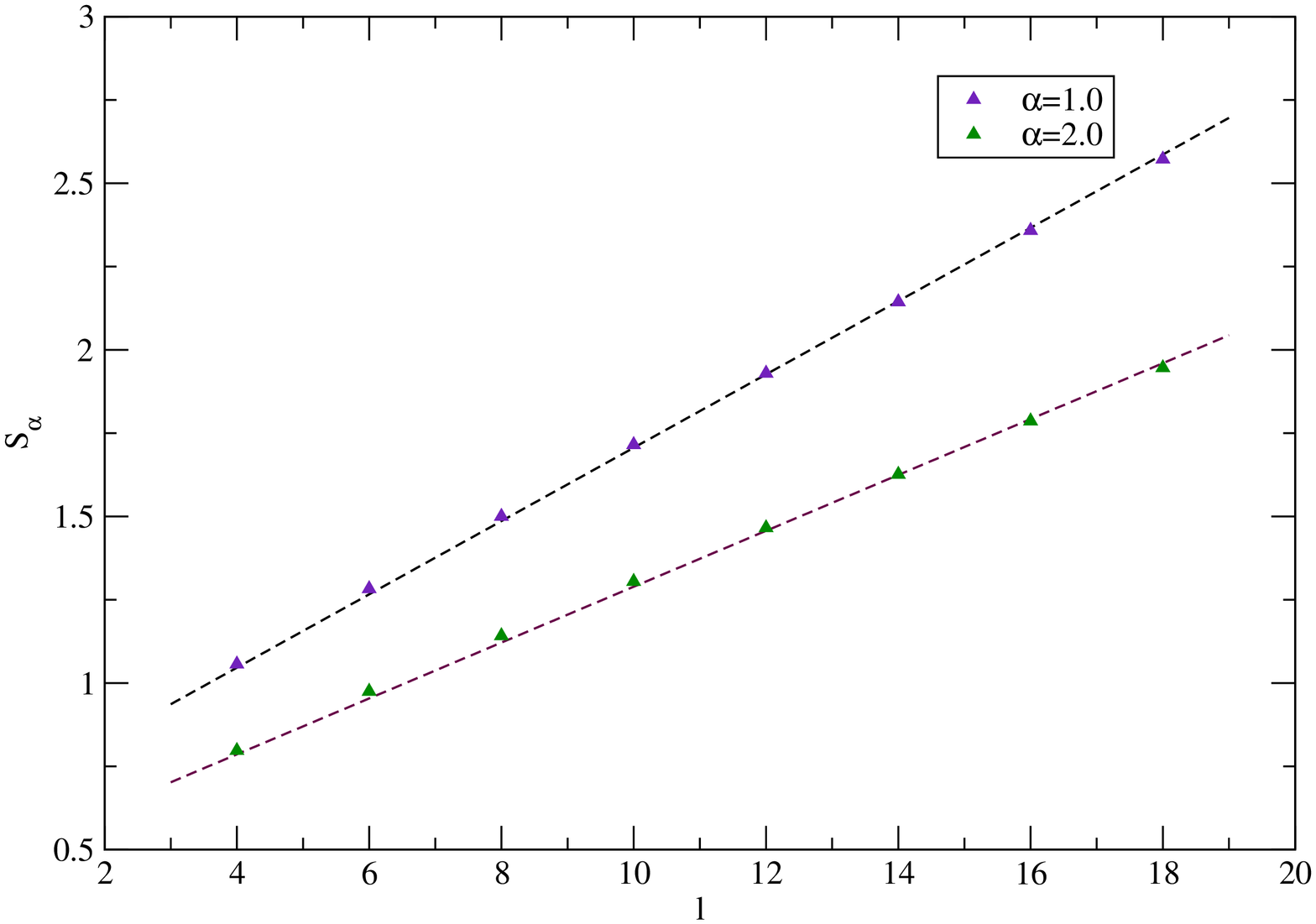}
\caption{(Color online) The high temperature R\'enyi entropy for the critical XX 
 chain for $\alpha=1$ and $2$ in the non-connected setup I. Top) The R\'enyi entropy with respect to the temperature
with fixed $l=10$. Down) The R\'enyi entropy with respect to the length
with fixed temperature $T=0.1$. } 

\label{fig:ft_pmchtxx}
\end{figure}

We then calculated the R\'enyi entropy for the  non-connected case in the setup I. We first considered the regime of the small
temperature with the constraint $\frac{\pi s}{\beta}\gg 1\gg \frac{\pi l}{8\beta}$,
where the entropy increases like a power-law with respect to the measurement region, see equation (\ref{power-law decay finite temperature2}).
The numerical results demonstrated in the Figure (\ref{fig:ft_pmdxx1}) are consistent with the CFT results.
Finally, we made some numerical computations in the large-temperature regime for the setup I. In this regime the R\'enyi entropy should increase 
linearly with respect to the temperature and size of the sub-region. Our numerical results
shown in the Figure (\ref{fig:ft_pmchtxx}) are compatible with the CFT formula (\ref{SB for infinite temperature final}). It is worth 
mentioning that although for the non-conformal configurations  we expect a similar linear increase in the R\'enyi entropy with respect to 
the temperature and the size of the subsystem we do not expect the ratio of the slops for different $\alpha$'s 
respects the equation (\ref{SB for infinite temperature final ratio}).

\section{Remarks on the possible experimental setup}

In this section we will briefly make some remarks on the possible method to produce the desired post measurement wave functions.
The setup studied in this paper was the following: take a wave function of the ground sate
of a quantum chain and then choose an observable (basis). Then make a partial
 projective measurement
of that observable in a subsystem $A$. The rest of the system collapses to a new wave function. The bipartite entanglement entropy of the remaining subsystem is the desired quantity. However, to use
the powerful techniques of the CFT the observable and the result of the measurement should be chosen appropriately. In the experiment
one can choose the observable as she wishes but the result of the measurement is something one can not control. On top of that after the measurement the system 
will evolve by time and again all the three parts of the system will get entangled one more time. To have the exact desired post measurement wave function one can do as
follows: prepare a system with the desired hamiltonian and let the system relax to the ground state. Then turn-off the interactions between the particles (for example spins). 
Choose a conformal observable (basis) and with an external
field force the desired conformal configuration in the subsystem $A$. For example, in the spin chains
this can be done by a magnetic field acting on the spins in the particular directions. The final wave function of $\bar{A}$ is the desired post measurement wave function. 
Then one can try to study the bipartite entanglement entropy of this wave function by one of many different methods that have been introduced recently, see \cite{experimentalEE1,experimentalEE2} . Notice that in the above procedure it is important to turn-off the interactions after 
preparing the system in the ground state. This method can be obviously used to prepare many body entangled states that are spatially disconnected.






\section{Conclusions}

In this paper we studied different aspects of the post measurement entanglement entropy in the critical and the non-critical quantum chains. We first derived 
different formulas for the post measurement entanglement entropy in the conformal field theories. We studied systems with boundaries and also  conformal field theories at the finite temperature.
In addition, we studied the role of the boundary entropy in the post measurement entanglement entropy. Some exact results were also presented for the entanglement Hamiltonian and 
the distribution of the eigenvalues of the reduced density matrices. Based on some physical arguments we also presented some predictions regarding the
post measurement entanglement entropy in the massive systems. The above analytical results are in principle valid for all the projective measurements that
respect the conformal symmetry of the bulk. However, in reality one needs to check what bases and configurations respect this symmetry in actual discrete models. To check the validity of our results
we first provided a method to study the post measurement entanglement entropy in the generic free fermion models. The method is based on Grassmann variables and can be used in any dimension.
We then used the technique to study the post measurement entanglement entropy in the XY-chain. In particular, we studied the transvese field Ising chain and the XX-chain.
Because of the presence of the $U(1)$ symmetry in the XX-chain the model is strikingly different from the Ising chain. Many subtilities appear during the study of
the discrete models which makes the applications of the CFT formulas to the discrete models very tricky. These subtilities encourage further analytical and numerical calculations on the discrete models.
In particular, it is very imporatnt to study the effect of the basis of the measurement on the post measurement entanglement entropy in different discrete models. Concerning the massive systems all of 
our results were based on huristic arguments some analytical results and further numerical calculations are surely necessary to put the results on the firm ground. In particular,
calculations based on boundary integrable models can in principle shed light in this direction. Most of the results presented in this paper can be more or less strightforwardly
generalized to higher dimensions \cite{Rajabpourarea} we leave more throuh analysis to a future work.

Finally, it is worth mentioning that the method used in this paper to calculate the post measurement entanglment entropy has a very intimte connection to the Casimir energy of floating objects on the Reimann
surfaces. In other words one can calculate the entanglement entropy by knowing the Casimir energy. Since the reverse is not true it is quite encouraging to think more seriously about the many implications that
this approach might have in the fundamental level.

 \paragraph*{Acknowledgments.} 
 We thank A Bayat and M M Sheikh-Jabbari  for useful discussions and comments on the draft. 
 The work of MAR
was supported in part by CNPq. KN acknowledges support from the National Science Foundation under grant
number PHY-1314295. MAR also thanks ICTP for hospitality during a period which part of this work was completed.

\begin{appendices}
\section{Conformal maps}

In this appendix we list the conformal maps derived in the \cite{Rajabpour2016}. Their exact form is needed to derive
the entanglement hamiltonians. 

\subsection{Infinite system}

The conformal map from the plane with two slits on a line with lengths $s_1$ and $s_2$ and a branch cut with the length
$l$ to an annulus with the inner 
and outer radiuses $r=e^{-h_{\alpha}}$ and $r=1$ with $h_{\alpha}=\frac{h}{\alpha}$ has the following form:

\begin{eqnarray}\label{conformal map1}
w_{\alpha}(z)&=&\Big{(}e^{-\frac{h}{2}}e^{h \frac{\mbox{sn}^{-1}(\tilde{z},k^2)}{2\mathcal{K}(k^2)}}\Big{)}^{\frac{1}{\alpha}},\\
h&=&2\pi\frac{\mathcal{K}(k^2)}{\mathcal{K}(1-k^2)},
\end{eqnarray}
where $\mathcal{K}$ and $\mbox{sn}^{-1}$ are
the elliptic  and inverse Jacobi functions \footnote{Note that in all of the formulas we adopt
the Mathematica convention for all the elliptic functions.} respectively and
\begin{eqnarray}\label{conformal map1 details}
\tilde{z}&=&\frac{2a}{k}\frac{z}{bz+1}-\frac{1}{k},\nonumber\\
a&=&\sqrt{\frac{s_2(s_2+l)}{s_1(s_1+l)}}\frac{1}{l+s_1+s_2},\nonumber\\
b&=&\frac{\sqrt{s_1s_2(l+s_1)(l+s_2)}-s_2(l+s_1)}{(l+s_1)(s_1s_2-\sqrt{s_1s_2(l+s_1)(l+s_2)})},\nonumber\\
\end{eqnarray}
with the parameter $k$ given by
\begin{eqnarray}\label{conformal map1 details2}
k=1+2\frac{s_1s_2-\sqrt{s_1s_2(l+s_1)(l+s_2)}}{l(l+s_1+s_2)}.
\end{eqnarray}

Having the above formulas we can calculate the geometric part of the partition function as

\begin{eqnarray}\label{geometric general full plane}
\frac{\delta \ln Z^{geom}_{\alpha}}{\delta l}=
\frac{c\alpha}{6}\frac{\Big{(}(-2a+b)^2-b^2k\Big{)}\Big{(}2\pi^2-(1+k(6+k))\alpha^2\mathcal{K}^2(1-k^2)\Big{)}}{16ak(1+k)\alpha^2\mathcal{K}^2(1-k^2)}.\hspace{0.5cm}
\end{eqnarray}
Different limits of the above formula have been discussed in \cite{Rajabpour2016}.

\subsection{Finite system}

The conformal map from the cylinder with two aligned slits and a branch cut to
annulus  with the inner 
and outer radiuses $r=e^{-h_{\alpha}}$ and $r=1$ with $h_{\alpha}=\frac{h}{\alpha}$ has the following form:

\begin{eqnarray}\label{conformal map2}
w_{\alpha}(z)&=&\Big{(}e^{-\frac{h}{2}}e^{h \frac{\mbox{sn}^{-1}(\tilde{z},k^2)}{2\mathcal{K}(k^2)}}\Big{)}^{\frac{1}{\alpha}},\\
h&=&2\pi\frac{\mathcal{K}(k^2)}{\mathcal{K}(1-k^2)},
\end{eqnarray}
where the conformal 
map $\tilde{z}(z)$, which takes the system from infinite cylinder with two slits
to the whole plane with two symmetric aligned slits on the real line has the following form:

\begin{eqnarray}\label{conformal map2 details}
\tilde{z}&=&\frac{e^{2i\pi\frac{z}{L}}+a_0}{b_1e^{2i\pi\frac{z}{L}}+b_0},\\
a_0&=&\frac{e^{2i\pi\frac{s_1}{L}}}{N}\Big{(}1-k-2e^{2i\pi\frac{l+s_1}{L}}+(1+k)e^{2i\pi\frac{l}{L}}\Big{)},\nonumber\\
b_1&=&\frac{-1}{N}\Big{(} (1-k)e^{2i\pi\frac{l+s_1}{L}}+2k-(1+k)e^{2i\pi\frac{s_1}{L}}\Big{)},\nonumber\\
b_0&=&\frac{e^{2\pi\frac{s_1}{L}}}{N}  \Big{(}1-k+2ke^{2i\pi\frac{l+s_1}{L}}-(1+k) e^{2i\pi\frac{l}{L}}\Big{)},\nonumber\\
N&=&-2-e^{2i\pi\frac{l+s_1}{L}}(-1+k)+e^{2i\pi\frac{s_1}{L}}(1+k),\nonumber
\end{eqnarray}
with the $k$ given by
\begin{eqnarray}
 k&=&1+2\frac{\sin[\frac{\pi s_1}{L}]\sin[\frac{\pi s_2}{L}]-\sqrt{\sin[\frac{\pi s_1}{L}]\sin[\frac{\pi s_2}{L}]\sin[\frac{\pi (s_1+l)}{L}]
\sin[\frac{\pi (s_2+l)}{L}]}}{\sin[\frac{\pi l}{L}]\sin[\frac{\pi (l+s_1+s_2)}{L}]}.
\end{eqnarray}
Then geometric part of the partition function can be derived as
\begin{eqnarray}\label{finite size geometric}
\frac{\delta \ln Z^{geom}_{\alpha}}{\delta l}=-i\pi c \frac{P-\alpha^2Q\mathcal{K}^2(1-k^2)}{\alpha R \mathcal{K}^2(1-k^2)}
\end{eqnarray}
with 

\begin{eqnarray}
P=2\pi^2\Big{(}-4k(e^{2\pi i\frac{l+s_1}{L}}-1)+(1+k)^2e^{2\pi i\frac{s_1}{L}}(e^{2\pi i\frac{l}{L}}-1)^2\Big{)},\nonumber\\
Q=(1+6k+k^2)\times\nonumber\\
\Big{(}-2(k-1)^2e^{2\pi i\frac{l+s_1}{L}}-4k-4ke^{4\pi i\frac{l+s_1}{L}}+(1+k)^2e^{2\pi i\frac{s_1}{L}}+(1+k)^2e^{2\pi i\frac{2l+s_1}{L}}\Big{)},\nonumber\\
R=48Lk(1+k)^2(-1+e^{\frac{2i\pi l}{L}})(-1+e^{\frac{2i\pi s_1}{L}})(-1+e^{\frac{2i\pi (s_1+l)}{L}}).\nonumber
\end{eqnarray}

Different limit of the above formula has been discussed in \cite{Rajabpour2016}.

\subsection{Infinite system in the finite temperature}

When the system is infinite but at finite temperature the slits are in the direction of the axes of the cylinder. This means that
one can derive the formulas in this case by just substituting $L$ with $i\beta$ in the formulas of the previous section.



\end{appendices}

\end{document}